\documentclass [prb,preprint,notitlepage,superscriptaddress,floatfix] {revtex4-1}
\usepackage{amsmath}
\usepackage{graphicx}
\usepackage{lmodern}
\usepackage{amsmath}
\usepackage{hyperref}

\usepackage{color}
\usepackage{amssymb}
\usepackage{bm}
\usepackage{ulem}
\newcommand{\beq} {\begin{equation}}
\newcommand{\eeq} {\end{equation}}
\newcommand{\bea} {\begin{eqnarray}}
\newcommand{\eea} {\end{eqnarray}}
\newcommand{\be} {\begin{equation}}
\newcommand{\ee} {\end{equation}}
\renewcommand{\(}{\left(}
\renewcommand{\)}{\right)}

\DeclareMathOperator{\sgn}{sgn}

\DeclareMathOperator{\Ree}{Re}
\DeclareMathOperator{\Imm}{Im}

\begin{document}

\title {The special role of the first Matsubara frequency for superconductivity near a quantum-critical point --
the non-linear gap equation below $T_c$ and spectral properties in real frequencies}
\author{Yi-Ming Wu}
\affiliation{School of Physics and Astronomy and William I. Fine Theoretical Physics Institute, University of Minnesota, Minneapolis, MN 55455, USA}
\author{Artem Abanov}
\affiliation{Department of Physics, Texas A\&M University, College Station,  USA}
\author{Yuxuan Wang}
\affiliation{Department of Physics, University of Florida, Gainesville, FL 32611,  USA}
\author{Andrey V. Chubukov}
\affiliation{School of Physics and Astronomy and William I. Fine Theoretical Physics Institute,
University of Minnesota, Minneapolis, MN 55455, USA}
\date{\today}

\begin{abstract}
 Near a quantum-critical point in a metal a
 strong fermion-fermion interaction, mediated by a soft boson, destroys fermionic coherence and also gives rise to an attraction in one or more pairing channels. The two tendencies compete with each other, and in a class of large $N$ models, where the tendency to incoherence is parametrically stronger, one would naively expect an incoherent (non-Fermi liquid)  normal state behavior to persist down to $T=0$. However, this is not the case for quantum-critical systems described by Eliashberg theory. In such systems, non-Fermi liquid part of the self-energy $\Sigma (\omega_m)$ is large for a generic  Matsubara frequency $\omega_m = \pi T(2m+1)$, but vanishes for fermions with  $\omega_m = \pm \pi T$, while the pairing interaction between fermions with these two frequencies remains strong.
    It has been shown [Y. Wang {\it et al} PRL 117, 157001 (2016)] that this peculiarity gives rise to a non-zero $T_c$, even at large $N$, when superconductivity is not expected
     from scaling analysis.  We consider the system behavior below $T_c$ and contrast the conventional case, when $\omega_m = \pm \pi T$ are not special, and the case when the pairing is induced by fermions with $\omega_m = \pm \pi T$.   We obtain the solution of the non-linear  gap equations in Matsubara frequencies  and then convert to real frequency axis and obtain the spectral function $A(\omega)$ and the density of states $N(\omega)$.  In a conventional BCS-type superconductor $A(\omega)$ and $N(\omega)$ are peaked at the gap value $\Delta (T)$, and the peak  position shifts to a smaller $\omega$ as temperature increases towards $T_c$, i.e. the gap ``closes in".  We show that in a situation when superconductivity is induced by fermions with $\omega_m = \pm \pi T$, the peak  $N(\omega)$ remains at a finite frequency even at $T =T_c-0$, the gap just ``fills in".  The spectral function $A(\omega)$ either shows almost the same ``gap filling" behavior as the density of states, or its peak position shifts to zero frequency already at a finite $\Delta$ ("emergent Fermi arc" behavior), depending on the strength of the thermal contribution. We compare our results with the data for the cuprates and argue that ``gap filling" behavior holds in the antinodal region, while the ``emergent Fermi arc" behavior holds in the nodal region.
\end{abstract}
\maketitle

\section{ Introduction.}

The pairing near a quantum-critical point (QCP) in a metal is a fascinating subject due to highly non-trivial  interplay between superconductivity  and non-Fermi liquid (NFL) behavior ~\cite{combescot,Bergmann,*Bergmann2,*ad,Marsiglio_88,*Marsiglio_91,Karakozov_91,nick_b,acf,acs,*acs2,finger_2001,son,*son2,sslee,*sslee2,subir,*subir2,moon_2,
max,*max2,senthil,raghu,*raghu2,*raghu3,*raghu4,*raghu5,scal,*scal2,book1,review4,review3,max_last,efetov,*efetov,wang,*wang2,raghu_15,Wang2016,steve_sam,
tsvelik,*Rice,vojta,mack,varma,matsuda,
metzner,*Metzner18,berg,*berg_2,*berg_3,kotliar,*kotliar2,review3,*tremblay_2,georges,*georges2,khvesh,Kotliar2018,*we_last_D}.  In most cases, the dominant interaction between low-energy fermions near a QCP is   mediated by critical fluctuations of the order parameter. In dimensions $D \leq 3$, this interaction gives rise to a singular fermionic self-energy, and a coherent Fermi-liquid behavior get destroyed below a certain temperature $T_{coh}$, either on the full Fermi surface~\cite{pepin,max, raghu, q=0,*q=01,*q=02} or in the hot regions~\cite{acf, acs,*acs2,finger_2001, acn, max, wang,*wang2, 2kf,*2kf2,*2kf3}.
 The same interaction, however, also mediates fermion-fermion interaction in the particle-particle channel.
  The electron-mediated interaction is positive (repulsive), but it depends on both momentum and frequency and generally has at least one attractive component  ($d-$wave for antiferromagnetic QCP,  $p-$wave for a ferromagnetic QCP, $s, p, d$-wave for a nematic QCP, Ref. \cite{review,*review2,*review3,*review4})
  If this  system becomes superconducting below some finite $T_c$, the range of  NFL behavior shrinks to $T_{coh} > T > T_c$, and even vanishes when $T_c > T_{coh}$~\cite{max_last}. A naked quantum-critical $T=0$ behavior can only be observed either if the pairing interaction is repulsive, or if
  fermionic incoherence prevents superconductivity to develop down to  $T=0$.

\begin{figure}
	\begin{center}
		\includegraphics[width=12cm]{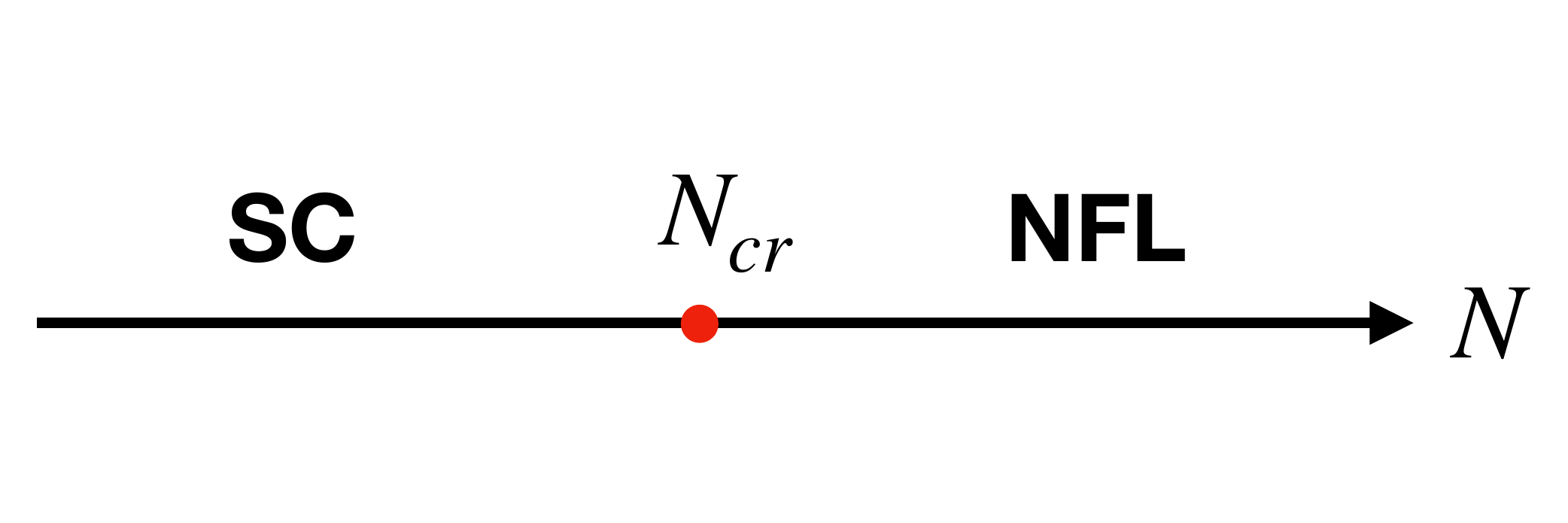}
		\caption{The $T=0$  phase diagram of an itinerant QC  model with fermion-fermion interaction mediated by a critical boson with dynamical propagator $\chi (\Omega_m) = (g/|\Omega_m|)^\gamma$, where $0<\gamma <1$.  The original model with $N=1$ has been extended to $N >1$ in such a way that the pairing interaction is reduced by $1/N$, while the interaction in the particle-hole channel (the one which gives rise to NFL behavior in the normal state) remains intact. The critical $N_{cr}  = N_{cr} (\gamma) >1$ separates the regions of superconductivity at $N < N_{cr}$  and NFL normal state behavior at $N > N_{cr}$.}\label{fig:phasedigram_T=0}
	\end{center}
\end{figure}

\begin{figure}
  \begin{center}
    \includegraphics[width=12cm]{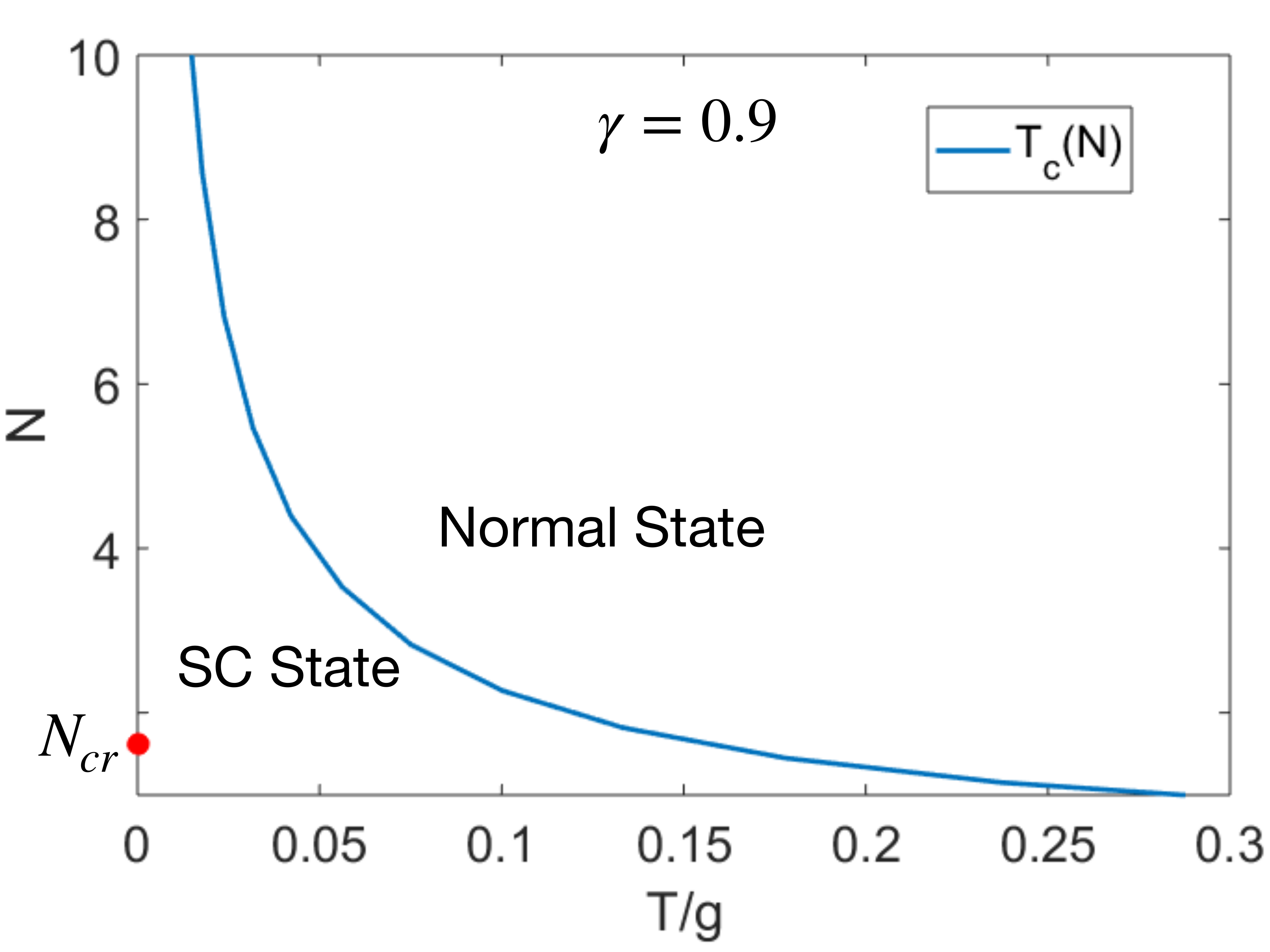}
    \caption{The onset temperature of superconductivity, $T_c (N)$, in the QC model, extended to $N >1$.  We set $\gamma =0.9$.
The line $T_c (N)$  by-passes $N_{cr}$ (the red dot).  At large $N$,  $T_c (N) \propto 1/N^{1/\gamma}$.}\label{fig:Tc_1}
  \end{center}
\end{figure}

In all known physical quantum-critical (QC) models of fermions, superconducting $T_c$ is finite~\cite{nick_b,acf,max_last,raghu_15,Wang2016,khvesh,Kotliar2018,*we_last_D}.
This can be interpreted as an evidence that the
 tendency to pairing is stronger than  towards incoherent, NFL behavior.  The situation can potentially be reversed
 if the interaction in the pairing channel is somehow reduced compared to that in the particle-hole channel. This can be achieved by either
  modifying the momentum dependence of the interaction mediated by critical fluctuations   to reduce the partial pairing component  in
    the cannel, where it is attractive, or by keeping the interaction  intact but extending the model to
 an $SU(N)$ global symmetry~\cite{raghu_15}  (the original model corresponds to $N=1$).  Under this extension,  the pairing interaction get reduced by $1/N$, but the self-energy stays intact~\cite{raghu_15}.  In both cases, the functional form of equation for the (frequency dependent) pairing vertex in the attractive channel  does not change, but
  the magnitude of the eigenvalue needed for superconductivity gets larger.
  The analysis of a large-$N$ QC  model at $T=0$ shows~\cite{raghu_15,Wang2016}
   that there exists a critical $N_{cr}$, separating a superconducting region  at $N < N_{cr}$ and a region of a $T=0$ NFL normal state behavior  at $N > N_{cr}$ (see Fig.~\ref{fig:phasedigram_T=0}).
    A conventional reasoning in this situation would be that
   the superconducting $T_c(N)$
      terminates at $T=0$, $N = N_{cr}$,
      and vanishes for $N>N_{cr}$.
  However, numerical studies of
   large-$N$ QC models yield a different result~\cite{Wang2016} -- $T_c$ remains finite at any $N$, and the critical line $T_c (N)$ by-passes $N = N_{cr}$, and $T_c (N)$ remains finite at all $N$ (see Fig.~\ref{fig:Tc_1}).

   This unusual behavior was argued  in Ref.~\onlinecite{Wang2016} to be the consequence of the special form of Matsubara fermionic self-energy  $\Sigma (\omega_m)$ at the two lowest Matsubara frequencies: $\omega_m = \pi T$ and $\omega_m = - \pi T$.  Namely, in Eliashberg theory $\Sigma (\pm \pi T)$ only contains the  self-action term (thermal contribution to $\Sigma (\omega_m)$ from zero bosonic Matsubara frequency), all other contributions cancel out. The thermal piece
     in $\Sigma (\omega_m)$  comes from scattering with zero frequency and finite momentum transfer and mimics the scattering by impurities. The same thermal scattering also contributes to the pairing vertex $\Phi (\omega_m)$.   For
   spin-singlet pairing, the two contributions cancel out in  equation  for the gap function $\Delta (\omega_m) = \Phi (\omega_m)/(1 + \Sigma (\omega_m)/\omega_m)$ by Anderson's theorem~\cite{anderson, abr-gor}.  As a consequence, fermions with   $\omega_m = \pm \pi T$ can be treated for the pairing as free quasiparticles.  Meanwhile the pairing interaction between fermions with $\omega_m = \pi T$ and $\omega_m = - \pi T$ remains strong. This strong interaction, not countered by the self-energy, gives rise to the emergence of $\Delta (\pm \pi T)$ below a certain $T_c (N)$, which remains finite for all values of $N$. A finite  $\Delta (\pm \pi T)$ then induces non-zero $\Delta (\omega_m)$ at other Matsubara frequencies, for which the self-energy without self-action is strong.

\begin{figure}
	\begin{center}
		\includegraphics[width=12cm]{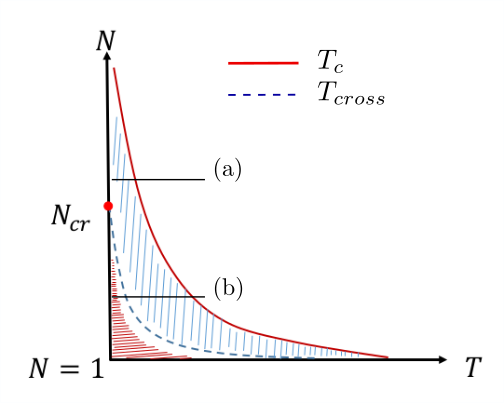}
		\caption{A schematic phase  diagram of our QC model, extended to $N >1$, for some  $\gamma<1$.
   The solid line is the onset temperature for superconductivity, $T_c (N)$. The dashed line  marks the crossover from the behavior similar to a BCS superconductor at a lower $T$ to the novel behavior at a higher $T$, in which superconducting order does not provide a substantial feedback effect on the fermionic self-energy, and it largely remains the same as in the normal state.   In this region, the spectral function $A(\omega)$ and the  DOS $N(\omega)$ are functions of $\omega/T$ rather than of $\omega/\Delta (T)$. The critical $N_{cr}$ separates  superconducting and normal states at $T=0$. This phase diagram has been obtained within the Eliashberg theory and does not include phase fluctuations.
    The latter  likely destroy long-range superconducting  order in some $T$ range below $T_c (N)$. Our results for $N(\omega)$ and $A(\omega)$ above the crossover line should survive in this range as they do not rely on the existence of a long-range superconducting order.  }\label{fig:phasedigram}
	\end{center}
\end{figure}

 In this communication we extend the analysis of superconductivity induced by first fermionic Matsubara frequencies to $T < T_c (N)$.   We  argue that, although $T_c (N)$ by-passes $N= N_{cr}$, there is a crossover  in the system behavior at  $T_{cross} (N) < T_c (N)$. The crossover line $ T_{cross} (N)$ originates at $T=0$ for $N = N_{cr}$ and ends at $T_{cross} \leq  T_c$  for the physical case $N=1$. At $T_{cross} (N) < T < T_c (N)$, superconductivity can be viewed as induced by fermions with $\omega_m = \pm \pi T$, at smaller $T < T_{cross} (N)$ fermions with all $\omega_m$ contribute to superconductivity, and the ones with $\omega_m = \pm \pi T$ are no longer special.  We show the schematic phase diagram in  Fig. \ref{fig:phasedigram}.

    We analyze the evolution of the gap $\Delta (\omega_m)$ below $T_c (N)$ at $N > N_{cr}$ and $N < N_{cr}$ and then convert from Matsubara to real frequencies and analyze the behavior of $\Delta (\omega)$, the spectral function at the Fermi surface $A(\omega)$, and the  density of states  (DOS) $N(\omega)$.   We argue that the system behavior below $T_c (N)$ is different for $N > N_{cr}$ and $N < N_{cr}$
    (see the paths (a) and (b) in Fig.\ref{fig:phasedigram}). 
    At $N > N_{cr}$ it is qualitatively different from BCS.  At $N < N_{cr}$, the system behavior is similar to a BCS superconductor for $T < T_{cross} (N)$ and to that at $N > N_{cr}$ for $T > T_{cross} (N)$.

  Along the Matsubara axis, we find that  at large $N > N_{cr}$, the pairing vertex $\Phi (\omega_m)$ is smaller than  $\Sigma (\omega_m)$ for all temperatures and all Matsubara frequencies, including $\pm \pi T$. In fact, $\Sigma (\omega_m)$ with $m \neq 0, -1$ remains essentially the same as in the normal state, i.e., the feedback effect from superconductivity on this self-energy is weak.  The  self-energy  $\Sigma (\pm \pi T)$ becomes finite below $T_c (N)$, but remains smaller by $1/N$ than $\Sigma (\omega_m)$ at other Matsubara frequencies. Still, it is larger by $\sqrt{N}$ than $ \Phi (\pm \pi T)$.   We show that in this situation,
  $\Delta (\omega_m, T)$ is monotonic as a function of $\omega_m$, with the largest value at $\pm \pi T$, but non-monotonic as a function of temperature, i.e.,
    $\Delta (\pi T)$   first increases when $T$ decreases below  $T_c (N)$, then passes through a maximum and eventually vanishes at $T=0$.
      At $N < N_{cr}$,  $\Delta (\pi T)$ becomes  non-zero at $T=0$, and  its magnitude increases as $N$ gets progressively smaller than $N_{cr}$.  At $N \leq N_{cr}$,  the temperature dependence of $\Delta (\pi T)$ is still non-monotonic, with the maximum at a finite $T$.  At smaller $N$, the maximum becomes more shallow, and at $N \gtrsim 1$,  $\Delta (\pi T)$ monotonically increases with decreasing $T$.

We use the results along the Matsubara axis as an input and obtain the behavior of $\Phi (\omega)$ and $\Sigma (\omega)$  along real frequency axis. Using these $\Phi (\omega)$ and $\Sigma (\omega)$, we obtain the DOS
 \beq
  N(\omega) = N_0 \Ree\left[\frac{1}{\sqrt{1 - \left(\Phi (\omega)/(\omega + \Sigma (\omega))\right)^2}}\right]
  \label{dos}
  \eeq
  The thermal contributions to $\Phi (\omega)$ and to $\omega + \Sigma (\omega)$ are the same and they cancel out in the DOS, i.e., in the calculations one can replace
 $\Phi (\omega)$ and $\Sigma (\omega)$ by $\Phi^* (\omega)$ and $\Sigma^* (\omega)$, which are the solutions of the Eliashberg equations with thermal contributions explicitly taken out.
 We show that, for  $N > N_{cr}$,
  $N(\omega)$ is finite for all frequencies, including $\omega =0$, and its  dependence on $\omega$  is determined by a universal  scaling function of $\omega/T$.   As the consequence,
  the frequency at which $N(\omega)$  has a maximum, linearly increases with increasing $T$. As
   $T$ approaches $T_c$ from below,  DOS ``fills in", i.e.,  the $N(\omega)$ approaches $N_0$,  but the position of the maximum in $N(\omega)$
    remains at a finite frequency.

\begin{figure}
      \begin{center}
        \includegraphics[width=12cm]{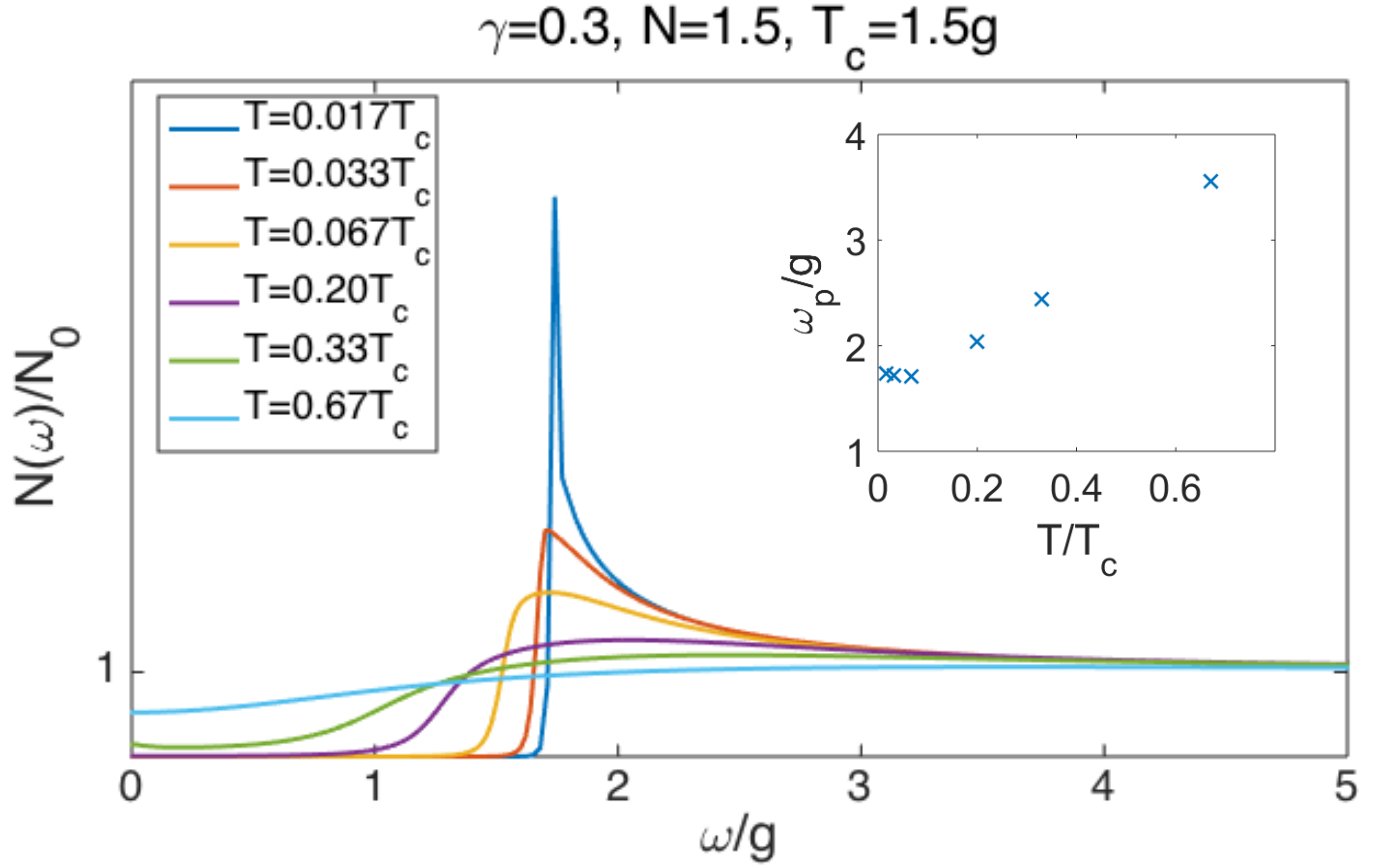}
        \caption{A representative of our results for the DOS.  We set $\gamma =0.3$ and $N=1.5$, which is smaller than $N_{cr}$ for this $\gamma$. 
        At low $T < T_{cross} \sim 0.1 T_c$, the DOS has a peak at $\omega \approx \Delta (T)$, and the peak frequency decreases as temperature increases, i.e. the gap in the DOS closes.  At $T > T_{cross}$ the DOS flattens up  with increasing $T$ (the gap fills in).
         In this $T$ range the  maximum in the DOS is located at $\omega_p \sim T$, which increases with increasing $T$. }\label{fig:DOSsummary}
      \end{center}
    \end{figure}

     At $N <N_{cr}$, the DOS $N(\omega)$ at the lowest $T < T_{cross} (N)$ displays a sharp gap, , i.e., it nearly vanishes at $\omega < \omega_0$, where $\omega_0$ is roughly equal to $\Delta (0)$ (more exactly, $\omega_0$ is the solution of $\Delta (\omega_0) = \omega_0$).
       As $T$ increases, the position of the maximum in the DOS initially shifts to a lower frequency, as  in a BCS superconductor, because $\Delta (0)$ gets smaller with increasing $T$, i.e.,  the gap in the DOS ``closes in" with increasing temperature. However, once temperature exceeds  $T_{cross} (N)$, this behavior changes and becomes the same as for larger $N$, i.e., at these $T$ the
    the position of the maximum in DOS shifts to a higher frequency with increasing $T$ and remains finite at $T = T_c(N) -0$, i.e., the DOS ``fills in" with increasing $T$.   We emphasize that these two distinct regimes of system behavior are present also in the original physical model with $N=1$.
    In this respect, the extension to $N >1$ is just a convenient way to understand the origin of such behavior by extending the regime in which superconductivity is generated by fermions with $\omega = \pm \pi T$.  A representative of our results for the DOS is shown in Fig.\ref{fig:DOSsummary}
\begin{figure}
       \begin{center}
         \includegraphics[width=15cm]{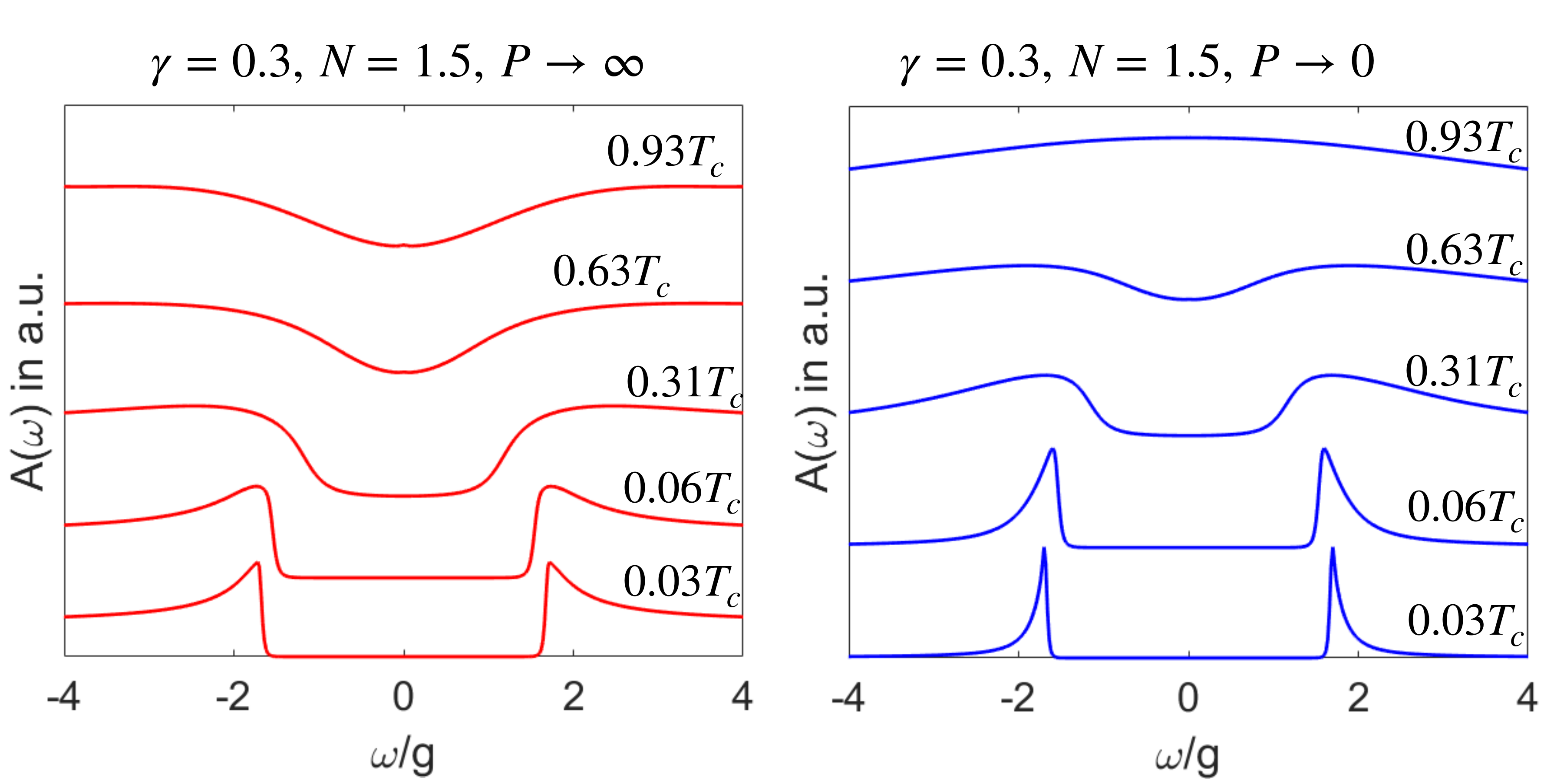}
         \caption{A representative of our results for the spectral function $A(\omega)$ for $\gamma =0.3$ and $N =1.5$ ($N < N_{cr}$).
     Left panel is for the case when thermal contribution to $A(\omega)$  is strong, right panel is for the case when it is weak (in our notations, the cases $P \to \infty$ and $P \approx 0$, respectively).  In both panels,
         $A(\omega)$ at low $T< T_{cross}$ has well pronounced peaks at $\omega =\pm \Delta (T)$. The peak frequency decreases with increasing $T$.  At $T > T_{cross}$, the peaks disappear, and the spectral function shows a dip, when $P$ is large, and a single peak at $\omega =0$, when $P$ is small. For cuprate superconductors, we associate the spectral function in the right panel with that of antinodal fermions, and the one in the left panel with the spectral function of fermions in near-nodal, Fermi arc region.}\label{fig:Asummary}
       \end{center}
     \end{figure}
     The behavior of the spectral function  is more involved because in $A(\omega)$  the thermal contribution does not  cancel out.
    The expression for $A(\omega) = - (1/\pi) \Imm[G(k_F, \omega)]$ at $\omega >0$ is (see Eq.\eqref{sf_1_a} below)
   \begin{equation}
  \begin{aligned}
    A(\omega) &=   \frac{1}{\pi} \Imm \left[\frac{ \omega+\Sigma^* (\omega)}{(\omega+\Sigma^* (\omega))^2 -\Phi^*(\omega)^2} L(\omega)\right]\\
    L(\omega)&=\frac{\sqrt{(\omega+\Sigma^* (\omega))^2-\Phi^*(\omega)^2}}{iP + \sqrt{(\omega+\Sigma^* (\omega))^2-\Phi^*(\omega)^2}}\label{sf_1}
  \end{aligned}
\end{equation}
   where frequency-independent $P =P(T)$ describes the thermal contribution to self-energy,
   and $\Sigma^*(\omega)$, $\Phi^*(\omega)$ are obtained from $\Sigma(\omega)$, $\Phi(\omega)$ by excluding thermal contributions (see \eqref{eq:anacon} and the Appendix for more details).
     For large $P$, $A(\omega) \propto N(\omega)$, i.e., the spectral function displays the same crossover from ``gap closing" to ``gap filling" as the DOS.  For smaller $P$, when the term next to $P$ in (\ref{sf_1}) is larger than $P$, $A(\omega)$ at $T < T_{cross}$ shows two sharp peaks at $\omega = \pm \omega_0$. At temperatures above $T_{cross}$, the two peaks merge, and $A(\omega)$ develops a maximum at $\omega=0$, like in the normal state.  A representative of  our results for $A(\omega)$ is shown in   Fig.\ref{fig:Asummary}.

   The transformation from  ``gap closing" to ``gap filling" behavior in the DOS  has been observed in several superconducting materials, most notably the cuprates~\cite{DOS,dessau,kaminski,*Kaminski2,kanigel,*kanigel2,*kanigel3,norman_review,shen,*shen2,shen3,shen4,*kordyuk2,hoffman,Peng2013}
    The spectral function in the cuprates shows the same behavior as the DOS in the antinodal regions, where the fermionic incoherence is the strongest, and the $d-$wave gap is the largest.    In the regions near the Brillouin zone diagonals, the symmetrized spectral function  has peaks at a finite frequency $\pm \omega_0$ at low temperatures, and a single maximum at $\omega =0$ at higher temperatures.  The angular range in which the system displays a single peak above a certain $T$ is termed as a Fermi arc~\cite{norman_review}.

\begin{figure}
	\begin{center}
		\includegraphics[width=8.5cm]{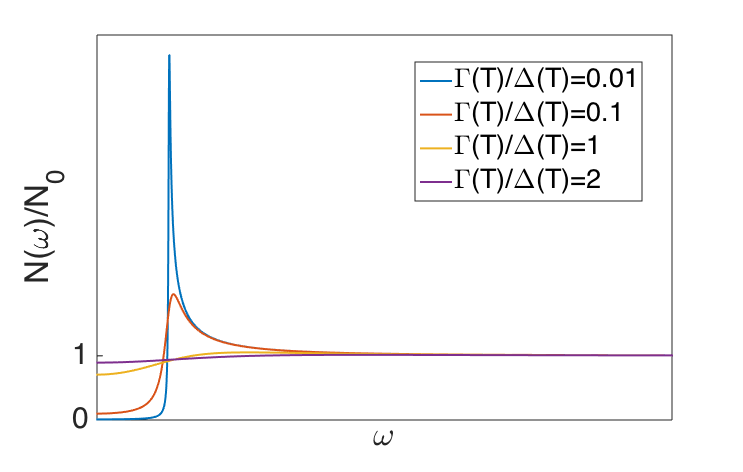}\includegraphics[width=8.5cm]{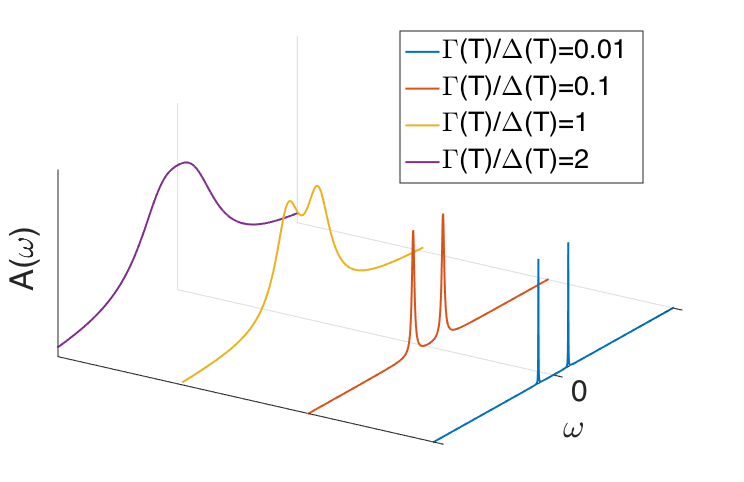}
		\caption{The DOS $N(\omega)$ and the spectral function $A(\omega)$ in a dirty BCS superconductor,  from Eq. \eqref{a_1} and Eq. \eqref{sf_2}.}\label{fig:simpleDOS}
	\end{center}
\end{figure}

  The crossover from ``gap closing" to ``gap filling" in the DOS and in $A(\omega)$ in the antinodal regions, and the crossover from two peaks to a single peak in $A(\omega)$ in the nodal regions, have been phenomenologically described
  by assuming that the pairing vertex $\Phi^* (\omega) = \Delta (T)$, as in a BCS superconductor, and  $\Sigma^* (\omega) = i\Gamma (T)$ (Refs. \cite{imp_general,*imp_general2,*imp_general3,imp_cuprates,*imp_cuprates2,*imp_cuprates3,*imp_cuprates4,*imp_cuprates5,elihu})
  Under these approximation,  the DOS becomes
 \beq
 N(\omega) = N_0 \Ree \left[\frac{1}{\sqrt{1- \(\frac{\Delta(T)}{\omega + i \Gamma (T)}\)^2}}\right]
 \label{a_1}
 \eeq
 Without $\Gamma (T)$, the DOS vanishes at $\omega <\Delta$ and is singular at $\omega = \Delta +0$. A non-zero $\Gamma (T)$ makes $N(\omega)$ continuous and non-zero down to $\omega=0$.  Furthermore, the position of the peak in $N(\omega)$ shifts to a higher frequency from $\omega = \Delta (T)$
    (see Fig.\ref{fig:simpleDOS})
   At vanishing $\Delta (T)$  the peak in
 $N(\omega) \approx N_0 \left(1 + \frac{1}{2}\Delta^2 \Ree\left[\frac{1}{(\omega + i \Gamma)^2}\right]\right)$
 remains at a finite $\omega = \sqrt{3} \Gamma$.
  In other words, the magnitude of the deviation of $N(\omega)$ from $N_0$ is set by $\Delta^2$, while its frequency dependence is set by $\Ree\frac{1}{(\omega + i \Gamma)^2}$ and does not depend on $\Delta$.  If one additionally sets $\Gamma = O(T)$, as in marginal FL theory,
  one obtains that the position of the maximum in the DOS increases linearly with $T$ near $T = T_c$, when $\Gamma > \Delta(T)$.  The same phenomenolgical model with $\Gamma (T) \propto T$ was used~\cite{elihu} to explain Fermi arcs (assuming that the thermal contribution can be neglected).
   Indeed, at  $P=0$, we have from  (\ref{sf_1})
   \bea
   A(\omega) &=& -\frac{1}{\pi} \Imm\left[\frac{\omega + i \Gamma (T)}{\left(\omega + i \Gamma (T)\right)^2 -\Delta^2 (T)}\right] \nonumber \\
   & = &\frac{1}{\pi} \frac{\omega^2 + \Delta^2 (T) +\Gamma^2 (T)}{\left(\omega^2 -  \Delta^2 (T) -\Gamma^2 (T)\right)^2 + 4 \omega^2 \Gamma^2 (T)}
   \label{sf_2}
   \eea
This spectral function has two separate peaks at positive and negative $\omega$ at  $\Gamma(T) < \sqrt{3} \Delta (T)$, and a single maximum at $\omega =0$ at $\Gamma(T) > \sqrt{3} \Delta (T)$
(Fig.\ref{fig:simpleDOS})

This phenomenon when $N(0)$ becomes finite is known as ``gapless superconductivity".  It was originally found by Abrikosov and Gorkov in their analysis of an s-wave BCS  superconductor with magnetic impurities~\cite{abr_gor_1}. At $T=0$, gapless superconductivity exists in a finite parameter range before magnetic impurities destroy superconductivity.  Several researchers later argued~\cite{maki,*maki2,*maki3,*maki4}
  that any phonon-mediated $s$-wave superconductor at a finite $T$ is a gapless superconductor due to scattering on thermally excited phonons, although in practice $\Gamma$ due to such scattering is extremely small at small coupling. For  electronically-mediated superconductivity in a clean metal, self-energy $\Sigma (\omega)$ in the normal state contains the imaginary part. In  a superconducting state, the imaginary part of $\Sigma (\omega)$ is reduced at $\omega <\Delta$ due to the reduction of the phase space for low-energy scattering. This holds for any symmetry of the gap function and gives rise to peak-dip-hump feature of the spectral functions, studied extensively in the cuprates~\cite{norman_review,fink,abanov_ch,eschrig}
    As long as $T$ is finite, $\Sigma'' (\omega =0)$ remains non-zero, but at low $T$ it  gets substantially reduced compared to its value at $T = T_c$.  Numerical analysis of Eliashberg equations  for several models of magnetically-induced $d$-wave superconductivity~\cite{acs,acn,scal}
  and for strong coupling (small Debye frequency) limit of electron-phonon superconductivity~\cite{combescot,Bergmann,*Bergmann2,*ad,Marsiglio_88,*Marsiglio_91,Karakozov_91}
  did find
  that $\Sigma'' (0)$ rapidly increases at $T$ near $T_c$, and the maximum in the theoretical DOS  shifts up from $\Delta (T)$ and remains at a finite
 frequency at $ \sim T_c$, where $\Delta (T)$ vanishes in the Eliashberg theory.  This is roughly consistent with the phenomenology of Eq.~(\ref{a_1}), although temperature variation of the peak position in the DOS has not been explicitly verified.

    We view our results as the microscopic explanation of the rapid increase of $\Sigma'' (0)$ above a certain $T$ within the superconducting state and
      the related transformation from ``gap closing" to ``gap filling" behavior of the DOS and the spectral function at large $P$ (and the transformation from gap to Fermi arc behavior at smaller $P$).    To reiterate --  we argue that the conventional ``gap closing" behavior occurs at $T <T_{c,1}$, while the ``gap filling" behavior occurs at $T_{c,1} < T < T_c$, where in Matsubata formalism  the pairing is induced by two lowest Matsubata frequencies at which self-energy vanishes, and would not happen  if fermions with these two frequencies were eliminated from the gap equation.

 The issue which we do not address here is the role of pairing fluctuations. We remind the reader that Eliashberg theory neglects phase and amplitude fluctuations of the pairing vertex and in this respect should be treated as effectively a ``mean-field" theory. It is very likely  that in some range below Eliashberg $T_c$ fluctuations  destroy long-range superconducting order, and the actual $T_{c,act} < T_c$.  Our results, that
   the DOS $N(\omega)$ is non-zero at all $\omega$ and
   the position of its  maximum  increases with $T$,
    should survive at $T_{c,act} < T <T_c$  as our reasoning only explores  the fact that in this $T$ range the feedback from the pairing on the fermionic self-energy is weak. Gap fluctuations reduce this feedback even further.  The same holds for the spectral function both at large $P$  and at smaller $P$.
    In other words, our theory describes gap filling and Fermi arcs in the pseudogap region.
    Still, to fully address the issue of gap fluctuations one needs to go beyond Eliashberg theory and analyze the full Luttinger-Ward functional~\cite{lw}.

The paper is organized as follows. In Sec.~\ref{sec:model} we present the microscopic model of pairing mediated by a gapless boson with $\chi (\Omega_m) = (g/|\Omega_m|)^\gamma$ (the $\gamma$-model) and its extension to $N >1$.  We present the set of coupled Eliashberg equations along Matsubara axis for the pairing vertex $\Phi (\omega_m)$ and the fermionic self-energy $\Sigma (\omega_m)$ and  summarize, in Sec.~\ref{sec:linearized}  earlier results of the  analysis of the linearized equation for $\Phi (\omega_m)$. At $T=0$, these results show that there exists the critical $N_{cr}$, separating  the superconducting state at $N < N_{cr}$ and the normal state at $N > N_{cr}$. At $ T >0$, these  calculations show that  superconductivity emerges for all $N$, below a certain $T_s (N)$ which only vanishes at $N = \infty$.   In Sec.~\ref{sec:largeN} we discuss the system behavior at  $N > N_{cr}$, first in Matsubara frequencies, in Sec.~\ref{sec:largeNMatsubara}, and then in real frequencies, in Sec.~\ref{sec:largeNreal}.  We present the  analytical solution of the Eliashberg equations at large $N$ and  discuss the behavior of the gap, the Free energy and the specific heat, the DOS, and the spectral function.  In Sec.~\ref{sec:smallN} we discuss system behavior at $N < N_{cr}$, again first in Matsubara frequencies, in Sec.~\ref{sec:smallNMatsubara}, and then in real frequencies, in Sec.\ref{sec:smallNreal}.  In Sec.~\ref{sec:conclusions} we summarize our results and compare them with the experimental data.

\section{The model.}
\label{sec:model}

 We  consider a model of itinerant fermions  at the onset
of a  long-range order in either spin or charge channel.  At the critical point the propagator of a soft boson
 becomes massless and
 mediates singular interaction between fermions. We follow earlier works~\cite{acf,acs,moon_2,max,senthil,scal,efetov,max_last,raghu_15,haslinger,Wang2016,Kotliar2018}  and assume that this interaction as attractive in at least one pairing channel and that
  bosons can be treated as slow modes compared to fermions, i.e.,  the Eliashberg approximation is valid.
Within this approximation
 one can explicitly integrate over the momentum component perpendicular to the Fermi surface (for a given pairing symmetry) and reduce the
   pairing problem  to a set of coupled integral equations for frequency dependent self-energy $\Sigma (\omega_m)$
   and the pairing vertex $\Phi (\omega_m)$ for fermions  on the Fermi surface, with effective frequency-dependent dimensionless interaction $\chi(\Omega) = (g/|\Omega|)^\gamma$  (the $\gamma$-model, Refs. \cite{acs,acf,moon_2,Wang2016,Kotliar2018}). This interaction simultaneously gives rise to NFL form of the self-energy in the normal state and to pairing.
     The equations we analyze are
    \bea \label{eq:gapeq}
    \Phi (\omega_m) &=&
    \pi T  g^\gamma \sum_{m'} \frac{\Phi (\omega_{m'})}{\sqrt{{\tilde \Sigma}^2 (\omega_{m'}) +\Phi^2 (\omega_{m'})}}
    ~\frac{1}{|\omega_m - \omega_{m'}|^\gamma}, \nonumber \\
     {\tilde \Sigma} (\omega_m) &=& \omega_m
   +  g^\gamma
    \pi T \sum_{m'}  \frac{{\tilde \Sigma}(\omega_m)}{\sqrt{{\tilde \Sigma}^2 (\omega_{m'})  +\Phi^2 (\omega_{m'})}}
    ~\frac{1}{|\omega_m - \omega_{m'}|^\gamma} \nonumber\\
\eea
 where ${\tilde \Sigma}(\omega_{m}) = \omega_m + \Sigma (\omega_m)$.
  Note that we define $\Sigma (\omega_m)$ as a real function of frequency, i.e., without the overall factor of $i$.  The self-energy along Matsubara axis, related by Kramers-Krong (KK)  formula to $\Sigma^{''} (\omega)$ along the real frequency axis, does contain the factor $i$.
  The  superconducting gap $\Delta (\omega_m)$ is defined as a real variable
\beq
 \Delta (\omega_m) = \omega_m  \frac{\Phi (\omega_m)}{{\tilde \Sigma} (\omega_m)}
  \label{ss_1}
  \eeq
   The equation for $\Delta (\omega)$ is readily obtained from (\ref{eq:gapeq}):
   \beq
   \Delta (\omega_m) = \pi T  g^\gamma \sum_{m'} \frac{\Delta (\omega_{m'}) - \Delta (\omega_m) \frac{\omega_{m'}}{\omega_m}}{\sqrt{\omega^2_{m'} +\Delta^2 (\omega_{m'})}}
    ~\frac{1}{|\omega_m - \omega_{m'}|^\gamma}.
     \label{ss_11}
  \eeq
   This equation contains a single function $\Delta (\omega)$, but  for the prize that $\Delta (\omega_m)$ appears on both sides of the equation, which makes (\ref{ss_11}) less convenient for the analysis than Eqs. (\ref{eq:gapeq}).

   The r.h.s. of the equations for $\Phi (\omega_m)$ and $\Sigma (\omega_m)$ contain  divergent pieces from the terms with $m' =m$, i.e., from $\chi (0)$.
    The divergence can be regularized by moving slightly away from a QCP, in which case
   $\chi (0)$ is large but finite.  This term mimics the effect of non-magnetic impurities.  To get rid of the thermal piece in the equations for $\Phi (\omega)$ and $\Sigma (\omega)$, we follow~\cite{msv,acn} and use the same trick as for the derivation of the Anderson theorem for impurity scattering~\cite{agd}
    Namely, we  pull out the term with $m'=m$ from the sum, move it to the l.h.s., and introduce
   \bea
   \Phi^* (\omega_m) &=& \Phi (\omega_m)
   \left(1- Q (\omega_m)\right), \nonumber \\
   {\tilde \Sigma}^* (\omega_m) &=& {\tilde \Sigma} (\omega_m) \left(1- Q (\omega_m)\right) \nonumber \\
   Q (\omega_m) &=&    \frac{\pi T \chi (0)}{\sqrt{{\tilde \Sigma}^2 (\omega_{m}) +\Phi^2 (\omega_{m})}}
   \label{ss_2}
   \eea
    The ratio $\Phi (\omega_m)/ {\tilde \Sigma} (\omega_m) = \Phi^* (\omega_m)/ {\tilde \Sigma}^* (\omega_m)$, hence $\Delta (\omega_m)$, defined in (\ref{ss_1}),  is invariant under $\Phi (\omega_m) \to \Phi^* (\omega_m)$ and ${\tilde \Sigma} (\omega_m) \to {\tilde \Sigma}^* (\omega_m)$.  Using  (\ref{ss_2}), one can easily verify that the equations on $\Phi^* (\omega_m)$ and ${\tilde \Sigma}^* (\omega_m)$  are the same as in (\ref{eq:gapeq}), but without the thermal piece, i.e., the summation over $m'$ now excludes the divergent term with $m' =m$.   The gap function $\Delta (\omega_m)$, defined in (\ref{ss_1})  is equally expressed in terms of $\Phi^* (\omega_m)$ and ${\tilde \Sigma}^* (\omega_m)$, and the gap equation (\ref{ss_11}) preserves its form: the sum over $m'$ now excludes the term with $m'=m$, but this term vanishes anyway because the numerator in the r.h.s. of (\ref{ss_11}) vanishes at $m'=m$.  One can also
     solve  (\ref{ss_2}) backwards and express $\Phi (\omega_m)$ and ${\tilde \Sigma} (\omega_m)$  via
     $\Phi^* (\omega_m)$ and ${\tilde \Sigma}^* (\omega_m)$ as
     \bea
   \Phi (\omega_m) &=& \Phi^* (\omega_m) \left(1 + Q^* (\omega_m) \right), \nonumber \\
   {\tilde \Sigma} (\omega_m) &=& {\tilde \Sigma}^* (\omega_m) \left(1 + Q^* (\omega_m) \right) \nonumber \\
   Q^* (\omega_m) &=&
   \frac{\pi T \chi (0)}{\sqrt{({\tilde \Sigma}^* (\omega_{m}))^2 +(\Phi^* (\omega_{m}))^2}}
   \label{ss_22}
   \eea

Eq. (\ref{eq:gapeq}) describes color superconductivity~\cite{son} ($\gamma = 0_+$, $\chi (\Omega_m) \propto \log{|\omega_m|}$),  spin- and charge-mediated pairing in $D=3-\epsilon$ dimension~\cite{senthil,max_last,raghu_15} ($\gamma = O(\epsilon) \ll 1$),  a 2D pairing ~\cite{2kf}  with  interaction peaked at $2k_F$ ($\gamma =1/4$),   pairing at a 2D nematic/Ising-ferromagnetic QCP~\cite{nick_b,steve_sam,triplet,*triplet2,*triplet3} ($\gamma =1/3$),   pairing at a 2D $(\pi,\pi)$ SDW QCP~\cite{acf,acs,millis_05,wang} and an incommensurate CDW QCP~\cite{ital,*ital2,*ital3,wang_2,*wang_22,*wang23} ($\gamma =1/2$), a 2D pairing  mediated by an undamped  propagating boson ($\gamma =1$), and the strong coupling limit of phonon-mediated superconductivity~\cite{combescot,Bergmann,*Bergmann2,*ad,Marsiglio_88,*Marsiglio_91,Karakozov_91} ($\gamma =2$).  The pairing models with parameter-dependent $\gamma$ have also been considered (Refs. \onlinecite{subir,moon_2}).
 In this communication we consider the set of $\gamma$-models with  $\gamma <1$. The analysis for $\gamma >1$ requires a separate consideration because of the divergence of the normal state self-energy at $T=0$.

The full set of Eliashberg equations for electron-mediated pairing contains also the equation describing the feedback from the pairing on $\chi(\Omega)$, e.g., the emergence of a propagating mode (often called a resonance mode)  in the dynamical spin susceptibility for $d-$wave pairing mediated by antiferromagnetic spin fluctuations.  To avoid additional complications, we do not include this feedback into our consideration.  In general terms, the feedback from the pairing makes bosons less incoherent and can be modeled by assuming that
  $\gamma$ moves towards $\gamma=1$ as $T$ moves down from $T_c$.

The two equations in  (\ref{eq:gapeq}) describe the interplay between two competing tendencies -- the tendency towards superconductivity, specified by $\Phi$,  and the tendency towards incoherent non-Fermi liquid behavior, specified by $\Sigma$.  The competition between the two tendencies is encoded in the fact that $\Sigma$ appears in the denominator of the equation for $\Phi$ and $\Phi$ appears in the denominator of the equation for $\Sigma$.
 Accordingly, a large, non-FL self-energy is an obstacle to Cooper pairing, while once $\Phi$  develops, it reduces the strength of the self-energy, i.e., moves a system back into a FL regime.  Like we said in the Introduction, our goal is to analyze the special role of fermions with Matsubara frequencies $\omega_m  = \pm \pi T$ in the situation when the tendency towards pairing is reduced compared to that for NFL normal state. For this,
we extend the model to matrix $SU(N)$. Under this extension,  the interaction in the particle-hole channel, which gives rise to fermionic self-energy, remains intact, while the interaction in the particle-particle channel  acquires an additional factor $1/N$.
 We emphasize that we extend to $N \neq 1$ {\it after} we invoke the analog of the Anderson theorem and
 eliminate the thermal contributions to $\Phi (\omega_m)$ and $\Sigma(\omega_m)$. In this respect  our approach differs from the
    one in Ref. \cite{raghu_15}
    There, the extension to large $N$ was done without first subtracting the thermal contributions. As a result, at a finite $N$ there appeared additional terms, singular at a QCP, which gave rise to qualitative changes in the system behavior.  In our extension to $N >1$ these additional terms do not appear. Put it more simply,  in our case after the extension the eigenvalues in the pairing channel get multiplied by $1/N$, i.e., a larger magnitude of the original eigenvalue is needed for superconductivity.

The modified equations for $\Phi^* (\omega_m)$ and ${\tilde \Sigma}^* (\omega_m)$ become
 \bea \label{eq:gapeq_1}
   && \Phi^* (\omega_m) =
    \frac{\pi T}{N}  g^\gamma \sum_{m' \neq n} \frac{\Phi^* (\omega_{m'})}{\sqrt{({\tilde \Sigma}^* (\omega_{m'}))^2 +(\Phi^* (\omega_{m'}))^2}}
    ~\frac{1}{|\omega_m - \omega_{m'}|^\gamma}, \nonumber \\
   &&  {\tilde \Sigma}^* (\omega_m) = \omega_m
   +  g^\gamma
    \pi T \sum_{m' \neq m}  \frac{{\tilde \Sigma}^* (\omega_m)}{\sqrt{({\tilde \Sigma}^* (\omega_{m'}))^2  +(\Phi^* (\omega_{m'}))^2}}
    ~\frac{1}{|\omega_m - \omega_{m'}|^\gamma}, \nonumber\\
\eea
 and the equation on $\Delta (\omega_m)$ becomes
 \beq
   \Delta (\omega_m) = \frac{\pi T}{N}  g^\gamma \sum_{m' \neq m} \frac{\Delta (\omega_{m'}) -N  \Delta (\omega_m) \frac{\omega_{m'}}{\omega_m}}{\sqrt{\omega^2_{m'}) +\Delta^2 (\omega_{m'})}}
    ~\frac{1}{|\omega_m - \omega_{m'}|^\gamma}.
     \label{ss_111}
  \eeq

Below we will also need the expression for the Free energy $F_{sc}$ of a
 superconductor, described by the Eliashberg theory. The formula for $F_{sc}$ has been  obtained in Refs. \cite{lw,Bardeen,eliashberg}
 in the studies of phonon-mediated superconductivity ($\gamma =2$ case at finite $\omega_D$ and $N=1$). Extending the results to $\gamma <1$, QC regime, and $N \neq 1$, we obtain
   \begin{widetext}
 \beq
 F_{sc} = - N_0 \left(2\pi T \sum_m \frac{\omega^2_m}{\sqrt{\omega^2_m + \Delta^2_m }} + \pi^2 T^2 g^\gamma
  \sum_{m \neq m'} \frac{\omega_m \omega_{m'} + \frac{1}{N} \Delta_m \Delta_{m'}}{\sqrt{\omega^2_m + \Delta^2_{m} } \sqrt{\omega^2_{m'} + \Delta^2_{m'}}} \frac{1}{|\omega_m-\omega_{m'}|^\gamma}\right)
\label{a_5}
 \eeq
 where $\Delta_m = \Delta (\omega_m)$.  The gap equation (\ref{ss_111}) is obtained from the condition $\delta F_{sc}/\delta \Delta_n =0$
 In the normal state the expression for the Free energy reduces to
 \beq
 F_{n} = - N_0 \left(2\pi T \sum_m |\omega_m|  + \pi^2 T^2 g^\gamma \sum_{m \neq m'} \frac{\sgn \omega_m \sgn \omega_{m'}}{|\omega_m-\omega_{m'}|^\gamma}\right)
 \label{a_6}
 \eeq
 The difference between $F_{sc}$ and $F_n$ at $T=0$ is known as the condensation energy of a superconductor. At a finite $T$,
 \bea
 \delta F = F_{sc} - F_n &=& - 2\pi T N_0 \sum_m  |\omega_m| \left(\frac{1}{\sqrt{1 + D^2_m}}-1\right) \nonumber \\
    && -  N_0 \pi^2 T^2 g^\gamma  \sum_{m \neq m'} \frac{\sgn \omega_m \sgn \omega_{m'}}{|\omega_m-\omega_{m'}|^\gamma} \frac{1 + \frac{1}{N} D_m D_{m'} - \sqrt{1 + D^2_m} \sqrt{1 + D^2_{m'}}}
    {\sqrt{1 + D^2_m } \sqrt{1 + D^2_{m'}}}
\label{a_7}
 \eea
where $D_n = D(\omega_n) = \Delta (\omega_n)/\omega_n$.
 Near $T = T_c$, one can expand $\delta F$ in powers of $\Delta_m$:
 \begin{align}
\delta F  =&  \pi T N_0 \sum_m  |\omega_m| D^2_m - N_0 \pi^2 T^2 g^\gamma  \sum_{m \neq m'} \frac{\sgn \omega_m \sgn \omega_{m'}}{|\omega_m-\omega_{m'}|^\gamma} \left(\frac{1}{N} D_m D_{m'} - \frac{D^2_m + D^2_{m'}}{2}\right) \label{a_8} \nonumber\\
 &+ \frac{3}{4} \pi T N_0 \sum_m  |\omega_m| D^4_m - N_0 \pi^2 T^2 g^\gamma  \sum_{m \neq m'} \frac{\sgn \omega_m \sgn \omega_{m'}}{|\omega_m-\omega_{m'}|^\gamma}\nonumber \\
 &~~~~~~~~~\times\left(\frac{1}{4} D^2_m D^2_{m'} + \frac{3}{8} \left(D^4_m + D^4_{m'}\right) - \frac{1}{2N} D_m D_{m'}  \left(D^2_m + D^2_{m'}\right)\right)
\end{align}
\end{widetext}

  \subsection{Linearized gap equation}
\label{sec:linearized}

  To obtain $T_c$ it is sufficient to consider the linearized gap equation. It is obtained from \eqref{eq:gapeq_1} by setting $\Phi^*$ to be infinitesimally small. Then $\Phi^*(\omega_{m'})$ in the denominators of \eqref{eq:gapeq_1} can be ignored and the self energy $\Sigma^*(\omega_m)$ is approximated by its normal state value. The resulting equations are:
\begin{equation}\label{eq:lineargap}
  \begin{aligned}
    \Phi^*(\omega_m)&=\frac{g^\gamma}{N}\pi T \sum_{m'\neq m}\frac{\Phi^*(\omega_{m'})}{|\omega_{m'}+\Sigma^*(\omega_{m'})|}\frac{1}{|\omega_m-\omega_{m'}|^\gamma}\\
    \Sigma^*(\omega_m)&=g^\gamma \pi T\sum_{m'\neq m}\frac{\sgn(\omega_{m'})}{|\omega_m-\omega_{m'}|^\gamma}.\\
  \end{aligned}
\end{equation}
\begin{figure}
	\begin{center}
		\includegraphics[width=12cm]{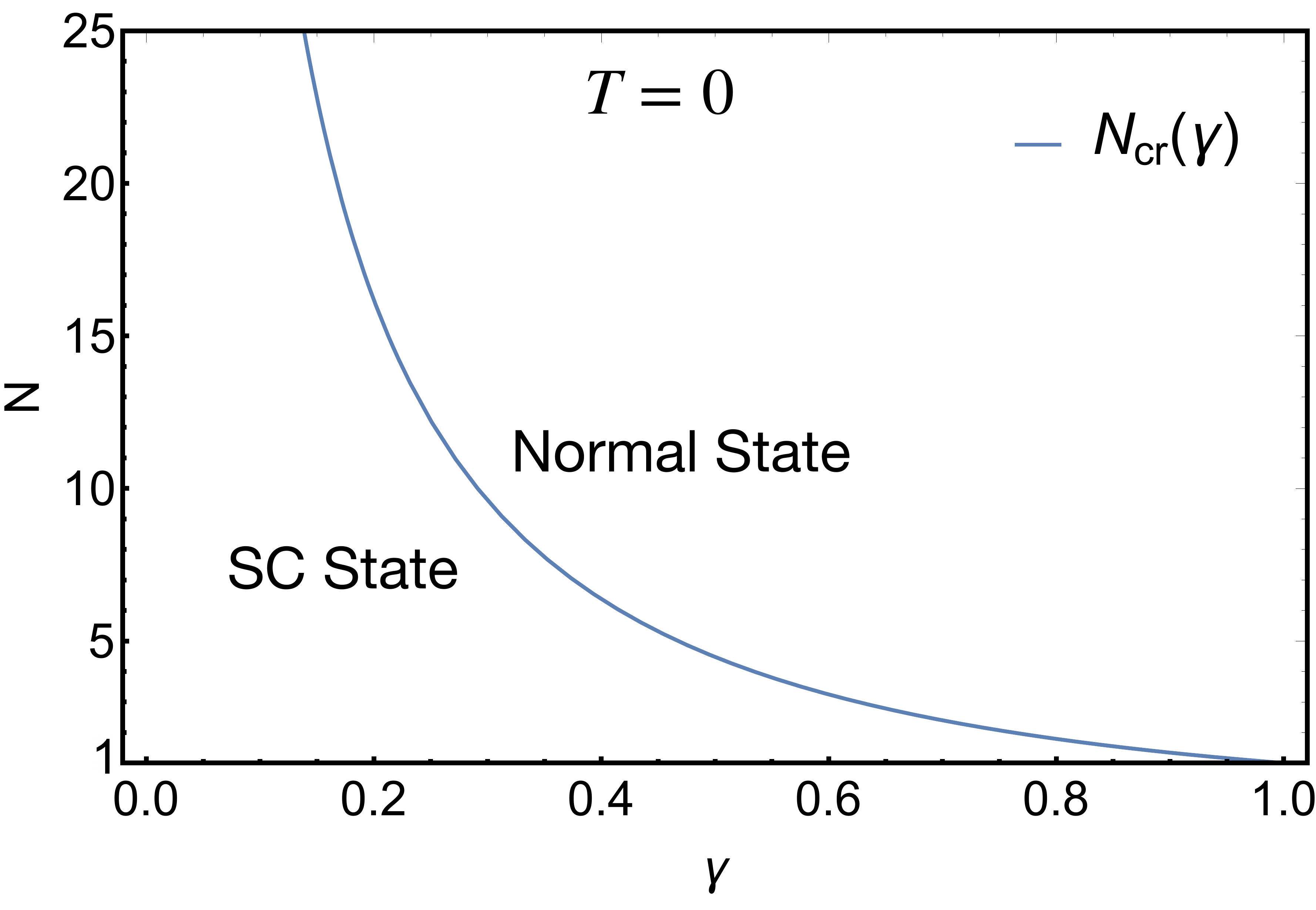}
		\caption{The behavior of $N_{cr} (\gamma)$,  given by Eq. \eqref{s_2a}. At $T=0$, this  critical $N$ separates superconducting and normal states
 at $N < N_{cr} (\gamma)$ and $N > N_{cr} (\gamma)$, respectively.}\label{fig:Ncr}
	\end{center}
\end{figure}
   By power-lay counting we expect $\Sigma^* (\omega_m) \propto g^\gamma \omega^{1-\gamma}$.  Substituting this into the equation for $\Phi$ in (\ref{eq:lineargap}) we obtain that at $|\omega_{m'}| > |\omega_m|$, the pairing kernel $K  = (g^\gamma/N)/(|\omega_{m'}+\Sigma^*(\omega_{m'})|)/|\omega_{m'}|^\gamma$ is marginal at $|\omega_{m'}| <g$: $K \propto 1/|\omega_{m'}|$ (with prefactor independent on $g$),  and  decays as $K \propto g^\gamma /|\omega_{m'}|^{1+\gamma}$ at $|\omega_{m'}|> g$  This implies that $T_c$, if it exists, should be generally of order $g$. The marginal form of the kernel
    is similar to the BCS case and it gives rise to logarithmical growth of the pairing susceptibility within the perturbation theory.  However, in distinction to BCS,  the marginal form of $K$  holds only if $|\omega_{m'}| > |\omega_m|$, i.e., at each order of perturbation the logarithm is cut by the running frequency in the next cross-section in the Cooper ladder.  As the consequence,  the summation of the logarithms alone does not lead to the divergence of the pairing susceptibility. In this situation,  the conventional wisdom is that the pairing is the threshold phenomenon, i.e., it occurs if the pairing vertex exceeds some finite value.  The pairing strength in Eq. (\ref{eq:lineargap}) is controlled by $1/N$, hence by this logics there should be a critical $N_{cr}$ separating superconducting state at $N < N_{cr}$ and non-superconducting naked critical non-FL state at $N > N_{cr}$.
     At larger $N$ the tendency towards pairing is  stronger than the tendency towards a non-FL behavior;  at smaller $N$ the situation is the opposite.
      The analysis of the pairing problem at $T=0$ does yield exactly this king of behavior~\cite{raghu_15,Wang2016}.  Namely, there exists
      \beq
      N_{cr} = (1-\gamma) \Gamma (\gamma/2) \left[\frac{\Gamma(\gamma/2)}{2\Gamma(\gamma)} + \frac{\Gamma(1-\gamma)}{\Gamma(1-\gamma/2)}\right],
       \label{s_2a}
       \eeq
       separating superconducting and non-superconducting states ($\Gamma (...)$ is a Gamma function). We plot $N_{cr} (\gamma)$ in Fig.\ref{fig:Ncr}

\begin{figure}
  \begin{center}
    \includegraphics[width=8cm]{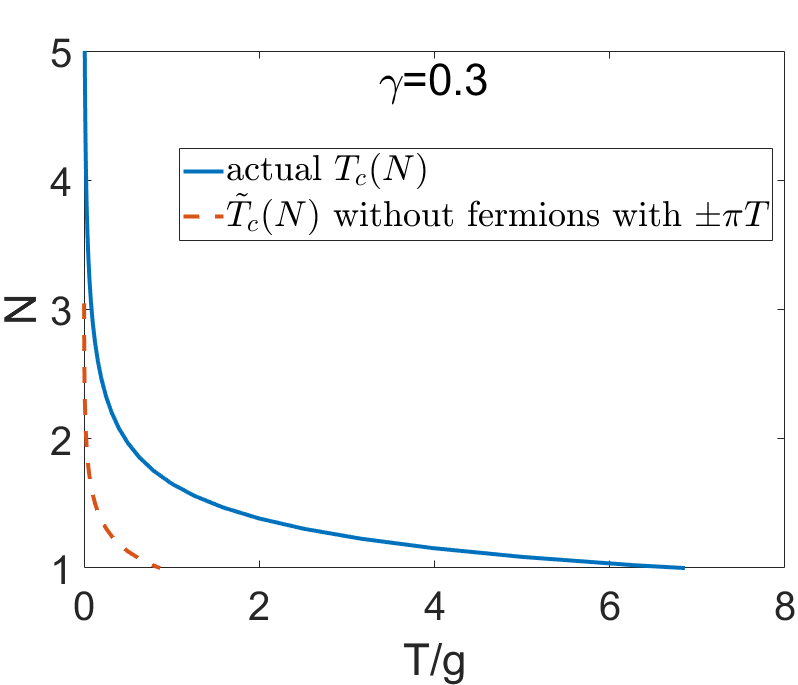}\quad\includegraphics[width=8cm]{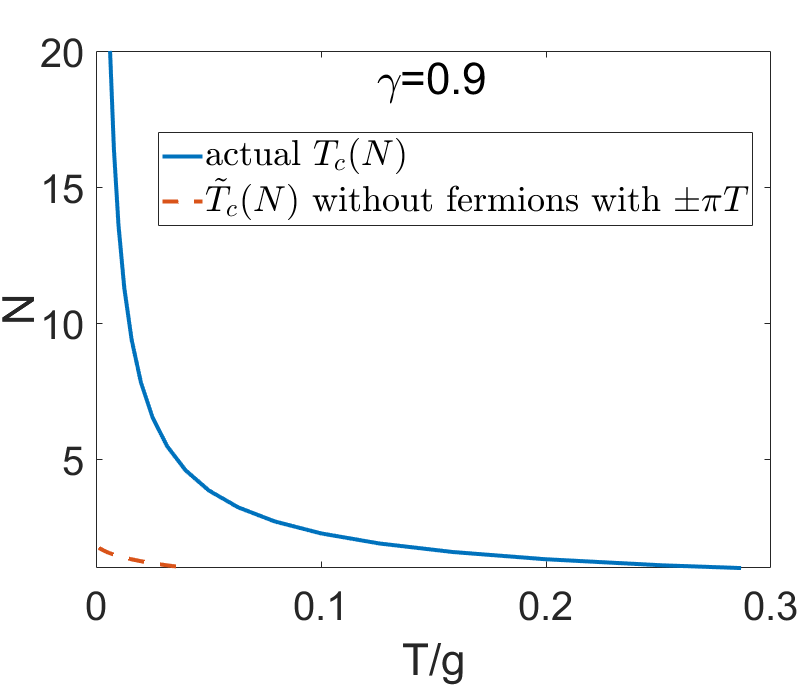}\\
    \caption{The pairing instability temperature $T_c(N)$, obtained by solving the linearized gap equation \eqref{eq:lineargap} as an eigenvalue/eigengunvction problem for $M =4000$ Matsubara frequencies, with $N$ playing the role of an eigenvalue. Upper and lower panels are for $\gamma=0.3$ and $\gamma=0.9$, respectively.  At large $N$, $T_c (N) \approx (g/2\pi) 1/N^{1/\gamma}$.   For comparison, we also show $\tilde{T}_c(N)$, which we obtained by solving the linearized gap equation without fermions with Matsubara frequencies  $\pm\pi T$. The temperature $\tilde{T}_c(N)$  terminates at $T = 0$ at the critical $N = N_{cr}$.
    }\label{fig:Tc}
  \end{center}
\end{figure}

     The existence of $N_{cr}$ at $T=0$ would normally imply that this is the termination point of the  line $T_c (N)$. However, the numerical solution of    (\ref{eq:lineargap}) yields qualitatively different result:  $T_c$ is non-zero at any $N$, and the line $T_c (N)$  by-passes $N_{cr}$ and approaches zero only at $N \to \infty$ (see Fig.\ref{fig:Tc}).
    The reason for this behavior has been  clarified in Ref. \cite{Wang2016}.  It turns out that power counting argument that $\Sigma^* (\omega_m) \propto \omega^{1-\gamma}_m$, does not work for the first two Matsubara frequencies $\omega_m = \pm \pi T$, for which Eq. (\ref{eq:lineargap}) yields $\Sigma^* (\pm \pi T) =0$. The reason is the presence of the sign-changing factor $\sgn(\omega_{m'})$ in the r.h.s. of the formula for $\Sigma^* (\omega_m)$.
     For $\omega_m = \pm \pi T$, contributions from positive and negative $\omega_{m'}$ cancel out. To see the consequence of $\Sigma^* (\pm \pi T) =0$,
     consider the limit $N \gg 1$ and set external $\omega_m = \pi T$.   For $\omega_{m'}  = O(T)$, but $\omega_{m'} \neq - \pi T$, the product $\pi T K(\omega_{m'})$ is independent of $T$ and is small in $1/N$.  However, for $\omega_{m'} = -\pi T$, this product is  $\pi T K = (1/N) (g/(2\pi T))^\gamma$, and it becomes large at small enough $T$.  A simple experimentation shows that in this situation the gap equation reduces to
     \bea
     &&\Phi^* (\pi T) \approx \frac{1}{N} \left(\frac{g}{2\pi T}\right)^\gamma \Phi^* (-\pi T) \nonumber \\
     &&\Phi^* (\omega_m) =  \frac{1}{N} \left(\frac{g}{2\pi T}\right)^\gamma \left(\frac{\Phi^* (\pi T)}{|\frac{1}{2}- \frac{\omega_{m'}}{2\pi T}|^\gamma} + \frac{\Phi^* (-\pi T)}{|\frac{1}{2}+ \frac{\omega_{m'}}{2\pi T}|^\gamma}\right)
     \label{s_1}
     \eea
     The last equation is for $\omega_m \neq \pm \pi T$.
     We will be searching for even-frequency solutions of the gap equation: $\Phi^* (\omega_m) = \Phi^* (- \omega_m)$.  Then the first equation in (\ref{s_1})  sets $T_c = (g/2\pi) 1/N^{1/\gamma}$, and the second shows that a non-zero $\Phi^* (\omega_m)$ is  induced by $\Phi^* (\pm \pi T)$.

 The functional form $T_c \propto 1/N^{1/\gamma}$ at large $N$ has been verified numerically in Ref. \cite{Wang2016} for a particular choice of $\gamma =0.1$. In Fig.\ref{fig:Tc} we show that the same behavior  holds for $\gamma =0.3$ and $0.9$.  We now go beyond Ref. \cite{Wang2016} and verify that this behavior of $T_c$ (i.e., that $T_c (N)$ line by-passes $N_{cr}$) is indeed due to vanishing of the self-energy at the first two Matsubara frequencies.  For this, we exclude $\omega_m = \pi T$ from the set of Matsubara frequencies and then solve again the linearized gap equation. The result is shown in Fig.\ref{fig:Tc}  We clearly see that ${\tilde T}_c$, obtained this way, tends to zero above some critical value of $N$, which numerically is close to  $N_{cr} (\gamma)$ in Eq. (\ref{s_2a}).  The outcome is that, without the first two Matsubara frequencies, the system would display a conventional behavior with ${\tilde T}_c (N)$ line terminating at a QCP at $N = N_{cr}$.  At larger $N$, superconductivity would be absent because of stronger tendency towards a (competing) non-FL ground state.   That the actual $T_c(N)$ by-passes $N_{cr}$ and vanishes only at $N = \infty$ is then entirely  due to the vanishing of the self-energy for  fermions with $\omega_m = \pm \pi T$.

The discrepancy between $T_c (N)$ and ${\tilde T}_c (N)$ suggests  that physical properties below the actual onset temperature for the pairing $T_c (N)$  depend on whether $N$ is smaller or larger than $N_{cr}$.  When $N > N_{cr}$,
   the pairing is induced by fermions with $\omega_m = \pm \pi T$ and, the order parameter $\Phi (\omega_m)$  emerges at $T_c (N)$ and vanishes at $T=0$, i.e., it is is non-monotonic as a function of temperature. For $N < N_{cr}$, there are two regimes of qualitatively different behavior  -- in between $T_c (N)$ and ${\tilde T}_c (N)$, the pairing is induced by fermions with $\omega_m = \pi \pi T$, while at $T <  {\tilde T}_c (N)$, fermions with all Matsubara frequencies contribute to the pairing.  This last behavior is a conventional one, in the sense that it holds in a non-critical, BCS superconductor, while the behavior at ${\tilde T}_c (N) < T <  T_c (N)$ is of non-BCS type as it is due to strong non-FL self-energy at all $\omega_m$ except for $\pm \pi T$.  At small $\gamma$, $N_{cr} \approx 4/\gamma$, and  the $T_c (N)$ and ${\tilde T}_c (N)$ lines remain close down to a very small $T \sim g (\gamma)^{1/\gamma} \ll g$. However, for $\gamma \leq 1$, the two lines separate already at $T \leq g$.
   We note in this regard that the range between ${\tilde T}_c (N) < T <  T_c (N)$ exists for the physical case of $N=1$, and the lower boundary of this range rapidly decreases as $\gamma$ approaches the value equal to $1$.  In other words, even for $N=1$, there exists an intermediate $T$ range where the pairing is induced by fermions with $\pm \pi T$, and would not exist if these fermions were excluded from the gap equation.
   The behavior of a system in this intermediate $T$ range at $N=1$ should be, at least qualitatively, the same as that at large $N$.

    Below we study superconductivity induced by fermions with $\omega_m = \pm \pi T$ in some detail by solving non-linear gap equation at $T < T_c$. We first solve the gap equation in Matsubara
       frequencies  and obtain the gap, the Free energy, and the specific heat, and  then convert to real frequencies and obtain the spectral function and the DOS.

\section{Non-linear gap equation, $N > N_{cr}$}
\label{sec:largeN}
We begin with the case $N > N_{cr}$ when ${\tilde T}_c =0$, i.e. the pairing would be impossible if the self-energy did not vanish at $\omega_m = \pm \pi T$.
The limit $N \gg 1$ can be treated analytically and we consider it in some detail below.

\subsection{Non-linear gap equation in Matsubara frequencies.}
\label{sec:largeNMatsubara}

The non-linear equation for the pairing vertex $\Phi^* (\omega_m)$  along with the equation for the fermionic self-energy $\Sigma^* (\omega_m)$  with the feedback from the pairing are given in (\ref{eq:gapeq_1}). We recall that at large $N$ the pairing temperature $T_c (N)$ is obtained by solving the linearized  equation for $\Phi^* (\omega_m)$ for fermions with only two Matsubara frequencies $\omega_m = \pm \pi T$; the pairing vertex $\Phi^* (\omega_m)$ for other $\omega_m$ is then expressed via $\Phi^* (\pi T) = \Phi^* (-\pi T)$.  We assume and then verify that this holds also for $T < T_c$, i.e., that the non-linear gap equation can be approximated by restricting to $\omega_{m'} = \pm \pi T$ in the r.h.s. of Eq. (\ref{eq:gapeq_1}).
Re-labeling $\Phi^* (\pi T) = \Phi^ (- \pi T) = \Phi^*_0,  \Sigma^* (\pi T) = - \Sigma^* (- \pi T) = \Sigma^*_0$, and ${\tilde \Sigma}^*_0 = \pi T + \Sigma^ (\pi T)$ to shorten notations,  we obtain from (\ref{eq:gapeq_1})
\bea
\Phi^*_0 &=& \pi T \left(\frac{T_c}{T}\right)^\gamma \frac{\Phi_0}{\sqrt{(\Phi^*_0)^2 + ({\tilde \Sigma}^*_0)^2}} \nonumber \\
{\tilde \Sigma}^*_0 &=& \pi T \left[1 + N \left(\frac{T_c}{T}\right)^\gamma \left(1 - \frac{{\tilde \Sigma}^*_0}{\sqrt{(\Phi^*_0)^2 + ({\tilde \Sigma}^*_0)^2}}\right)\right]
\label{s_2b}
\eea
The solution of (\ref{s_2b}) to leading order in $1/N$ is
\bea
\Phi^*_0 &=& \pi T \left(\frac{2}{N}\right)^{1/2} \left(\frac{T_c}{T}\right)^{\gamma} \left(1 - \left(\frac{T}{T_c}\right)^\gamma\right)^{1/2} \nonumber \\
{\tilde \Sigma}^*_0 &=& \pi T \left(\frac{T_c}{T}\right)^\gamma, {\text or}~  \Sigma^*_0 = \pi T \left(\left(\frac{T_c}{T}\right)^\gamma -1\right)
\label{s_3}
\eea
The superconducting gap  $\Delta (\pm \pi T) = \Delta_0$ is
\beq
\Delta_0 = \pi T \left(\frac{2}{N}\right)^{1/2} \left(1 - \left(\frac{T}{T_c}\right)^\gamma\right)^{1/2}
\label{s_4}
\eeq
The gap $\Delta_0$ vanishes both at $T=0$ and at $T= T_c$. In between, it is finite, but for any $T$,  $D_0 = \Delta_0/(\pi T)$ is small and at most of order $1/N^{1/2}$.
In other words, the gap at $N \gg 1$ remains smaller than  the temperature.

\begin{figure}
   \begin{center}
     \includegraphics[width=17cm]{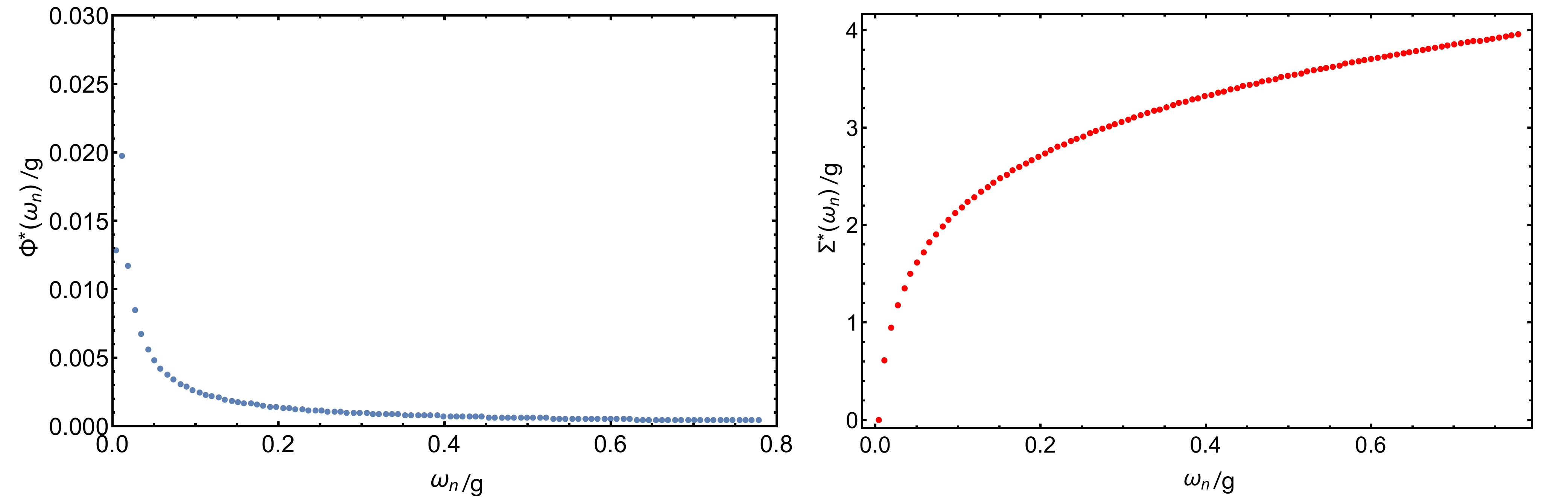}\\
     \caption{The pairing vertex $\Phi^*(\omega_m)$ and the self-energy $\Sigma^*(\omega_m)$  from Eq. \eqref{s_5}.
     For definiteness we set  $\gamma=0.9$, $N=10$,  and $T=0.1T_c$.
      }\label{fig:PhistarSigstar}
   \end{center}
 \end{figure}

 Solving next the set of Eliashberg equations for other $\omega_m \neq \pm \pi T$ we obtain at large $N$
 \bea
 &&\Phi^* (\omega_m) \approx \Phi^*_0 \left[\left(\frac{2 \pi T}{|\omega_m - \pi T|}\right)^\gamma + \left(\frac{2 \pi T}{|\omega_m + \pi T|}\right)^\gamma\right] \nonumber \\
&& \Sigma^* (\omega_m) \approx  2 N  {\tilde \Sigma}^*_0  H\left(\frac{|\omega_m|-\pi T|}{2\pi T}, \gamma\right) \sgn(m+\frac{1}{2}) \nonumber  \\
 \label{s_5}
 \eea
 where $H(a,b) = \sum_{1}^a n^{-b}$ is a Harmonic number.
   We plot $\Phi^* (\omega_m)$ and $ \Sigma^* (\omega_m)$ in Fig.\ref{fig:PhistarSigstar}.  Note that at $\omega_m \sim T$, $\Sigma^* (\omega_m) \sim N \omega_m$, i.e., $\Sigma^* (\omega_m) \approx {\tilde \Sigma}^* (\omega_m)$.

  At large $ m$ (but still when $\Sigma^* (\omega_m) \gg \omega_m$)
\beq
 \Phi^* (\omega_m) \approx \frac{2\Phi_0}{|m|^\gamma}, ~~{\tilde \Sigma}^* (\omega_m) \approx 2N \frac{|{\tilde \Sigma}_0|}{1-\gamma}  |m|^{1-\gamma} \sgn(m)  \label{s_5_1}
 \eeq
 Note that below $T_c$, the self-energy at all $\omega_m$, including $\omega_m = \pm \pi T$, behaves as   $\Sigma^* (\omega_m)  \propto T^{1-\gamma}$, consistent with the scaling $\Sigma^* (\omega_m) \propto (\omega_m)^{1-\gamma}$. Still, the self-energy at  $\pm \pi T$ is smaller in $1/N$ than $\Sigma^* (\omega_m)$ at other Matsubara frequencies.

\begin{figure}
  \begin{center}
    \includegraphics[width=12cm]{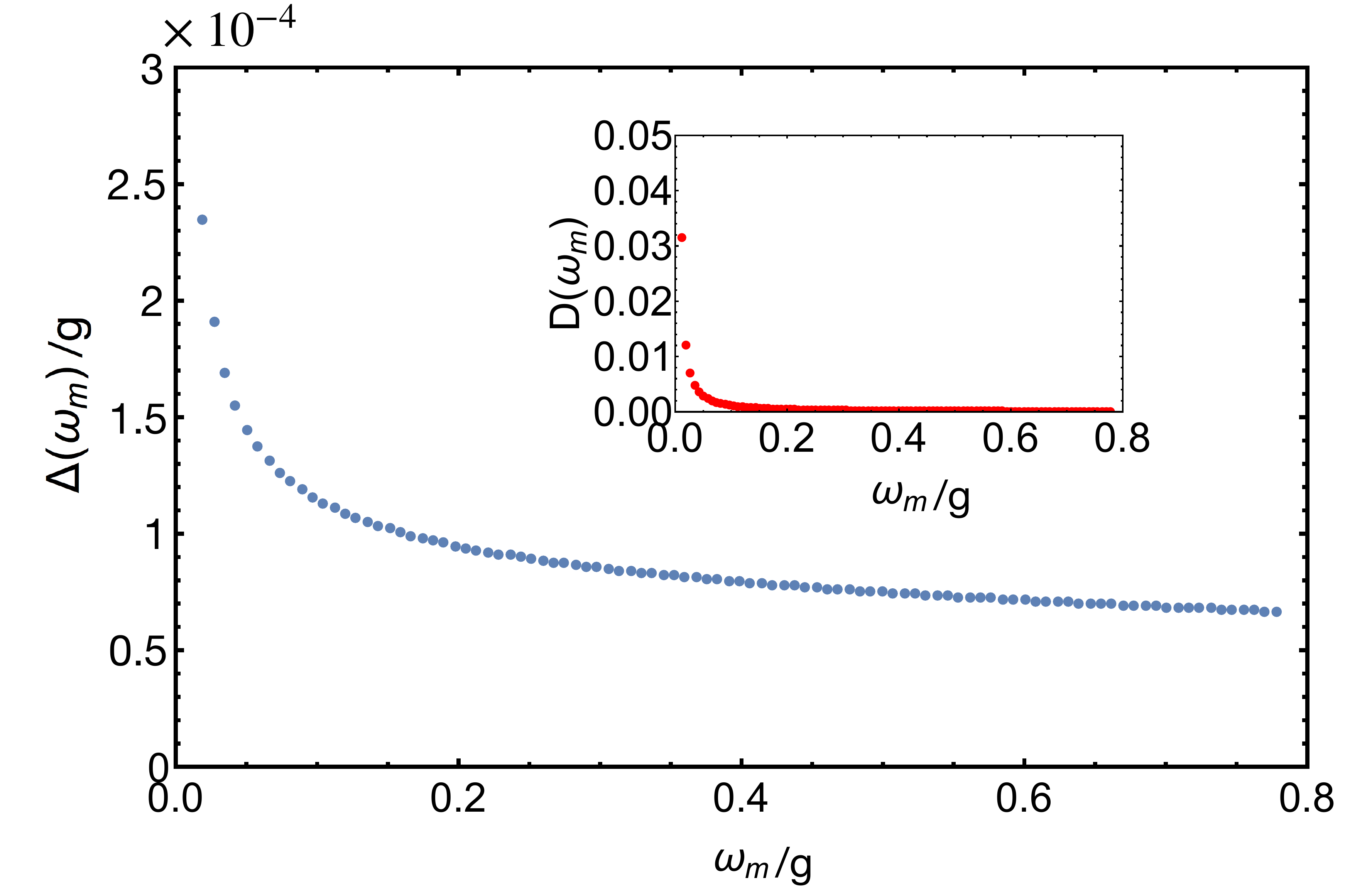}
    \caption{The pairing gap $\Delta(\omega_m) = \Phi^* (\omega_m)  \omega_m/{\tilde \Sigma} (\omega_m)$ and $D(\omega_m) = \Delta (\omega_m)/\omega_m$ for the same parameters as in Fig. \ref{fig:PhistarSigstar}.}\label{fig:DeltaD}
  \end{center}
\end{figure}

   From (\ref{s_5}) we have
 \beq
 \Delta (\omega_m) \sim \frac{\Delta_0}{N} \sim \pi T \left(\frac{2}{N}\right)^{3/2} \left(1 - \left(\frac{T}{T_c}\right)^\gamma\right)^{1/2}
 \label{s_4a}
 \eeq
  both at $m = O(1)$ and at $m \gg 1$.  We see that at any $T < T_c$, $\Delta (\omega_m)$ at any Matsubara frequency is parametrically smaller than $T$.  Put it differently,   $D (\omega_m) = \Delta (\omega_m)/\omega_m$ is small, of order $1/N^{3/2}$, at $m = O(1)$, and even smaller at larger $m$.  We plot $\Delta (\omega_m)$ and $D(\omega_m)$ in Fig.\ref{fig:DeltaD}.

 Taking  $-iD_0$ as an estimate for small frequency limit of $D^R(\omega) \equiv \Delta^R( (\omega)/\omega$ in real frequencies,  we find that $D^R(\omega \to 0)$ tends to a finite imaginary value, i.e., at large $N$ is a gapless superconductivity in the sense that $\Delta^R (\omega) \propto i \omega$.
For notational simplicity for all functions of real frequencies below we will drop the superscript ``$R$".\cite{footnote}
 Using then $N(\omega) = N_0 \Ree[1/\sqrt{1-  D^2 (\omega)}]$ for the DOS ($N_0$ is the normal state value), we find that the DOS at zero frequency  $N(\omega = 0) = N_0 /\sqrt{1 + D^2_0} \approx N_0 \left(1 - \frac{1}{2}D^2_0\right)$  is reduced below $T_c$ compared to the normal state value, but remains finite for any $T$, like it is expected in a gapless superconductor.

 To verify this result and to get the full form of $N(\omega)$ we need to obtain $\Delta (\omega)$ as a function of a real frequency $\omega$. This is what we will do in Sec.~\ref{sec:largeNreal}. Before that, we use the result for $D(\omega_m)$ and obtain the Free energy $F_{sc} (T)$  and the specific heat $C(T)$   at $N > N_{cr}$.

 \subsubsection{The Free energy and the specific heat}
 \label{sec:cond_energy}
The Free energy $F_{sc}$ and $\Delta F = F_{sc} - F_n$ are given by Eqs. (\ref{a_5})-(\ref{a_8})
 At large $N$, we keep only contributions which contain $D_m$, $D_{m'}$  with $m, m' = 0, -1$. Contributions from $D_m$ with other $m$ are smaller in $1/N$, as we explicitly verified.
   Using that $\sum_m \frac{\sgn~ m}{|\pi T \pm  \omega_m|^\gamma} =0$, we obtain from (\ref{a_8})
\beq
\delta F  \approx  - 2 \pi^2 T^2 N_0 \left(\frac{T_c}{T}\right)^\gamma \left[ D^2_0 \left(1 - \left(\frac{T}{T_c}\right)^\gamma\right) - \frac{N D^4_0}{4}\right]
\label{a_9}
\eeq
Varying $\delta F$ by $\Delta_0$, one reproduces Eq. (\ref{s_4}).  Substituting $D_0$ from (\ref{s_4}) into (\ref{a_9}), we obtain
\beq
\delta F  \approx  - \frac{2}{N} \pi^2 T^2 N_0 \left(\frac{T_c}{T}\right)^\gamma \left(1-  \left(\frac{T}{T_c}\right)^\gamma\right)^2
\label{a_10}
\eeq
The specific heat variation between the superconducting and the normal state  $\delta C_v = -T \partial^2 \delta F/\partial T^2$ is
\beq
\delta C_v =  \frac{2}{N} \pi^2 T_c N_0 C_\gamma{\left(\frac{T}{T_c}\right)}
\label{a_11}
\eeq
where
\bea
&&C_\gamma{(x)} =  2 \gamma^2 x^{\gamma+1} -2 \gamma (3-\gamma) x (1-x^\gamma) \nonumber \\
&& + (2-\gamma)(1-\gamma) x^{1-\gamma} (1-x^\gamma)^2
\label{a_12}
\eea
At $T \to 0$, $C_\gamma (0) \to 0$, i.e., $\delta C_v$ vanishes and $C_v$ recovers its normal state limiting behavior $C_V \propto T^{1-\gamma}$.  At $T= T_c -0$, $C_\gamma = 2 \gamma^2$, i.e.,  the magnitude of the specific heat jump at $T_c$ is
\be
\delta C_v = (4\gamma^2/N) \pi^2 T_c N_0.
\ee

\begin{figure}
	\includegraphics[width=16.5cm]{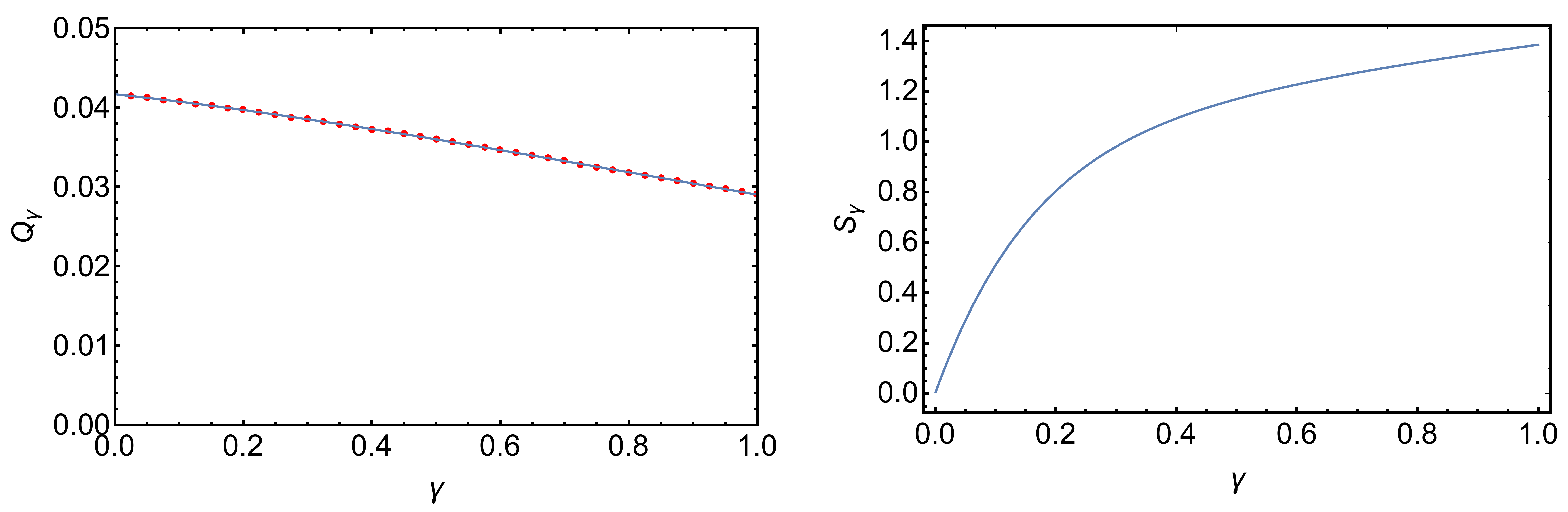}
	\caption{The plots of the scaling functions $Q_\gamma$ from Eq. \eqref{yy_1} and $S_\gamma$ from Eq. \eqref{s_10}. }\label{fig:Qgamma}	
\end{figure}

 The specific heat in the normal state is obtained from (\ref{a_6}).
 The first term in (\ref{a_6}) gives the conventional free-fermion contribution to Free energy  $F_{n,free} (T) = F_{n,free} (0) - N_0 \pi^2 T^2/3$.
  The second term gives
  \begin{equation} \label{a_6_1}
  	\begin{aligned}
  		&F_{n,int} (T) =
   -N_0 N \pi^2 T^2 \left(\frac{T_c}{T}\right)^\gamma \sum_{m \neq m'} \frac{\sgn (m+1/2) \sgn (m'+1/2)}{|m-m'|^\gamma}\\
  	\end{aligned}
  \end{equation}
 At $T \sim T_c$, this second term is larger by $N$ than the free-fermion contribution.  The calculation of the double sum in (\ref{a_6_1})
  requires care as one needs to extract the universal constant on top of formally ultra-violet divergent contribution, which actually is the factor in  $F_{n,int} (0)$.  To extract the universal constant, we note that the summation over $m-m'$ can be done explicitly. The result is
  \beq
  \sum_{m \neq m'} \frac{\sgn(m+1/2) \sgn (m'+1/2)}{|m-m'|^\gamma}= 4 \sum_{m=0}^\infty H(m,\gamma),
  \label{a_6_2}
  \eeq
  where, we remind, $H(m,\gamma) = \sum_1^m 1/p^\gamma$ is the Harmonic number.  For the remaining summation we use the Euler-Maclaurin formula
  \bea
 &&\sum_{m=0}^\infty f(m+1/2) = \int_0^\infty f(x)dx  + Q \nonumber\\
 && Q =  -\int_{0}^{1/2}  f(x)dx  + \frac{1}{2} f(1/2) - \sum_{n=2}^\infty \frac{B_n}{n!} \frac{d^{n-1} f}{dx^{n-1}}_{|~x=1/2},
  \label{a_6_3}
  \eea
where $B_n$ are Bernoulli numbers.  The first term in the upper line in (\ref{a_6_3}) contributes to $F_{n,int} (T=0)$, the second term determines  the universal prefactor in the temperature-dependent piece in the Free energy.   It is essential that the argument of the function under the sum is $m+1/2$ because this is how Matsubara frequency $\omega_m$ depends on $m$.  Accordingly, we re-define $H(m+1/2,\gamma) = \sum_1^{m+1/2-1/2} 1/p^\gamma$ and extend it to a function $H(x,\gamma)$ of a continuous variable $x$.  Evaluating then the integral and the derivatives in the second line in (\ref{a_6_3}) numerically, we obtain
\beq
\sum_{m=0}^\infty H(m, \gamma) = \int_0^\infty H(x,\gamma) +  Q_\gamma.
\label{yy_1}
\eeq
We plot $Q_\gamma$  in Fig.\ref{fig:Qgamma}.

\begin{figure}
  	\begin{center}
  		\includegraphics[width=12cm]{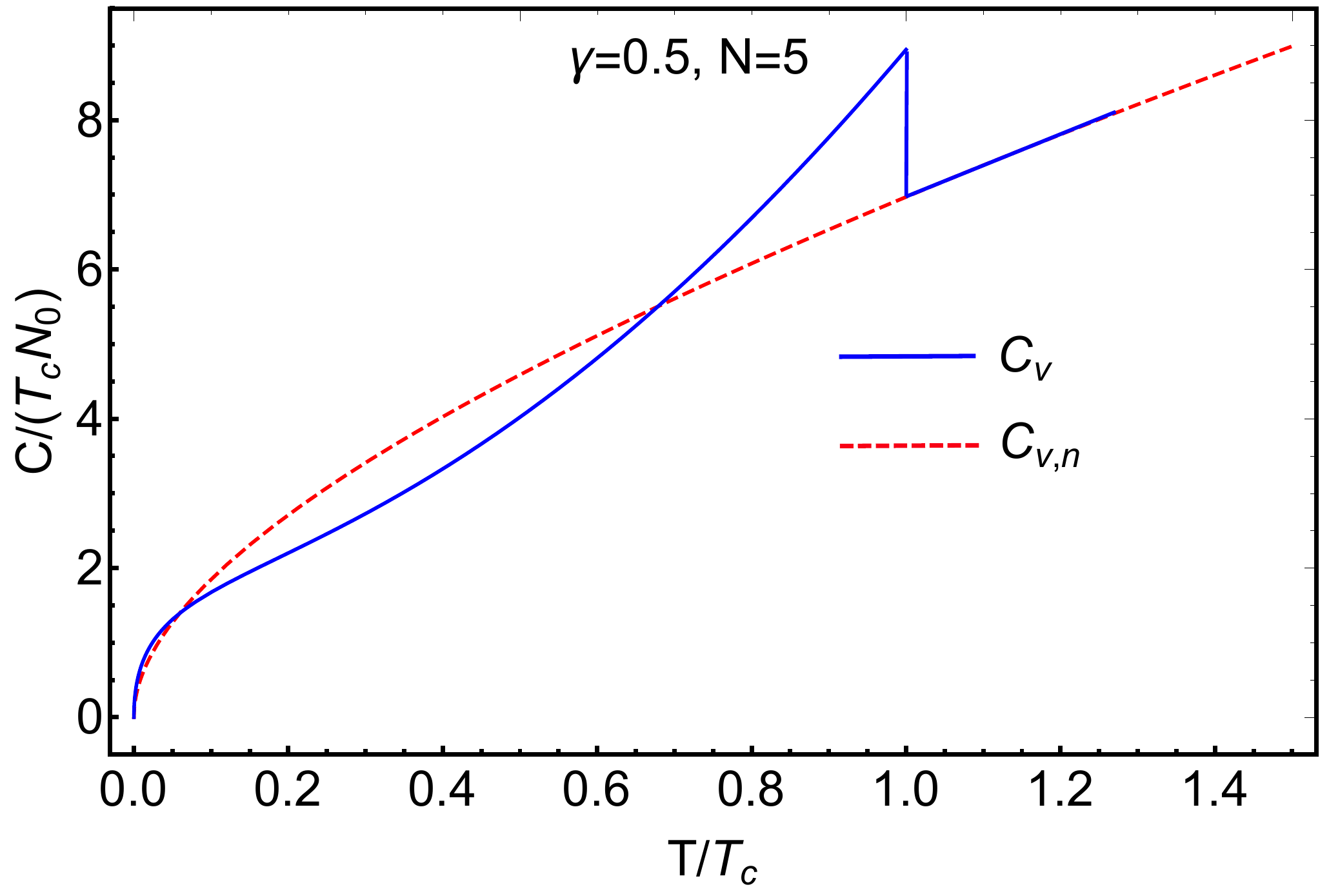}
  		\caption{The specific  heat (in units of $T_c N_0$) vs $T/T_c$.  The dashed line is the normal state result. We set $\gamma=0.5$ and $N=5$.
   Observe that the jump of $C(T)$ at $T_c$ is small, and that at low $T$ specific heat returns back to its normal state value. }\label{fig:sHeatJump}
  	\end{center}
  \end{figure}

 Substituting the result into (\ref{a_6_1}) and differentiating the free energy over $T$, we obtain
 \beq
 C_{v,n} =  N (4\pi^2 N_0 T_c) (2-\gamma) (1-\gamma) Q_{\gamma} \left(\frac{T}{T_c}\right)^{1-\gamma}
 \label{a_14}
 \eeq
  The ratio of the specific heat jump to its value at $T = T_c +0$ is then
  \beq
  \frac{\delta C_v}{C_{v,n}} = \frac{1}{N^2} \frac{ \gamma^2}{(2-\gamma)(1-\gamma) Q_\gamma}
  \label{a_16}
  \eeq
  We see that the relative jump of $C_v$ at $T_c$ is by $1/N^2$ smaller than in a BCS superconductor.
  In Fig.\ref{fig:sHeatJump} we plot $C_v (T) = C_{v,n} (T) + \delta C_v (T)$ in the full temperature range below $T_c$. At sufficiently small $T$, both $C_v$ and $C_{v,n}$ scales as $T^{1-\gamma}$.

\subsection{Beyond leading order in $1/N$}

We now  go beyond the leading order in $1/N$. The goal here is to analyze how fermions with other $\omega_m$
 affect the magnitudes of $\Phi (\pi T) = \Phi_0$ and $D(\pi T) = D_0$ at a small but finite temperature.  We recall that at large $N$,
  $\Phi_0 \approx (2/N)^{1/2} \pi T (T_c/T)^\gamma$ and $D_0 \approx (2/N)^{1/2}$.  We show that both $\Phi_0$ and $D_0$ increase as $N$ get smaller.

   For the analysis to next order in $1/N$ we use the fact that $D_0 \propto 1/N^{1/2}$, while for other Matsubara frequencies  $D (\omega_m) \propto 1/N^{3/2}$ (Eqs (\ref{s_4}) and (\ref{s_4a}). Because $D$ appears in even powers in the equation for the self-energy in (\ref{eq:gapeq_1}),  the inclusion of these $D (\omega_m)$ with $m \neq 0, -1$  would lead to corrections  of at least of order $1/N^2$. To order $O(1/N)$ we then still have the same equation for $ {\tilde \Sigma}^*_0$ as in (\ref{s_2b}). Expanding in this equation in two orders of $D^2_0 \propto 1/N$ and
     setting $T \ll T_c$,
    we obtain
   \beq
   {\tilde \Sigma}^*_0 =  N \pi T \left(\frac{T_c}{T}\right)^\gamma \left(\frac{D^2_0}{2} - \frac{3 D^4_0}{8}\right)
\label{b_2}
\eeq
The expansion to next order in $1/N$  in the equation for $\Phi_0$  requires more care as the leading term (the one kept in the first equation in
 (\ref{s_2b})) is of order $1/N^{1/2}$, while other terms in the r.h.s. of the equation for $\Phi^* (\omega_m)$ in (\ref{eq:gapeq_1}) are of order $D (\omega_m) \propto 1/N^{3/2}$, i.e., they contain only one additional power of $1/N$.  These terms then should be kept in calculation to subleading  order in $1/N$.  Keeping these terms, we obtain from  (\ref{eq:gapeq_1}):
 \bea
\Phi_0 &=& N \pi T \left(\frac{T_c}{T}\right)^\gamma D_0 \left(1 - \frac{D^2_0}{2}\right) \nonumber \\
&& + \sum_{m=1}^\infty  D (\omega_m) \left(\frac{1}{m^\gamma} + \frac{1}{(m+1)^\gamma}\right)
\label{b_3}
\eea
\begin{figure}
	\begin{center}
		\includegraphics[width=12cm]{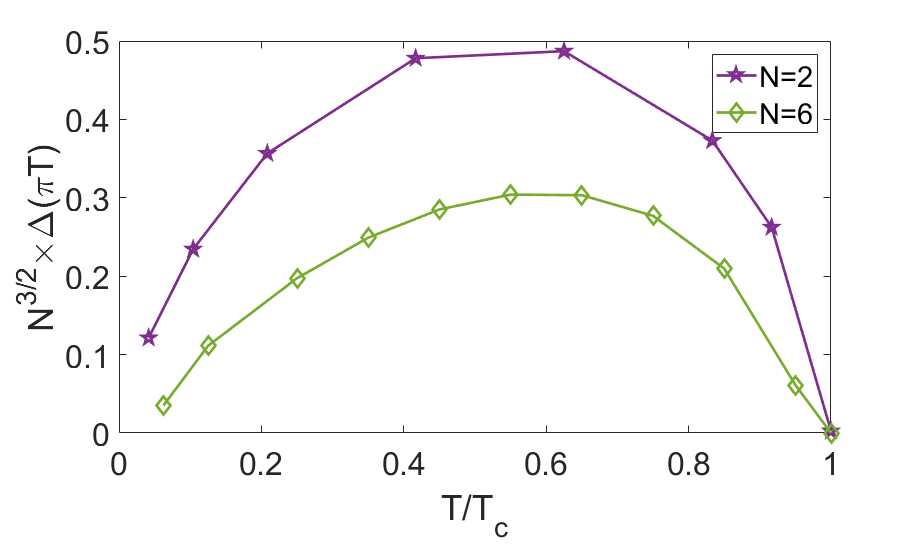}\\
		\caption{The gap at the first Matsubara frequency $\Delta (\pi T) = \Delta_0 $  as a function of temperature for $\gamma =0.9$ and two different $N > N_{cr}$. The slope of $\Delta_0 (T)$ at small $T$ increases as $N$ gets smaller. }\label{fig:Mgap_1}
	\end{center}
\end{figure}

\begin{figure}
  \begin{center}
    \includegraphics[width=12cm]{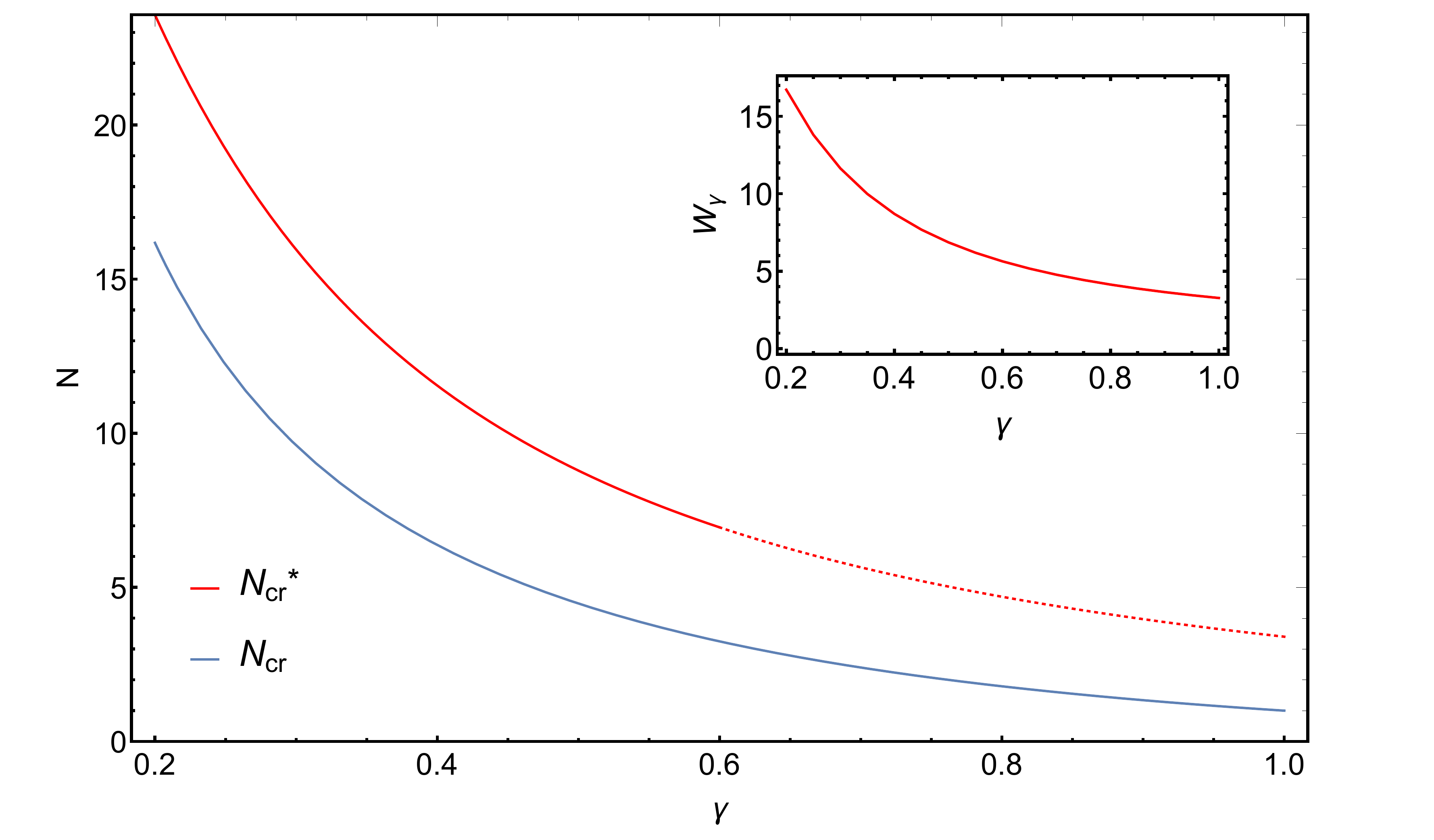}
    \caption{The approximate $N^*_{cr} (\gamma) = 3 (W_\gamma -1)/2$ vs the actual $N_{cr} (\gamma)$. The inset shows $W_\gamma$ given by Eq. \eqref{b_6}.}\label{fig:Nstar}
  \end{center}
\end{figure}

Substituting $D (\omega_m)$ from Eq. (\ref{s_5}):
\beq
D (\omega_m) = \frac{\Phi^* (\omega_m)}{{\tilde \Sigma}^* (\omega_m)} = \frac{1}{N} \frac{D_0}{H(m,\gamma)}  \left(\frac{1}{m^\gamma} + \frac{1}{(m+1)^\gamma}\right)
\label{b_4}
\eeq
 we obtain
 \beq
\Phi^*_0 \left(1 - \frac{W_\gamma}{2N}\right) = N \pi T \left(\frac{T_c}{T}\right)^\gamma D_0 \left(1 - \frac{D^2_0}{2}\right)
\label{b_5}
\eeq
where
 \beq
 W_\gamma = \sum_{m=1}^\infty \frac{1}{H(m,\gamma)}  \left(\frac{1}{m^\gamma} + \frac{1}{(m+1)^\gamma}\right)^2
\label{b_6}
\eeq
and, we remind, $H(m,\gamma) = \sum_1^m{1/n^\gamma}$ is a Harmonic number. We plot $W_\gamma$ in inset of Fig.\ref{fig:Nstar}.

Solving (\ref{b_2}) and (\ref{b_5}) to order $1/N$ we obtain at low $T \ll T_c$
\bea
&&\Phi^*_0 = \left(\frac{2}{N}\right)^{1/2} \pi T \left(\frac{T_c}{T}\right)^\gamma \left(1 + \frac{3(W_\gamma-1)}{4N}\right) \nonumber \\
&&{\tilde \Sigma}^*_0 =  \pi T \left(\frac{T_c}{T}\right)^\gamma \left(1 + \frac{W_\gamma-2}{2N}\right) \nonumber \\
 &&D_0 = \frac{\Delta_0}{\pi T}  \left(\frac{2}{N}\right)^{1/2} \left(1 + \frac{W_\gamma+1}{4N}\right)
 \label{b_7}
 \eea
The analysis at larger $T \leq T_c$ proceeds in the same way and we refrain from presenting the full formulas.
In Fig. \ref{fig:Mgap_1} we show $\Delta_0 = \Delta (\pi T)$ as a function of $T/T_c$ for $\gamma =0.9$ and two different values of $N > N_{cr}$  ($N_{cr} \sim 1.3$ for $\gamma =0.9$).  In both cases, $\Delta_0$ vanishes at $T=0$, but the slope of $\Delta_0 (T)$ at small $T$ gets larger when $N$ decreases.

 The result for $\Phi^*_0$  can be cast into $\Phi^*_0 \approx (2/(N-N^*_{cr}))^{1/2} \pi T \left(\frac{T_c}{T}\right)^\gamma $ where $N^*_{cr} = 3(W_\gamma-1)/2$ is some $\gamma$-dependent constant.
 Taking this approximate formula as an indication of the evolution of $\Phi^*_0$ with decreasing $N$, we find that $\Phi^*_0 \propto T^{1-\gamma}/(N-N^*_{cr})^{1/2}$. At $N > N^*_{cr} (\gamma)$, $\Phi_0$ vanishes at $T=0$ (we recall that we consider $\gamma <1$), but   $N = N^*_{cr} (\gamma)$ the slope of $\Phi^*_0 (T)/T^{1-\gamma}$ (and of $\Delta_0$)
  diverges. This divergence is consistent with the $T=0$ analysis, which indicates that at $N < N_{cr}$, given by Eq. (\ref{s_2a}), the system has superconducting order at $T=0$. This will change the system behavior at small temperature and frequencies compared to what we found above. We emphasize that the increase of $\Phi^*_0 (T \to 0)$ is due to the contribution from fermions with $|\omega_m| \neq \pi T$, which give rise to the $S$ term in $1/N$ correction.  This means that, as $N$ get reduced, fermions with Matsubara frequencies other than $\pm \pi T$ become progressively more involved in the pairing.

 The $N^*_{cr} (\gamma) = 3(W_\gamma-1)/2$ is an approximate form of critical $N$ and does not have to coincide with the actual $N_{cr}
 (\gamma)$, given by Eq. (\ref{s_2a}). We plot both functions in Fig.\ref{fig:Nstar}.
  Interestingly, $N^*_{cr} (\gamma)$ and $N_{cr} (\gamma)$ show quite similar variation with $\gamma$.

 We next consider the solutions for the pairing vertex and the self-energy in real frequencies.  This will allow up to compute the spectral function $A(\omega)$ and the DOS $N(\omega)$.

 \subsection{Non-linear gap equation in real frequencies}
\label{sec:largeNreal}

The transformation of Elishberg equations for electron-phonon interaction from Matsubara to real frequencies using spectral decomposition method and analytical continuation has been discussed in several publications~\cite{combescot,Marsiglio_88,Marsiglio_91,Karakozov_91}.
We extend these result to our case with $\chi (\Omega_m) = (g/|\Omega_m|)^\gamma$.  The details of the conversion procedure are presented in the Appendix.
The conversion procedure requires special care by two reasons. First, if one simply replaces $\omega_m$ by $-i\omega$, the  bosonic propagator $\chi (\omega_{m'} + i \omega)$ will have a set of branch cuts in the complex $\omega$ plane,  along $\omega = i \omega_m + b$, where $b$ is real.  One then need to add additional terms to the r.h.s. of the  equations for retarded functions $\Phi (\omega)$ and $\Sigma (\omega)$ to cancel these singularities and restore analyticity.  Second, we again need to eliminate singular contributions from the terms with zero bosonic Matsubara frequency.  This is done in the same way as in the calculations along the Matsubara axis. Namely, we introduce new functions $\Phi^* (\omega)$ and ${\tilde \Sigma}^* (\omega)$ related to
 $\Phi (\omega)$ and ${\tilde \Sigma} (\omega) = \omega + \Sigma (\omega)$ as
 \beq
 \Phi^* (\omega) = \Phi (\omega) \left(1 - Q (\omega)\right), {\tilde \Sigma}^* (\omega) = {\tilde \Sigma} (\omega) \left(1 - Q (\omega)\right),
 \label{new_new}
 \eeq
  where  $Q_\omega$ is singular (see Eq. (\ref{ex_1}) below),
   but
$\Phi^* (\omega)$ and ${\tilde \Sigma}^* (\omega)$ are free from singularities.  The gap function $\Delta (\omega) = \omega \Phi (\omega)/{\tilde \Sigma} (\omega)$ is equally expressed in terms of $\Phi^* (\omega)$ and ${\tilde \Sigma}^* (\omega)$:
\beq
\Delta (\omega) = \omega \frac{\Phi(\omega)}{{\tilde \Sigma} (\omega)} =  \omega \frac{\Phi^*(\omega)}{{\tilde \Sigma}^* (\omega)}
\label{bew_new_2}
\eeq
 The equations on
$\Phi^* (\omega)$ and ${\tilde \Sigma}^* (\omega)$ are the same as on $\Phi (\omega)$ and ${\tilde \Sigma} (\omega)$, but with additional terms which cancel out divergent contribution from $\chi (0)$.   We have (see Appendix for details)
\begin{widetext}
\begin{equation}\label{eq:anacon}
  \begin{aligned}
    &\Phi^*(\omega)=\frac{\pi T}{N} \sum_{m}\frac{\Phi^*(\omega_m)}{\sqrt{(\Phi^*(\omega_m))^2 + ({\tilde \Sigma}^*(\omega_m))^2}}
    \chi(\omega_m +i\omega)\\
    &+\frac{i}{N}\int dx \left[S_\Phi (\omega -x)  \chi^{''}(x)  \left(n_F(x-\omega)+n_B(x)\right)-S_\Phi (\omega) \chi^{''} (x) \frac{T}{x} \right] \\
     &{\tilde \Sigma}^* (\omega)= \omega + i\pi T \sum_{m}\frac{{\tilde \Sigma}^*(\omega_m)}
     {\sqrt{(\Phi^*(\omega_m))^2+({\tilde \Sigma}^*(\omega_m))^2}}  \chi(\omega_m +i\omega)\\
    & + i \int dx \left[S_\Sigma (\omega-x) \chi^{''}(x) \left(n_F(x-\omega)+n_B(x)\right) -S_\Sigma (\omega) \chi^{''} (x) \frac{T}{x} \right]
    \end{aligned}
\end{equation}
where
\bea
S_\Phi (\omega) &=&  \frac{\Phi(\omega)}{\sqrt{{\tilde \Sigma}^2(\omega) - \Phi^2(\omega)}} = \frac{\Phi(\omega)}{{\tilde \Sigma}(\omega)} \frac{1}{\sqrt{1- \left(\frac{\Phi(\omega)}{{\tilde \Sigma}(\omega)}\right)^2}} =
\frac{\Phi^*(\omega)}{{\tilde \Sigma}^*(\omega)} \frac{1}{\sqrt{1- \left(\frac{\Phi^*(\omega)}{{\tilde \Sigma}^*(\omega)}\right)^2}} =
 \frac{\Delta(\omega)}{\sqrt{\omega^2 - (\Delta(\omega))^2}} \nonumber \\
S_\Sigma (\omega) &=&  \frac{{\tilde \Sigma}(\omega)}{\sqrt{{\tilde \Sigma}^2(\omega) - \Phi^2(\omega)}} =\frac{1}{\sqrt{1- \left(\frac{\Phi(\omega)}{{\tilde \Sigma}(\omega)}\right)^2}} = \frac{1}{\sqrt{1- \left(\frac{\Phi^*(\omega)}{{\tilde \Sigma}^*(\omega)}\right)^2}} =
 \frac{\omega}{\sqrt{\omega^2 -(\Delta(\omega))^2}}
\label{aaa}
\eea
\end{widetext}
and  $ \chi^{''}(x)= \sgn(x)\frac{g^\gamma}{|x|^{\gamma}}\sin \frac{\pi\gamma}{2}$.
In these equations, the solution of the Eliashberg set in Matsubara frequencies, i.e., $\Phi^* (\omega_m)$ and ${\tilde \Sigma}^* (\omega_m)$ are considered as  inputs. The first term in each of the two equations is obtained by just replacing $\omega_m$ by $-i \omega$, and the second one cancels out non-analyticities. The last piece in the second term  cancels out the divergent contribution from $\chi (0)$. Note that the subtraction of the divergence   at $x=0$ has to be done before extending the model to large $N$.  The function $Q(\omega)$, which determines the relations between $\Phi^* (\omega)$ and
 ${\tilde \Sigma}^*(\omega)$ and the original $\Phi (\omega)$ and  ${\tilde \Sigma}(\omega)$, Eqs. (\ref{new_new}),  is
\beq
Q (\omega) = \frac{iP}{\sqrt{{\tilde \Sigma}^2(\omega) - \Phi^2(\omega)}}
\label{ex_1}
\eeq
where
\beq
P= \int dx \chi^{''} (x) \frac{T}{x} = \pi T \chi^{'} (0)
\label{aaaa}
\eeq
Equivalently we can express $\Phi (\omega)$ and  ${\tilde \Sigma}(\omega)$ via $\Phi^* (\omega)$ and
 ${\tilde \Sigma}^*(\omega)$ as
  \beq
 \Phi (\omega) = \Phi^* (\omega) \left(1 + Q^* (\omega)\right), {\tilde \Sigma} (\omega) = {\tilde \Sigma}^* (\omega) \left(1 + Q^* (\omega)\right),
 \label{new_new_1}
 \eeq
where
\beq
Q^* (\omega) = \frac{iP}{\sqrt{({\tilde \Sigma}^*)^2(\omega) - (\Phi^*)^2(\omega)}}
\label{exx_1}
\eeq
In Eqs. (\ref{aaa}-\ref{exx_1}) the branch cut of the square root is defined along positive real axis.

At $\omega=0$ we have
\bea
&&\Phi^* (0) = \frac{\pi T}{N} \sum_{m}\frac{\Phi^*(\omega_m)}{\sqrt{(\Phi^*(\omega_m))^2 + ({\tilde \Sigma}^*(\omega_m))^2}}
    \chi(\omega_m)  \label{s_7aa}   + \frac{i}{N}\int dx \chi^{''} (x) \left(\frac{S_\Phi(-x)}{\sinh{x/T}} - \frac{S_\Phi(0)}{x/T} \right) \nonumber \\
&&{\tilde \Sigma}^* (0) = i\pi T \sum_{m}\frac{{\tilde \Sigma}^*(\omega_m)}{\sqrt{(\Phi^*(\omega_m))^2 + ({\tilde \Sigma}^*(\omega_m))^2}}
    \chi(\omega_m )   + i\int dx \chi^{''} (x) \left(\frac{S_\Sigma (-x)}{\sinh{x/T}} - \frac{S_\Sigma (0)}{x/T} \right) \nonumber \\
\label{s_7bb}
\eea
The first term in the formula for ${\tilde \Sigma}^* (0)$ vanishes by symmetry, after summing up the contributions from positive and negative $\omega_m$.

We first consider large $N$. We assume and then verify that in this case  ${\tilde \Sigma}^*$ is parametrically larger than $\Phi^*$ not only along the Matsubara axis but also along the real axis.
 To leading order in $1/N$ we then have for the self-energy
 \beq
{\tilde \Sigma}^* (0) = i \int dx \chi^{''} (x)   \left(\frac{1}{\sinh{x/T}} - \frac{T}{x} \right)  =
 - i \pi T  \left(\frac{g}{\pi T}\right)^\gamma  S_\gamma
\label{s_9}
\eeq
where
\beq
S_\gamma = 2 \sin{\pi \gamma/2} \int_0^\infty \frac{dx}{x^\gamma} \left(\frac{1}{\pi x} - \frac{1}{\sinh{\pi x}}\right)
\label{s_10}
\eeq
We plot $S_\gamma$ in Fig.\ref{fig:Qgamma}

For $\Phi^* (0)$  we find from Eq. (\ref{s_7aa})
\beq
\Phi^* (0) \approx \frac{\pi T}{N} \sum_{m}\frac{\Phi^*(\omega_m)}{|{\tilde \Sigma}^*(\omega_m)|} \chi(\omega_m)
\label{s_11}
\eeq
Using the fact that at large N the dominant contribution to the Matsubara sum comes from $m=0,-1$ and substituting the expressions for $\Phi^* (\pm \pi T)$ and $\Sigma^* (\pm \pi T)$,  we obtain
\beq
\Phi^* (0) =  \left(\frac{2}{N}\right)^{3/2}  \pi T \left(\frac{g}{\pi T}\right)^\gamma  \left(1- \left( \frac{T}{T_c}\right)^\gamma\right)^{1/2}
 \label{s_12}
\eeq
Then
$D_0 = \Phi^*(0)/{\tilde \Sigma}^* (0)$ is
\beq
D_0 = i \left(\frac{2}{N}\right)^{3/2} \frac{1}{S_\gamma} \left(1 - \left( \frac{T}{T_c}\right)^\gamma\right)^{1/2}
\label{s_14}
\eeq
and the DOS at zero frequency is
\beq
N(0) = N_0 \left(1 - \left(\frac{2}{N}\right)^{3}  \frac{\left(1 - \left( \frac{T}{T_c}\right)^\gamma\right)}{2 S^2_\gamma}\right)
 \label{s_15}
 \eeq
 This agrees, up to a prefactor, with the result that we obtained along the Matsubara axis, by assuming that $D(\pi T)$ is comparable with $D(\omega =0)$.

 We emphasize that $N(0)$ differs  from the normal state value $N_0$ at all $T < T_c$, including $T=0$, where we expect superconductivity to disappear. We will show below that the limit $\omega \to 0$ and $T \to 0$ has to be taken carefully, and at any non-zero $\omega$ the DOS indeed transforms into $N_0$ at $T \to 0$.
 Still, strictly at $\omega =0$, $N (0) < N_0$.   This is similar, but indeed not identical, to behavior of $N(\omega)$ in an ideal  BCS superconductor, where
   $N(0) =0$ for all $T$ up to $T_c$, while $N(\omega \neq 0)$ approaches $N_0$ at $T \to T_c$.

\begin{figure}
	\begin{center}
    \includegraphics[width=16.5cm]{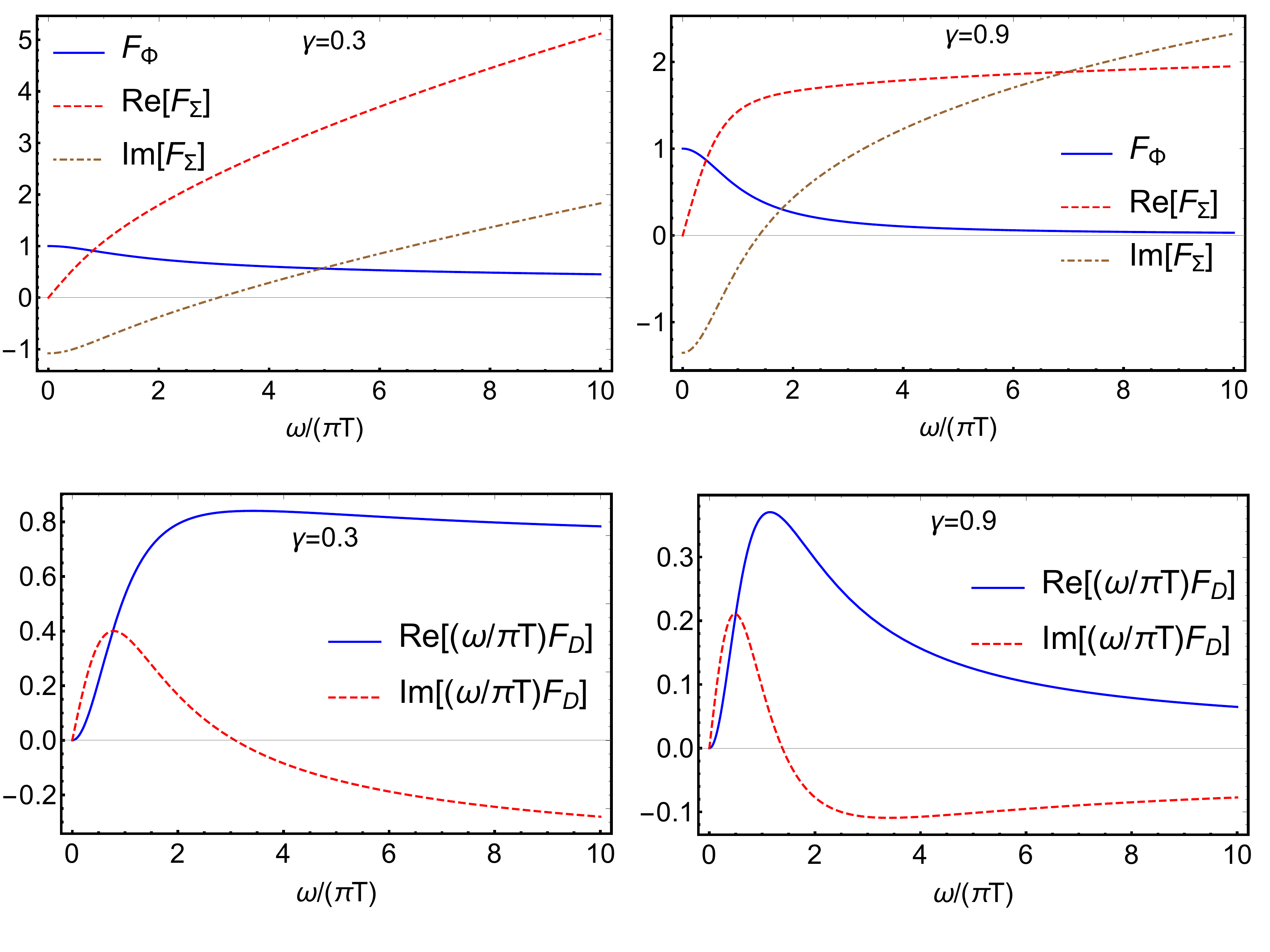}
		\caption{The scaling functions $F_{\Phi}(\frac{\omega}{\pi T})$,  $F_{\Sigma}(\frac{\omega}{\pi T})$ and
$\frac{\omega}{\pi T}F_D(\frac{\omega}{\pi T}) = \frac{\omega}{\pi T} F_{\Phi}(\frac{\omega}{\pi T})/F_{\Sigma}(\frac{\omega}{\pi T})$ for the pairing vertex, the self-energy and the gap function respectively, see Eqs. (\ref{s_17}), (\ref{s_19}), and \eqref{s_21_1}. We recall that $F_{\Phi}(\frac{\omega}{\pi T})$ and  $F_{\Sigma}(\frac{\omega}{\pi T})$ are computed without the thermal contribution.
The function $F_{\Phi}(x)$ is real, $F_{\Sigma}(x)$ and $F_D(x)$ are complex, i.e., the gap function $\Delta (x)$ is a complex function of frequency. The results are for $\gamma =0.3$ and $\gamma =0.9$. Observe that $\Imm F_{\Sigma}(x)$ changes sign at some frequency. This sign change is necessary to satisfy KK relation on $\Sigma^* (\pi T) =0$ (see Fig. \ref{fig:fsigma}). }\label{fig:F}
	\end{center}
\end{figure}

  We next move to finite $\omega$. For $\Phi (\omega)$, the second term in (\ref{eq:anacon}) still scales as $\Phi (\omega)/N$ and can be neglected. Evaluating the first term by summing up contributions from $m =0, -1$ at which $\Phi (\omega_m)/|{\tilde \Sigma} (\omega_m)|$ is the largest at large $N$, we obtain
  \beq
  \Phi^* (\omega) =  \left(\frac{2}{N}\right)^{3/2} \pi T  \left(\frac{g}{\pi T}\right)^\gamma  \left(1 - \left(\frac{T}{T_c}\right)^\gamma\right)^{1/2}
  F_\Phi \left(\frac{\omega}{\pi T}\right)
  \label{s_16}
  \eeq
  where
  \bea
  F_\Phi (x) = \frac{1}{2} \left( \frac{1}{\left(1 + ix\right)^\gamma} + \frac{1}{\left(1 - ix \right)^\gamma}\right)
    \label{s_17}
  \eea
 Note that in this  large $N$  approximation  $\Phi^* (\omega)$ is real and even in $\omega$.

 Because $\Phi^* (\omega)$ is small in $1/N^{3/2}$, the  self-energy at finite $\omega$ remains the same as in the normal state, up to $1/N^3$ corrections:
  \beq
 \Sigma^* (\omega)  = \pi T  \left(\frac{g}{\pi T}\right)^\gamma  F_\Sigma \left(\frac{\omega}{\pi T}\right)
 \label{s_18}
 \eeq
 where
 \begin{widetext}
 \bea
 &&F_\Sigma (x)= i \sum_{m=0}^\infty  \left( \frac{1}{\left(2m+1 + ix\right)^\gamma} - \frac{1}{\left(2m+1 - ix\right)^\gamma}\right)  \nonumber \\
 && - i \sin{\frac{\pi \gamma}{2}} \int_0^\infty \frac{dy}{y^\gamma} \left( \frac{2}{\pi y} - \coth{\frac{\pi y}{2}} + \frac{ \sinh{\pi y} }{ \cosh{\pi y} +\cosh{\pi x}}\right).
 \label{s_19}
 \eea
 \end{widetext}
  The first term in $F_\Sigma (x)$ is real, the second is imaginary.  At large $x$ (i.e., at $\omega \gg \pi T$), $F_\Sigma (x) \approx (x^{1-\gamma}/(1-\gamma)) e^{i \pi \gamma/2}$.  We plot the scaling functions $F_\Phi (x)$, $\Ree[F_\Sigma (x)]$, and $\Imm [F_\Sigma (x)]$ in Fig. \ref{fig:F}.

\begin{figure}
  \begin{center}
    \includegraphics[width=12cm]{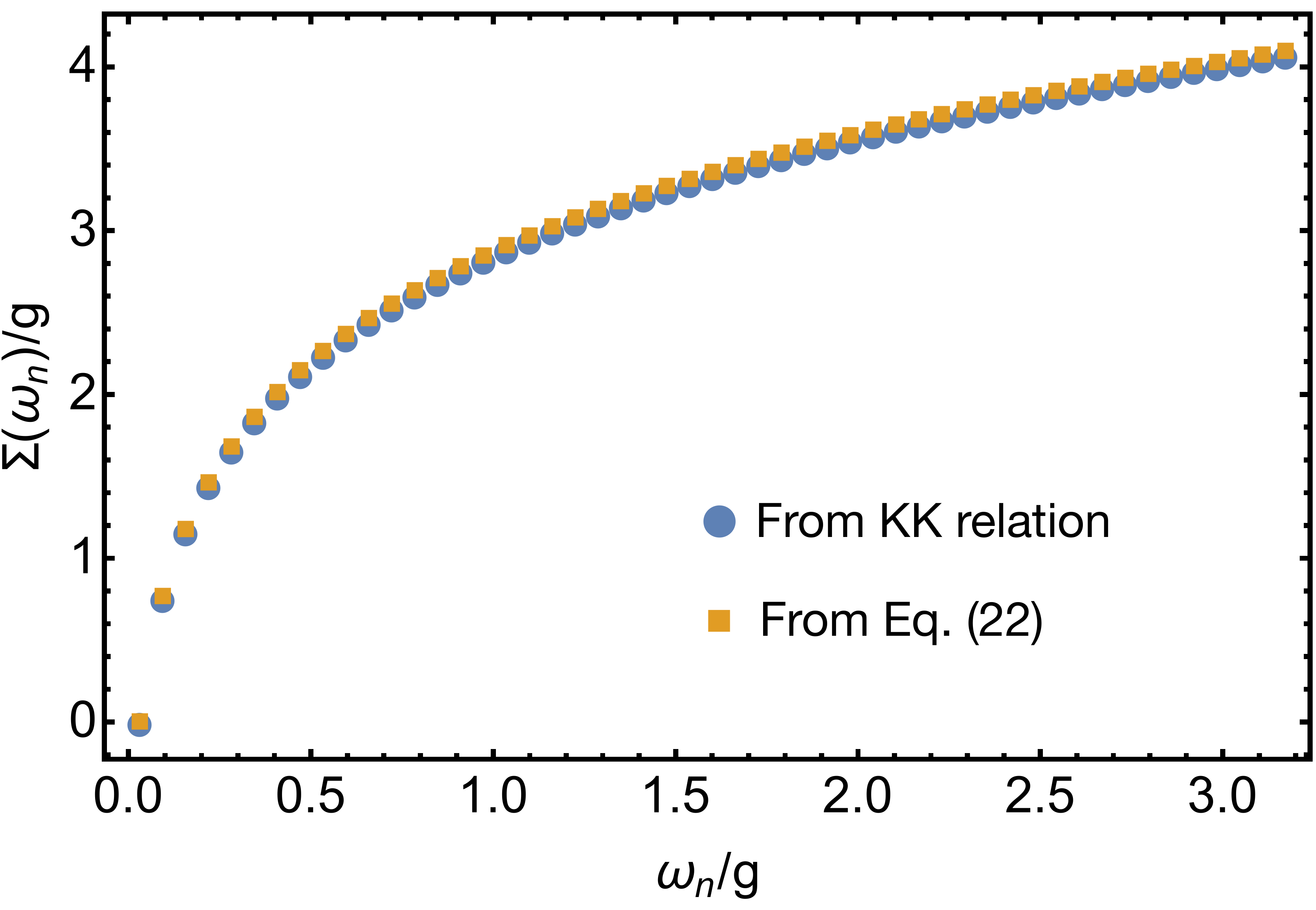}
    \caption{ The verification of the KK transformation.
     Yellow squares -- the self-energy obtained directly along the Matsubara axis: $\Sigma^* (i\omega_n) = 2\pi T (g/2\pi T)^\gamma H(n,\gamma)$, Eq. (\ref{s_5}).
    Blue circles -- the self-energy $\Sigma^* (i\omega_n) = -i \pi T (g/\pi T)^\gamma F_\Sigma (\omega_n)$, where $F_{\Sigma}(i\omega_n) = (2i \omega_n/pi) \int_0^\infty  d x \Imm F_{\Sigma}(x)/(x^2 + \omega^2_n)$ is obtained by KK transformation from
     $\Imm F_{\Sigma}(x)$ along the real axis, see \eqref{s_19}.
      The two expressions coincide. To better show this we manually split the two expressions for $\Sigma^* (i\omega_n)$  by multiplying the yellow curve by
       1.01.  Observe that $F_\Sigma (i\pi T) =0$, i.e., the self-energy $\Sigma^* (i\omega_n)$, extracted from KK transformation,  vanishes at the first Matsubara frequency.  We set  $\gamma=0.9$ and $T=0.01g$. }\label{fig:fsigma}
  \end{center}
\end{figure}

We see that  Im $\left[F_\Sigma \left(\frac{\omega}{\pi T}\right)\right]$ changes sign as a function of frequency (and then Im $[\Sigma^* (\omega)]$ also changes sign).  This sign change is necessary because $\Sigma^* (\pi T) =0$ and   $\Imm [\Sigma^* (\omega)]$ are related by Kramers-Kronig(KK) formula,
\begin{equation}\label{eq:KK}
  2 T \int_0^\infty d\omega \frac{\Imm\Sigma^*(\omega)}{\omega^2+(\pi T)^2} = \Sigma^* (\pi T) =0,
\end{equation}
and the vanishing of the integral in (\ref{eq:KK}) is only possible if
  $\Imm [\Sigma^* (\omega)]$ has different sign at small and large frequencies.   We verified numerically that the KK relation is indeed satisfied, see Fig.\ref{fig:fsigma}. We remind in this regard that $\Sigma^*$ is the self-energy without the thermal contribution.
  For the full self-energy $\Imm[\Sigma(\omega)]$ indeed remains positive for all frequencies.

Substituting the results for $\Phi^* (\omega)$ and ${\tilde \Sigma}^* (\omega)$ into $D(\omega) = \Phi^* (\omega)/{\tilde \Sigma}^ (\omega)$, we obtain
\beq
D(\omega)  =  \left(\frac{2}{N}\right)^{3/2}  \left(1 - \left(\frac{T}{T_c}\right)^\gamma\right)^{1/2}  F_D \left(\frac{\omega}{\pi T}\right),
\label{s_20}
\eeq
where at  $\omega \leq g$, when the bare $\omega$ term is smaller than $\Sigma^* (\omega)$, i.e., ${\tilde \Sigma}^* (\omega) \approx \Sigma^* (\omega)$,
\beq
F_D (x)  =  \frac{F_\Phi \left(\frac{\omega}{\pi T}\right)}{F_\Sigma \left(\frac{\omega}{\pi T}\right)}
\label{s_21_1}
\eeq
The DOS is
\bea
&& N(\omega) =  N_0 \Ree  \left[\frac{1}{(1 - D^2 (\omega))^{1/2}}\right] \approx N_0 \left(1 + \frac{1}{2}\Ree  \left[D^2 (\omega)\right]\right) \nonumber \\
&& = N_0 \left(1 + \frac{1}{2}\left(\frac{2}{N}\right)^{3} \left(1 - \left(\frac{T}{T_c}\right)^\gamma\right) \Ree  \left[F^2_D \left(\frac{\omega}{\pi T}\right)\right]\right)
\label{s_21}
\eea
We see that the magnitude of
$N(\omega)/N_0 -1 =\frac{1}{2} \Ree D^2 (\omega)$
is determined by the overall temperature-dependent factor in (\ref{s_20}) and depends on $T/T_c$ ratio. However, the frequency dependence of $D(\omega)$ and of the DOS is determined by $F_D (\omega/(\pi T))$, which for any given $\gamma$ is a universal function of $\omega/T$  and  does not depend on $T/T_c$.  This implies that the characteristic frequency, at which $N(\omega)$ deviates from $N_0$, is determined by the temperature rather than by the magnitude of the superconducting gap, as was the case for a BCS superconductor.

\begin{figure}
	\begin{center}
		\includegraphics[width=16cm]{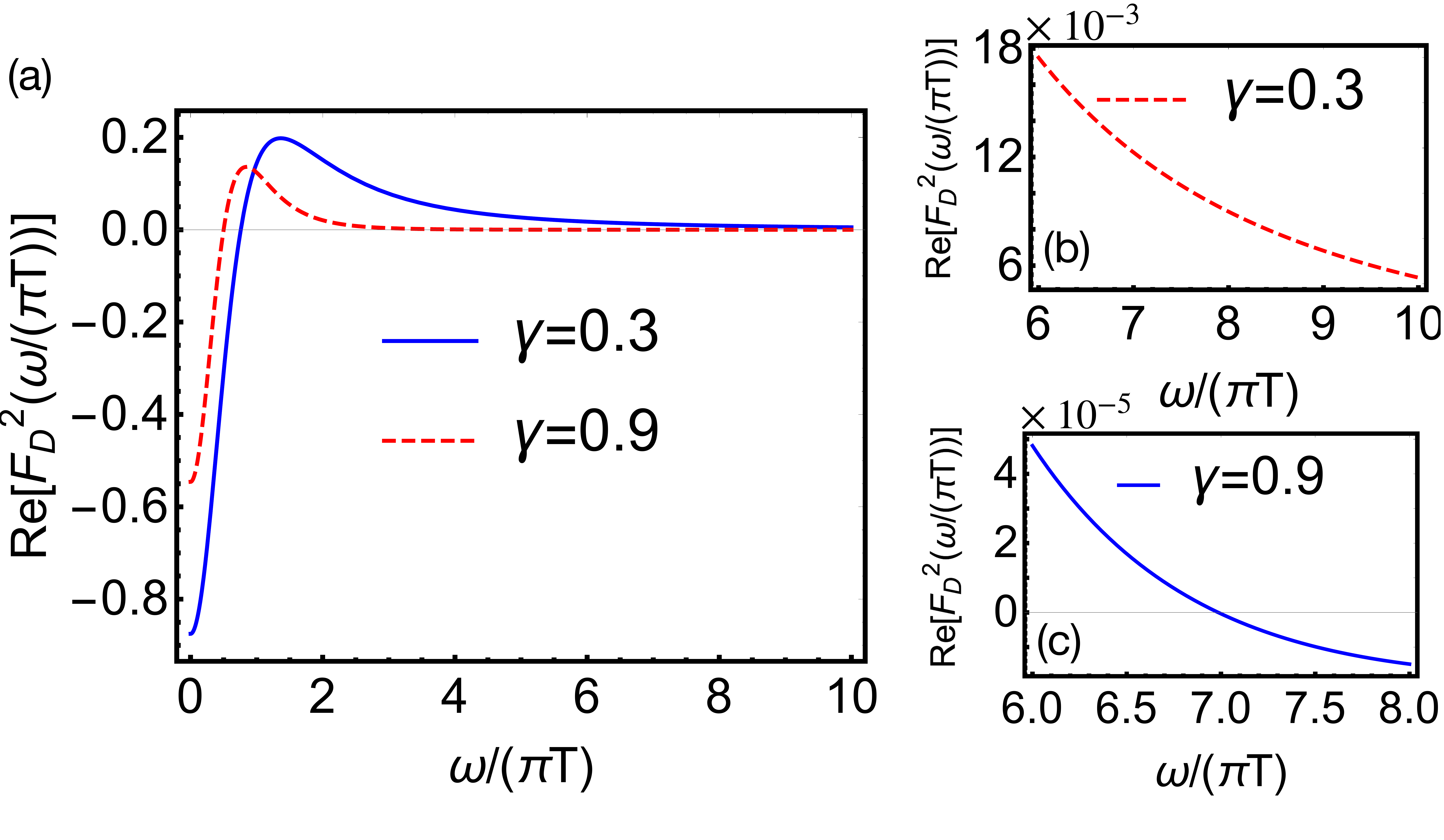}
		\caption{(a) The real part of the scaling function  $F_D^2(\frac{\omega}{\pi T})$, defined in Eq. (\ref{s_21_1}), for $\gamma=0.3$ and $\gamma=0.9$. The Re[$F_D^2(\frac{\omega}{\pi T})$] determines the frequency dependence of the DOS at large $N$,  Eq. (\ref{s_21}).   In the normal state
$F_D =0$. Observe that Re[$F_D^2(\frac{\omega}{\pi T})$] has a peak at $\omega \sim T$.   (b) and (c) The magnified plots of  Re[$F_D^2(\frac{\omega}{\pi T})$] at lager $\omega/(\pi T)$.
 For $\gamma =0.3$,  Re[$F_D^2(\frac{\omega}{\pi T})$] gradually decreases, for $\gamma =0.9$  it
  changes sign at $\frac{\omega}{\pi T}\sim7$. }\label{fig:FD}
	\end{center}
\end{figure}

\begin{figure}
  \begin{center}
    \includegraphics[width=10cm]{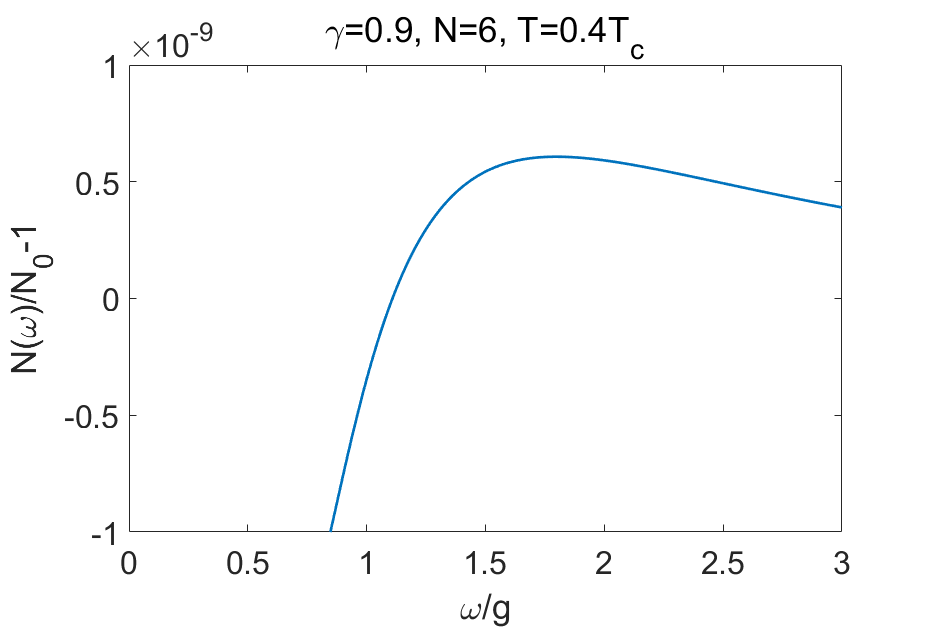}
    \caption{DOS at large $\omega \sim g$ for $\gamma=0.9$. We set $N=6$ and $T=0.4T_c$. At some $\omega\sim g$, $N(\omega)-N_0$ changes sign from negative to positive, and at even larger frequencies approaches zero from above. }\label{fig:DOSlarge_omega}
  \end{center}
\end{figure}
\begin{figure}
  \begin{center}
    \includegraphics[width=12cm]{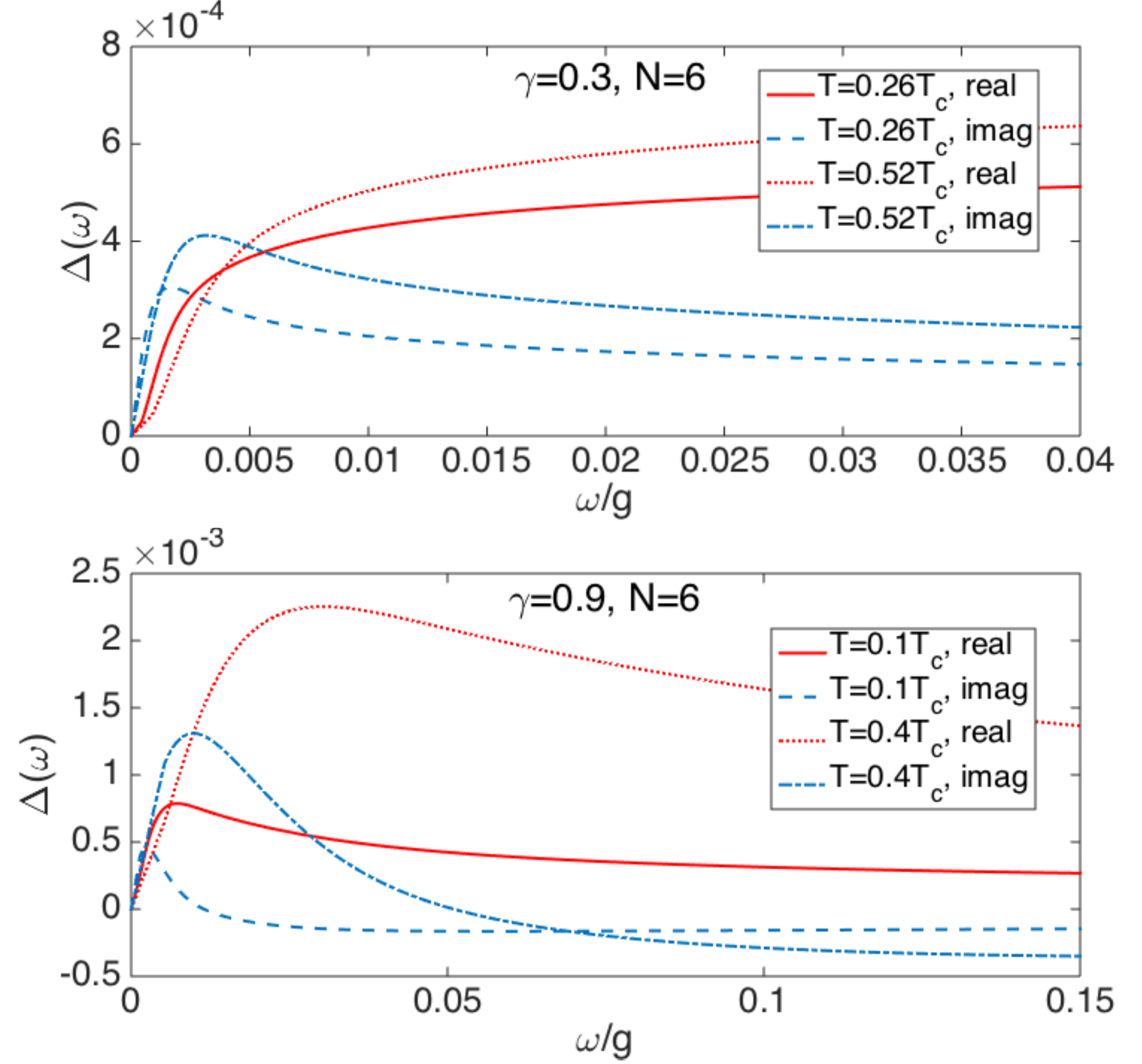}
    \caption{$\Delta(\omega)$ for various $T>T_{cross}$. Upper panel: $\gamma=0.3, N=6.$ Lower panel: $\gamma=0.9, N=6.$ Red lines are for the real part $\Delta'(\omega)$ and blue lines are for the imaginary part $\Delta''(\omega)$. At small but non zero $\omega$, both the real and imaginary parts are finite, in contrast to the BCS-like behavior where $\Delta''(\omega)$ is zero up to some $\omega_0\approx \Delta'$ at low temperatures.  }\label{fig:largeNgap}
  \end{center}
\end{figure}

\begin{figure}
  \begin{center}
    \includegraphics[width=16cm]{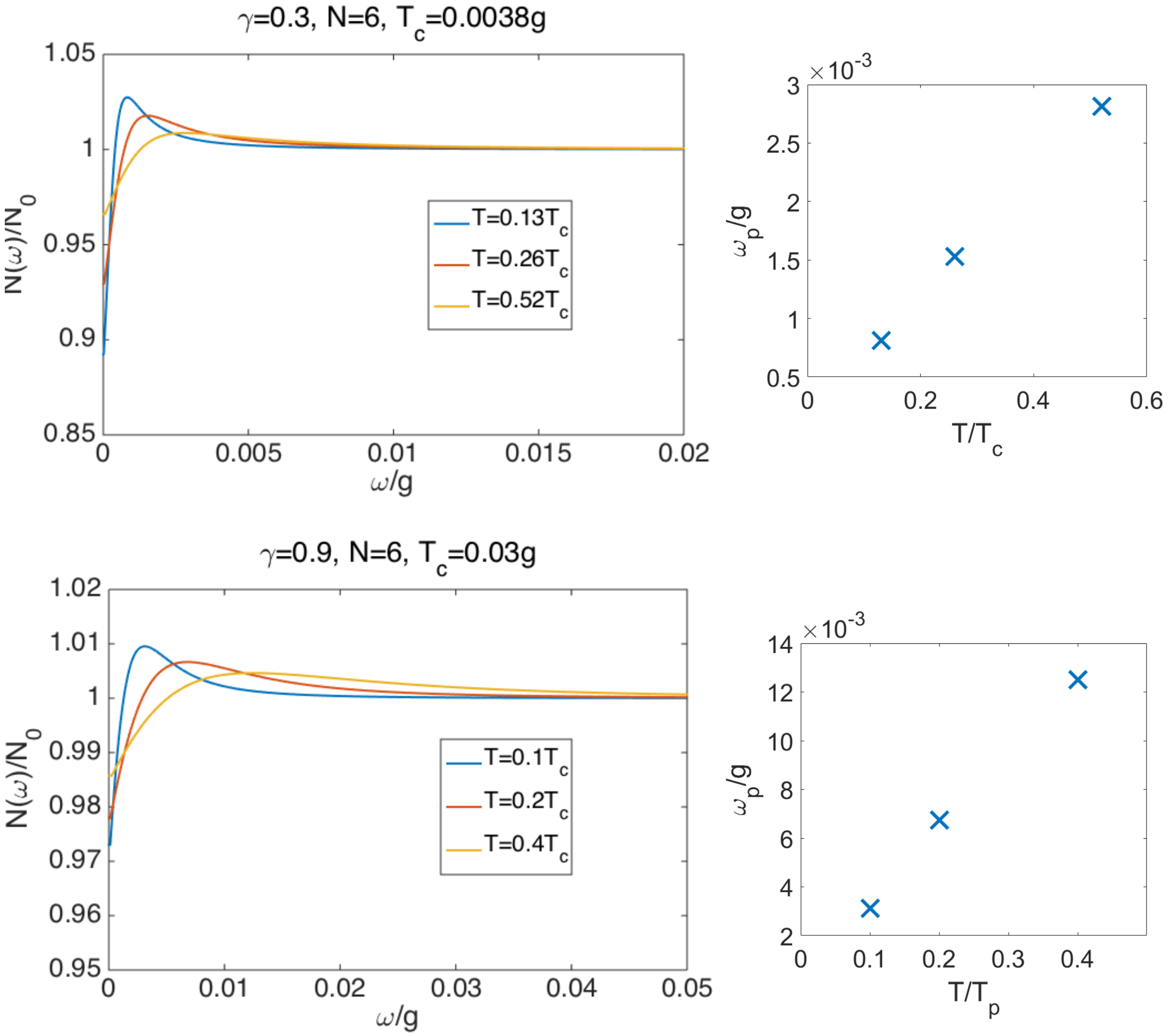}
    \caption{The DOS $N(\omega)$  for various $T > T_{cross}$. Upper panel: $\gamma =0.3, N=6$. Lower panel: $\gamma =0.9, N=6$.  Right panels: The temperature dependence of the characteristic frequency $\omega_p$, defined as the peak position of the $N(\omega)$.
    }\label{fig:largeNdos}
  \end{center}
\end{figure}

\begin{figure}
	\begin{center}
    \includegraphics[width=7cm]{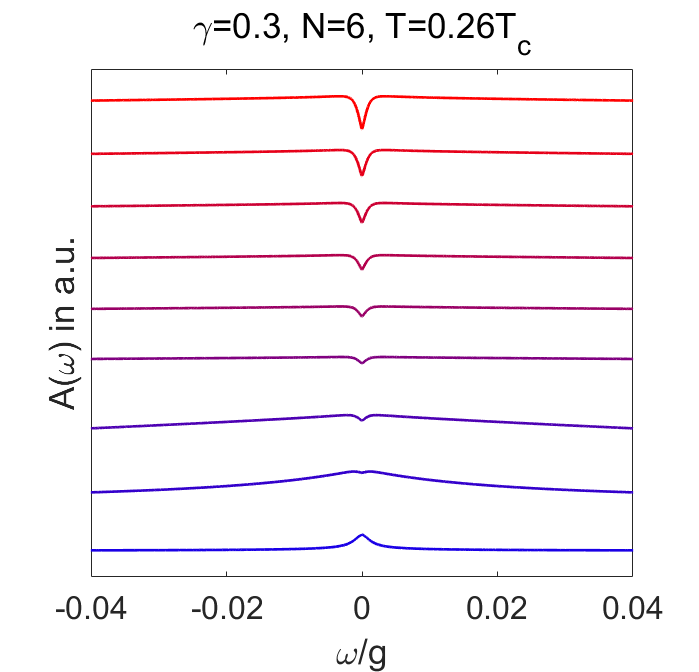}\includegraphics[width=7cm]{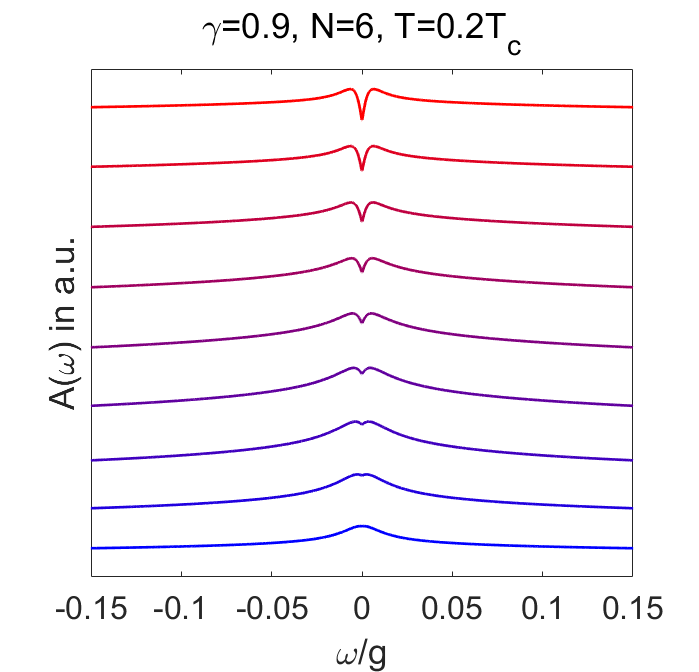}
		\caption{The spectral function $A(\omega)$ at a fixed $T > T_{cross}$, plotted as a function of $\omega$ for various values of parameter $P$,
 which measures the strength of thermal contributions to the self-energy and the pairing vertex. At large $P$,  $A(\omega)$ shows the same behavior as the DOS, with the dip at small $\omega$. At small $P$, it shows instead the maximum at $\omega =0$.  The plots are for
 $\gamma=0.3$ and $\gamma=0.9$. }\label{fig:largeNspectral}
	\end{center}
\end{figure}

Because $F_\Phi (x)$ is real,
\beq
\Ree F^2_D (x) = F^2_\Phi(x) \frac{(\Ree F_\Sigma (x))^2 - (\Imm F_\Sigma (x))^2}{\left((\Ree F_\Sigma (x))^2 + (\Imm F_\Sigma (x))^2\right)^2}
\label{very_new}
\eeq
At small $x = \omega/\pi T$, $\Ree F_\Sigma (x) \propto x^2$ and $\Imm F_\Sigma (x)$ is finite.  Then $\Ree F^2_D (x)$ is negative. At $x$ where
$\Imm F_\Sigma (x)$ changes sign, $\Ree F_\Sigma (x)$ is finite, hence for this $x$, $\Ree F^2_D (x)$ is positive.  In between $\Ree F^2_D (x)$ then necessary changes sign. This in turn implies that  $N(\omega)< N_0$ at small $x$ and exceeds $N_0$ at larger $x$.  At even larger $x \gg 1$, $N(\omega)$ approaches $N_0$. Then, for any $\gamma$,  $N(\omega)$ has a dip at $\omega =0$ and a hump at  a characteristic frequency set by temperature, rather than by the gap itself. This frequency then {\it increases with increasing $T$}, in qualitative difference with a BCS superconductor, in which the maximum in the DOS is located at $\omega = \Delta (T)$, and shifts to a lower frequency with increasing $T$ because  $\Delta (T)$ gets smaller.
We plot $\Ree F^2_D (x)$ in Fig.\ref{fig:FD} for two different $\gamma$. The hump at $\omega \sim T$ is clearly visible.  The position of the hump shifts to a lower frequency with increasing $\gamma$ but remains at a finite $\omega$ even at $\gamma =1$.

On a more careful look, we find that there is still a small difference in the behavior  of the DOS between $\gamma <1/2$ and $\gamma >1/2$. Namely, at
 $\omega \gg T$, $\Ree \Sigma^* (x) = \cos{\pi \gamma/2}  (x^{1-\gamma}/(1-\gamma))$ and $\Imm \Sigma^* (x) = \sin{\pi \gamma/2}  (x^{1-\gamma}/(1-\gamma))$. As a result, $\Ree F^2_D (x) \propto \cos{\pi \gamma}$, i.e., it is positive at $\gamma <1/2$ and negative at $\gamma >1/2$.  This implies that for $\gamma >1/2$  $N(\omega)$ crosses $N_0$ twice at $\omega = O(T)$ because $(\Imm F_\Sigma (x))^2$ is larger than
 $(\Ree F_\Sigma (x))^2$ at both large and small frequencies.  The second crossing at $x \sim 7$ is seen in Fig. \ref{fig:F}  for $\gamma =0.9$.
  Digging further into this issue, we find that for $\gamma >1/2$, $N(\omega)$ crosses $N_0$ one more time, now at $\omega \sim g \gg T$, when the bare $\omega$ term in ${\tilde \Sigma}^* (\omega)$ becomes relevant, and at highest $\omega$ approaches $N_0$ from above.  To see this, we extend the analysis of the DOS to $\omega \sim g$. The calculation is straightforward and we only cite the result:
   the difference $N(\omega)/N_0-1$ at $\omega \sim g$ is proportional to
     $\cos{\pi \gamma} + (1-\gamma)^2 (\omega/g)^{2\gamma} + 2  (1-\gamma) (\omega/g)^{\gamma} \cos{\pi\gamma/2} =0$.
      Solving this equation for $\gamma >1/2$,  we find the sign change of $N(\omega)/N_0-1$ at $\omega  = \omega_1 \sim g$.
       We show this in Fig. \ref{fig:DOSlarge_omega}.

In Fig.\ref{fig:largeNgap} and 
Fig.\ref{fig:largeNdos} we show the results of the full numerical calculation of the temperature evaluation of the gap $\Delta(\omega)$ and the DOS $N(\omega)$ for two values of $\gamma$: $\gamma =0.3$ and $\gamma =0.9$. For $\gamma =0.9$, $N=6$ is above $N_{cr}\sim 1.3$.
  For $\gamma=0.3$ we show the results for $N =6$, which is below  $N_{cr} \approx 9.6$ (the numerical analysis for $N > N_{cr}$ for such small $\gamma$ is challenging).  For  $N < N_{cr}$,  the behavior similar to the one at  large $N$  exists above the crossover temperature $T_{cross} (N)$ (see Sec. \ref{sec:smallN})  and we show the results only in this $T$ range. The value of  $T_{cross} (N=6)$ for $\gamma =0.3$ is only $0.01 T_c$,
  so the range of $T > T_{cross}$ is rather wide.
  The gap $\Delta(\omega)$ is complex even at very small $\omega$, in contrast to the conventional BCS-like behavior where $\Delta(\omega)$ is almost real up to some frequency which is approximately equal to this real value. 
  For the DOS we clearly see that there is a dip in $N(\omega)$ at small frequencies and a characteristic frequency $\omega_p$ at which $N(\omega)$ approaches $N_0$ is set by the temperature.

  A remark is in order here.  The  $\int d \omega N(\omega)$, with $N(\omega)$ as in Fig.\ref{fig:largeNdos} does have some $T$ dependence. At a first glance, this contradicts the requirement that the total number of particles is a conserved quantity. In fact, there is no contradiction.
The reasoning is that the momentum integration in Eliashberg equations is performed assuming particle-hole symmetry, i.e., neglecting contributions from energies of order $\mu$.  There are additional contributions to the DOS from energies of order $\mu$, both in the normal and the superconducting state.  They are not equal, because $\mu$ changes between normal and superconducting states~\cite{cee}. This additional contribution must be included to ensure particle conservation.

We next consider the spectral function $A(\omega)  = - (1/\pi) \Imm[G(k_F, \omega)]$. In terms of original $\Phi (\omega)$ and ${\tilde \Sigma} (\omega)$, we have
\beq
A(\omega) = -\frac{1}{\pi} \Imm \left[\frac{{\tilde \Sigma} (\omega)}{{\tilde \Sigma}^2 (\omega)- \Phi^2(\omega)}\right]
\label{b_1}
\eeq
Expressing  $\tilde{\Sigma}(\omega)$ and  $\Phi(\omega)$ via $\tilde{\Sigma}^*(\omega)$ and $\Phi^*(\omega)$, Eq. (\ref{exx_1}), we
 find
\begin{equation}
    A(\omega) =   -\frac{1}{\pi} \Imm \left[\frac{{\tilde \Sigma}^* (\omega)}{({\tilde \Sigma}^* (\omega))^2 - (\Phi^*(\omega))^2} L(\omega)\right]\\
\label{sf_1_b}
\end{equation}
where
\begin{equation}
  L(\omega)=\frac{1}{1 + Q^* (\omega)}=\frac{\sqrt{\tilde{\Sigma}^*(\omega)^2-\Phi^*(\omega)^2}}{iP + \sqrt{\tilde{\Sigma}^*(\omega)^2-\Phi^*(\omega)^2}}\label{sf_1_a}
\end{equation}
   To leading order in $1/P$, $A(\omega) \propto \frac{1}{P} \Ree\left[\frac{1}{\sqrt{1 - \left(\Phi^* (\omega)/{\tilde \Sigma}^* (\omega)\right)^2}} \right] \propto  N(\omega)/N_0$, i.e., the spectral function has the same dependence on $\omega$ as the DOS. Accordingly, at a finite $T$, $A(\omega)$ is non-zero for any frequency, and the position of the maximum in $A(\omega)$ scales with $T$ and remains at a finite frequency at $T_c$ (Fig.\ref{fig:largeNspectral}).  Like we said, this  behavior has been termed as ``gap filling".   If  $P$ is finite, either because the system is at some distance from a QCP, or we probe $A(\omega)$ for fermions not connected by momenta at which static $\chi$ diverges (like near-nodal fermions in the cuprates, if a pairing boson is an antiferromagnetic spin fluctuation), the behavior of $A(\omega)$ depends on the interplay between $P$ and the other term in $L(\omega)$ in (\ref{sf_1_a}).  If $P$ is smaller, $A(\omega)$ is given by (\ref{sf_1_b}), (\ref{sf_1_a}).
    Substituting the expressions for $\Phi^* (\omega)$ and ${\tilde \Sigma}^* (\omega)$ we find that in this situation $A(\omega)$ is peaked at zero frequency, as if the system was in the normal state (see Fig.\ref{fig:largeNspectral}).

The analysis beyond the leading order in $1/N$ proceeds in the same way as for Matsubara frequencies. As $N$ gets smaller,
 the maximum in the DOS becomes more pronounced, and, at the same time,  the DOS at zero frequency, $N(0)$ gets smaller.  These modifications get larger as $N$ decreases towards $N_{cr}$ and  eventually
 qualitatively change the system behavior at $N < N_{cr}$ and $T < T_{cross}$, as we show in the next Section.

\section{The case $N < N_{cr}$}
\label{sec:smallN}

At smaller $N <N_{cr}$ analytical solution is difficult to obtain because there is no obvious small parameter, so our discussion will be based on numerical results.

 \subsection{Non-linear gap equation in Matsubara frequencies}
\label{sec:smallNMatsubara}

\begin{figure}
	\begin{center}
		\includegraphics[width=8.1cm]{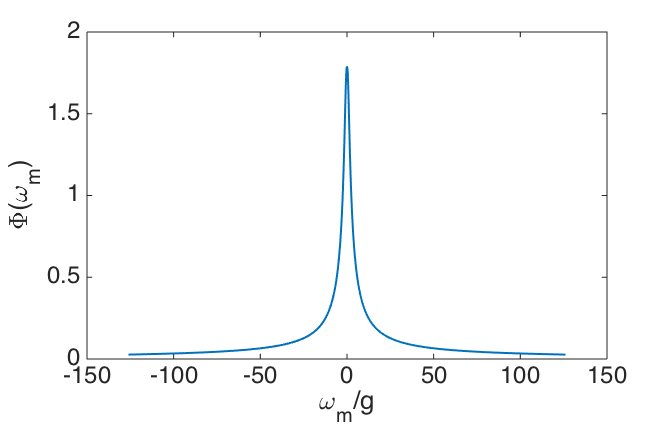}
		\includegraphics[width=8.1cm]{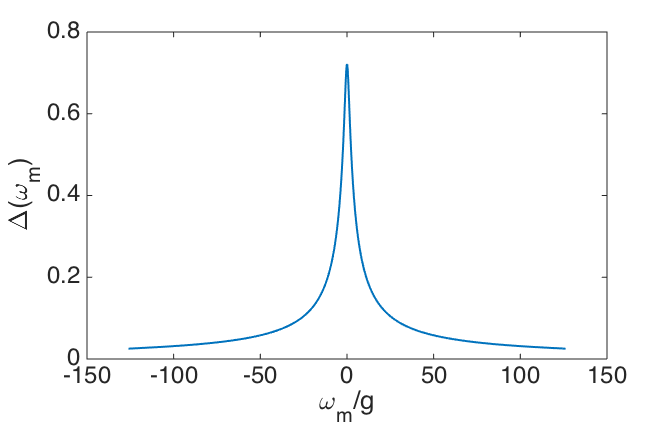}\\
		\caption{The pairing vertex $\Phi(\omega_m)$ and the gap $\Delta(\omega_m)$ as functions of Matsubara frequency for $\gamma=0.9$, $N=1$, and
 $T=0.18T_c < T_{cross}$. }\label{fig:Phi}
	\end{center}
\end{figure}
\begin{figure}
	\begin{center}
		\includegraphics[width=12cm]{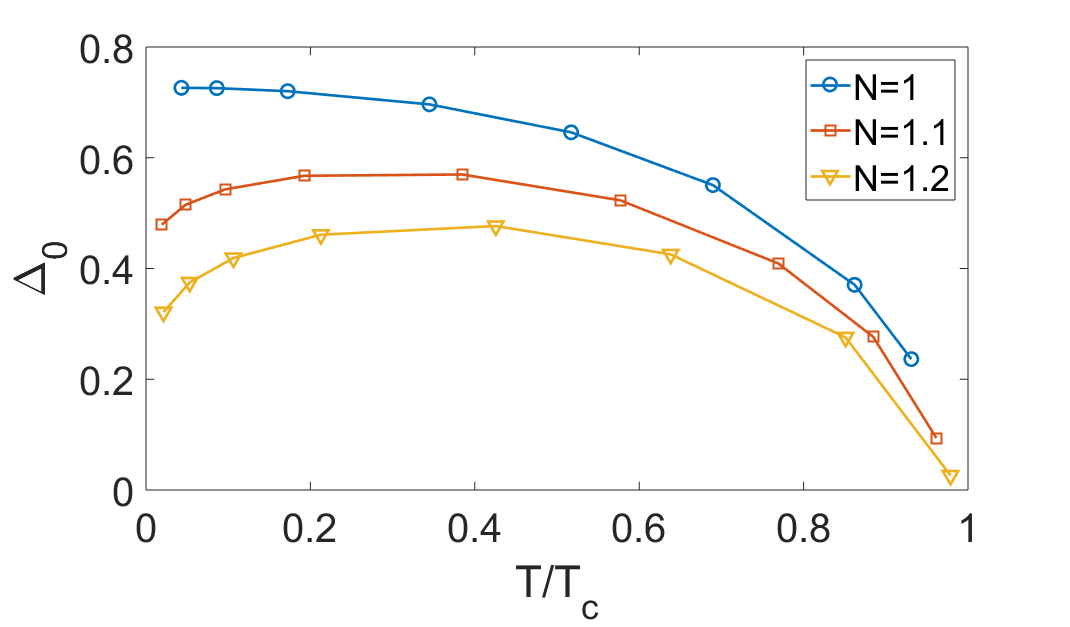}
		\caption{The gap $\Delta (\pi T) = \Delta_0$  as a function of temperature for $\gamma =0.9$ and three different $N < N_{cr} \approx 1.34$.
 The gap now tends to a finite value at $T=0$. For $N$ slightly below  $N_{cr}$, $\Delta_0 (T)$ is still non-monotonic, but for $N=1$, $\Delta_0$ monotonically increases with decreasing $T$. }\label{fig:Mgap}
	\end{center}
\end{figure}

In Fig.\ref{fig:Phi} we show the results for $\Phi^* (\omega_m)$ and $\Delta (\omega_m)$. We see that now $\Phi^*(\pi T) = \Phi^*_0$ and $\Delta (\pi T) = \Delta_0$ tend to finite values at $T \to 0$, i.e., show a ``conventional"  superconducting behavior.
Because at $T=0$ Matsubara frequency is a continuous variable, and it does not select between  $\pm \pi T$ and other frequencies, the development of
 a finite gap at $T=0$ implies that at $N < N_{cr}$ and a finite $T$, there should exist a $T$ range in which all Matsubara frequencies  equally contribute to the pairing, i.e., fermions with $\omega_m = \pm \pi T$ are no longer crucial to the pairing.  This is consistent with our earlier result that at $N < N_{cr}$, the transition temperature  remains finite even we exclude fermions with $\omega_m = \pm \pi T$ from Eliashberg equations (${\tilde T}_c (N)$ in Fig.\ref{fig:Tc}).

 In Fig.\ref{fig:Mgap} we show $\Delta (\pi T) =\Delta_0$ as a function of $T$. The temperature dependence of $\Delta_0$ is still non-monotonic, i.e.,
as $T$ is reduced below $T_c$, $\Delta_0$ first increases, and then drops below a certain $T$, before reaching a finite value at $T\to 0$. As $N \to 1$,
the maximum in $\Delta_0 (T)$ becomes shallow.   The frequency dependence of $\Phi^* (\omega_m)$ and of $\Delta (\omega_m)$ at a given $T$ is monotonic, with the maximum at $|\omega_m| = \pi T$.

In Fig.\ref{fig:Mgap} we compare the behavior of $\Delta_0 (T)$ at $N > N_{cr}$ and $N < N_{cr}$.  Near $T_c$, the behavior in the two cases is the same, but
   at low $T$ $\Delta_0 (T)$ at $N > N_{cr}$ continue decreasing, while $\Delta_0 (T)$ at $N < N_{cr}$ saturates.  The temperature at which the two curve separate marks the crossover between the conventional behavior at low $T$ and the behavior, undistinguishable from the one at $N > N_{cr}$, at higher $T$.   In the higher $T$ region, the pairing can still be viewed as induced by fermions with Matsubara frequencies $\pm \pi T$.
 The crossover line  $T_{cross} (N)$ ends at $T=0$ at $N = N_{cr}$, just like  ${\tilde T}_c (N)$, but these two temperatures  are not directly proportional to each other.

\subsection{Non-linear gap equation in real frequencies}
\label{sec:smallNreal}
\begin{figure}
	\begin{center}
    \includegraphics[width=12cm]{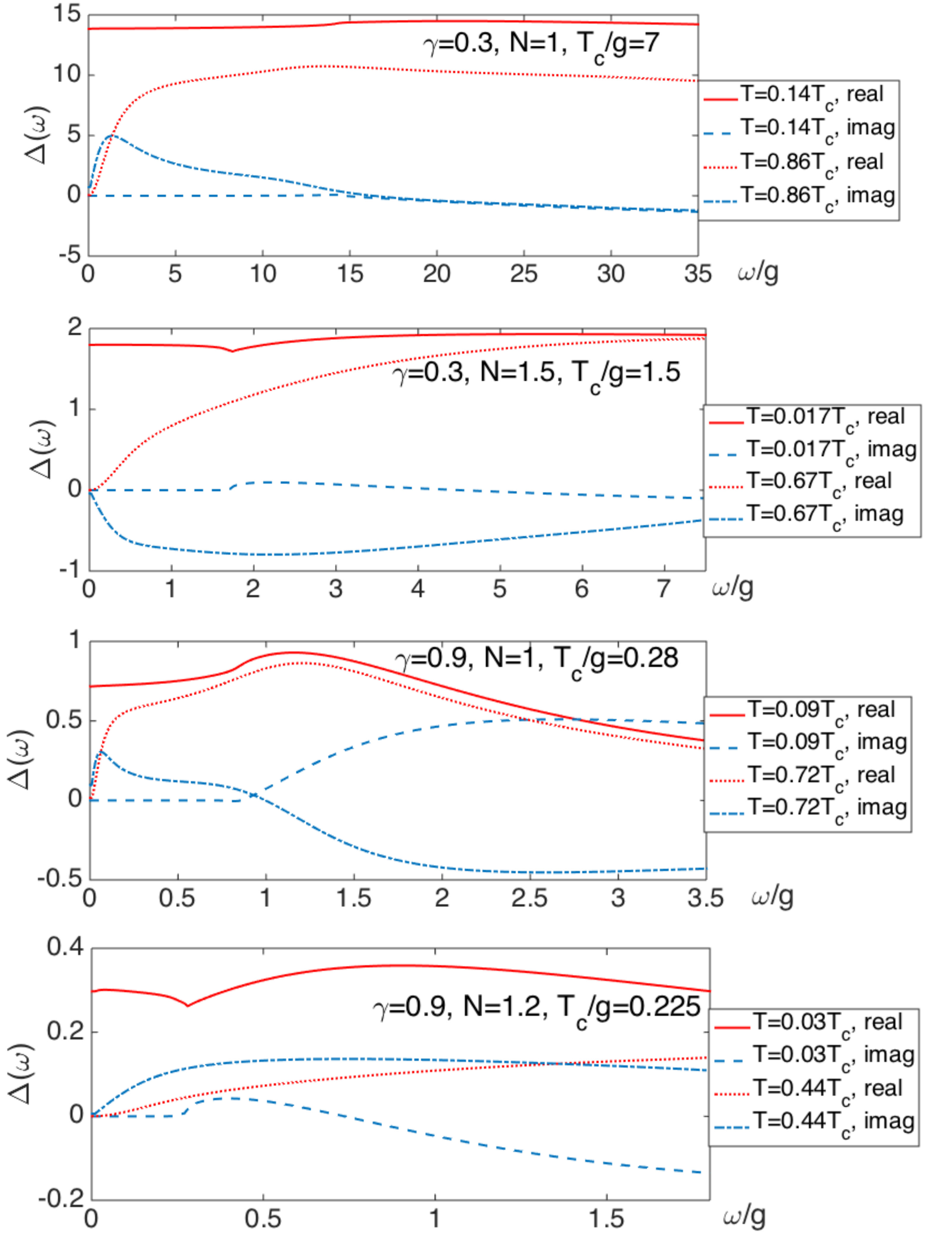}
		\caption{Real and imaginary parts of the gap $\Delta(\omega)$ as functions of $\omega$ for various $T$. The results are for $\gamma=0.3$ and $\gamma=0.9$, in both cases for $N<N_{cr}$.
    Red and blue lines 
     are for $\Delta'(\omega)$ and $\Delta''(\omega)$, respectively.  The data clearly show a crossover at $T \sim T_{cross}$ from BCS-like behavior at  smaller $T$ to the behavior similar to that at $N > N_{cr}$, at larger $T$.  }\label{fig:gap}
	\end{center}
\end{figure}

We used the same computational procedure as at large $N$ and obtained $\Phi^* (\omega)$, ${\tilde \Sigma}^* (\omega)$, and $\Delta (\omega)$ along the real frequency axis.  We present the results in Fig.\ref{fig:gap}.  We again see the crossover in the system behavior around $T_{cross} (N)$. At smaller $T$, the behavior of the gap function is conventional in the sense that $\Delta' (\omega =0)$ is finite and $\Delta^{''} (\omega)$ emerges only above a finite frequency $\omega_0 \approx \Delta' (0)$.  At higher $T$, at small frequencies $\Delta^{''} (\omega) \propto \omega$ and $\Delta^{'} (\omega) \propto \omega^2$, i.e., the systems displays gapless superconductivity. The self-energy $\Sigma^* (\omega)$ is strongly reduced below $\omega_0$ at small $T$ compared to that in the normal state, but almost recovers the normal state value in the regime of gapless superconductivity (Fig.\ref{fig:gap}).

\begin{figure*}
  \begin{center}
    \includegraphics[width=17cm]{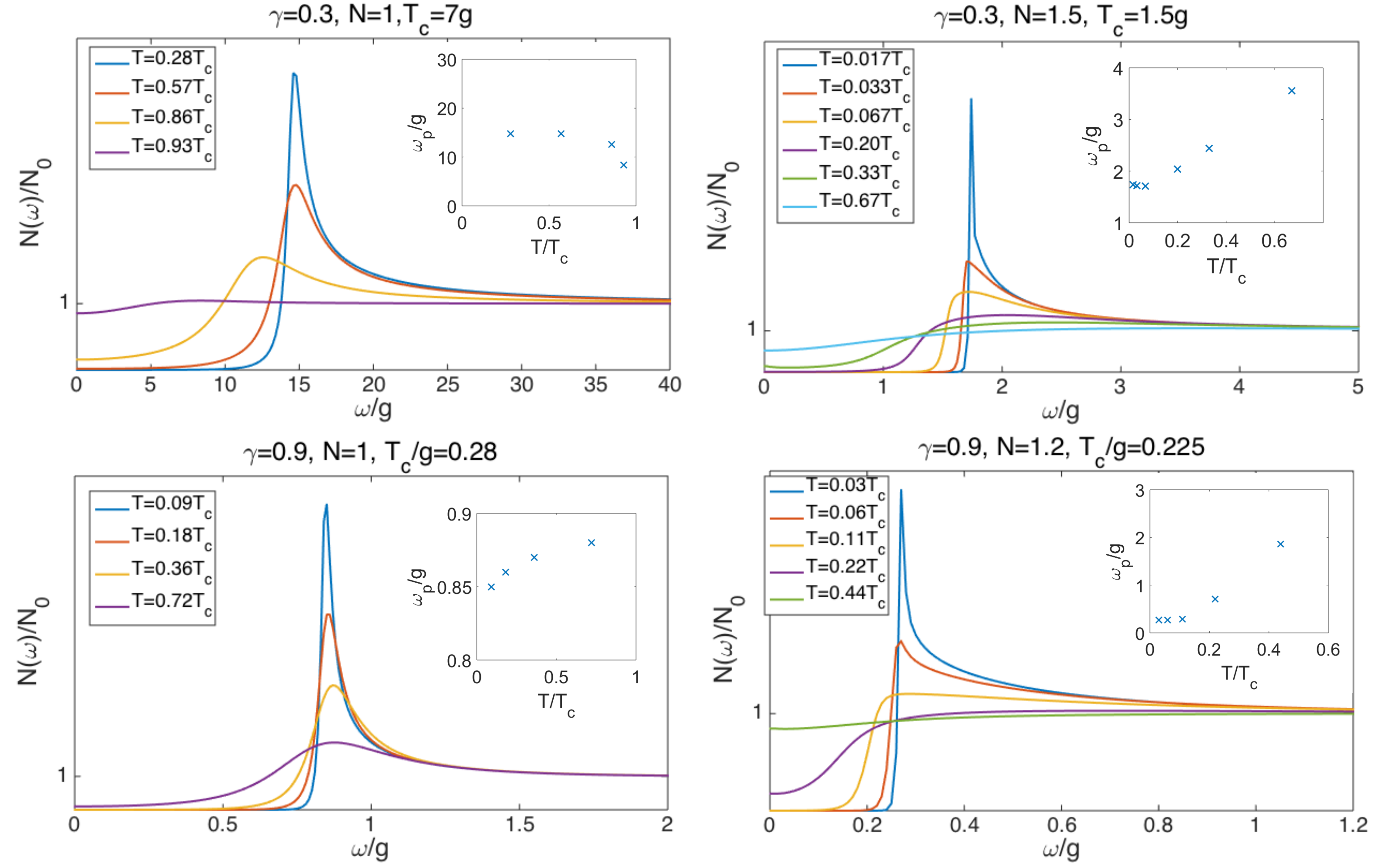}
    \caption{DOS $N(\omega)$ as a function of frequency for $\gamma=0.3$ and $\gamma=0.9$ and several $N < N_{cr} (\gamma)$. At low $T < T_{cross}$,
     the DOS has a sharp peak at $\omega = \Delta (T)$ and nearly vanishes below the peak.
 At higher $T > T_{cross}$  the DOS has qualitatively the same functional form as for large $N$, and the peak position shifts to a higher frequency with increasing temperature. The insets:  the peak position $\omega_p$ as function of $T/T_c$.  The crossover at $T_{cross}$ is clearly visible.}
 \label{fig:smallNdos}
  \end{center}
\end{figure*}

In Fig.\ref{fig:smallNdos} we show the behavior of the DOS $N(\omega)$.   We see qualitative change of the behavior between $T > T_{cross} (N)$ and $T < T_{cross} (N)$. At smaller $T$,
 the DOS is similar to that in a BCS superconductor: it has a sharp peak at $\omega \approx \Delta (0)$ and nearly vanishes below the peak frequency.
  At $T$ increases but remains smaller than $T_{cross} (N)$, the position of the maximum in  $N(\omega)$ moves to a smaller frequency because $\Delta (0)$ get reduced, i.e., the gap in the DOS ``closes in".   However, at higher $T > T_{cross} (N)$, the DOS becomes non-zero at all frequencies, and the position of its maximum moves to a higher frequency and remains finite at $T= T_c -0$, i.e., the gap in the DOS ``fills in".  We plot the variation of the position of the maximum in $N(\omega)$ with $T$ on the right side of each DOS in Fig.\ref{fig:smallNdos}.

\begin{figure*}
	\begin{center}
	
    \includegraphics[width=16.5cm]{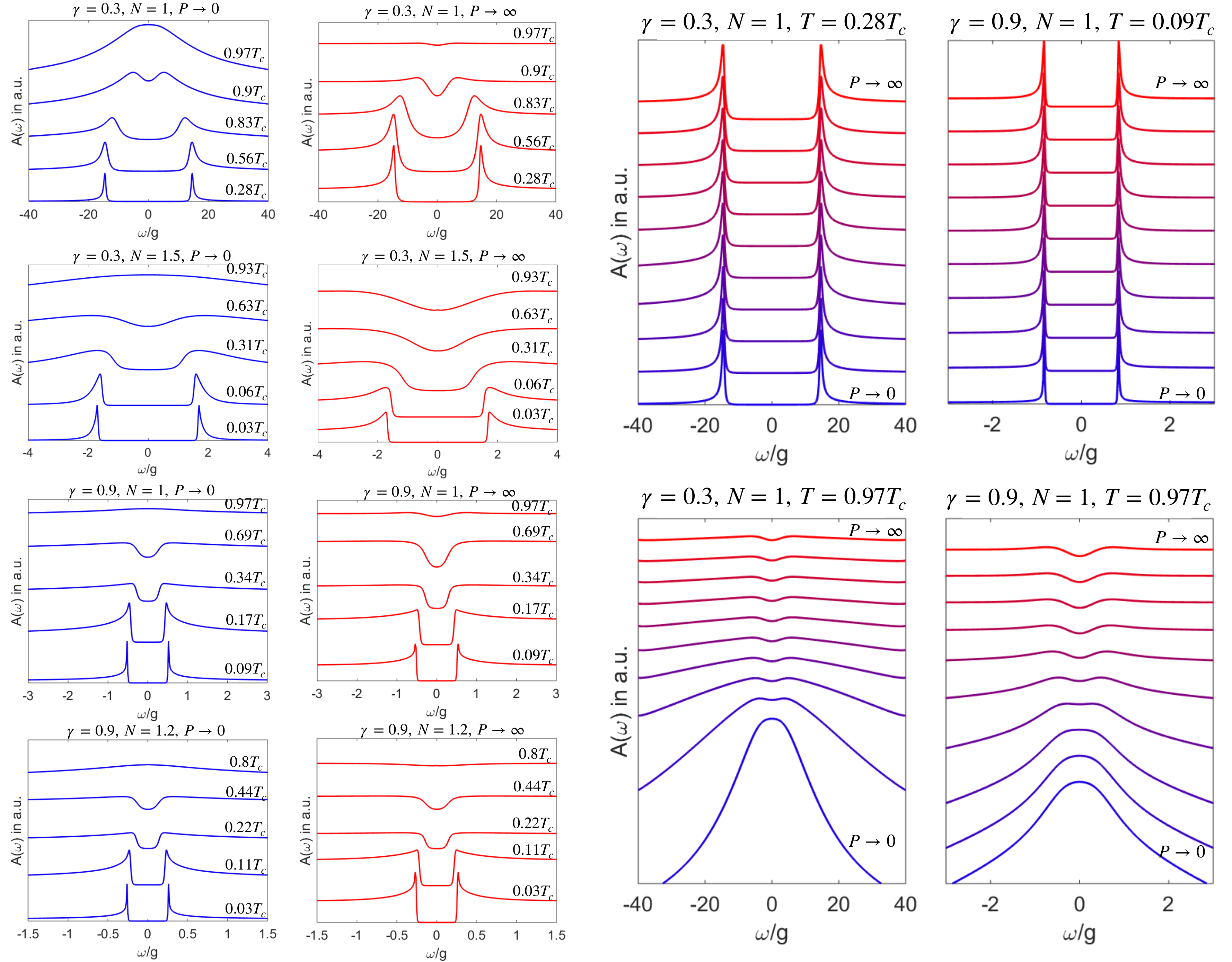}
		\caption{The spectral function $A(\omega)$ for $\gamma=0.3$ and $\gamma=0.9$ and several $N < N_{cr}$. Left panels: $A(\omega)$  for a set of  temperatures at either strong or weak thermal contribution  (the limits $P = \infty$ and $P=0$ in Eq. (\ref{sf_1_a})). At small $T < T_{cross}$ the spectral function has sharp peaks at $\omega = \pm \Delta (T)$, like in a BCS superconductor. At $T > T_{cross}$, $A(\omega)$ shows the same  behavior as the DOS in Fig. \ref{fig:smallNdos}, when the thermal contribution  is strong, and  develops a single peak at $\omega =0$ when the thermal contribution is weak. Right panels -- $A(\omega)$ at a fixed $T$ for different strengths of the thermal contribution. Upper panels -- $T < T_{cross}$, lower panels -- $T > T_{cross}$. }
\label{fig:smallNspectral}
	\end{center}
\end{figure*}

The spectral function $A(\omega)$ shows the similar crossover (Fig.\ref{fig:smallNspectral}). In the limit when the thermal contribution  is large, it shows the same behavior as $N(\omega)$. In the opposite limit, $A(\omega)$ at $T < T_{cross} (N)$ has two sharp  peaks at  frequencies close to $\pm \Delta (0)$, and at $T > T_{cross} (N)$ it has a single peak at $\omega =0$

\section{Discussion}
\label{sec:conclusions}

In this work we analyzed the interplay between the tendency towards fermionic incoherence and the tendency towards pairing
 near a quantum-critical point in a metal.  We used the $\gamma$ model of dynamical fermion-fermion interaction mediated by a critical boson with
  susceptibility $\chi (\Omega_m) \propto (g/|\Omega_m|)^\gamma$. We extended the model to $SU(N)$ global symmetry and used $N$ as a parameter. At large $N$,  the interaction in the pairing channel is smaller by $1/N$ than the one in the particle-hole channel, which gives rise to a fermionic incoherence.  Earlier work by some of us and others~\cite{Wang2016} found markedly different behavior at $T=0$ and at a finite $T$.
     Namely, the calculations at $T=0$ showed that superconductivity develops if $N$ is smaller than some $\gamma$-dependent $N_{cr}$, while at larger $\gamma$ the system remains in a NFL normal state.  On the other hand, computations of the onset temperature for the pairing  $T_c (N)$ showed that $T_c (N)$ remains finite at any $N$ and the line $T_c (N)$ by-passes $N_{cr}$ (Fig.\ref{fig:Tc_1}). The authors of~\cite{Wang2016} argued that this
      discrepancy is due to the fact that Eliashberg equations for spin-singlet pairing contain fermionic self-energy without thermal contribution (the self-action term), and this  self-energy is large (and has NFL) form for all frequencies except for $\omega_m = \pm \pi T$, at which it vanishes. The existence of a finite $T_c$ for any $N$ then follows from the fact that the pairing interaction between fermions with $\pi T$ and $-\pi T$ is not countered by the self-energy and opens the gap $\Delta$ at these two frequencies at $T = T_c -0$. A non-zero $\Delta (\pm \pi T)$  then  induces the pairing gap for fermions with other Matsubara frequencies.

      In this communication we extended the analysis of the pairing problem to $T < T_c (N)$ and solved the non-linear gap equation.  We analyzed the large $N$ limit analytically and solved the gap equation at smaller $N$ numerically.  We first obtained $\Delta (\omega_m)$ along Matsubara axis and used it to compute the Free energy and the specific heat. We found that the specific heat jumps at $T_c$, but at large $N$ the relative magnitude of the jump $\Delta C (T_c-0) /C_n (T_c)$ is smaller by a factor $1/N^2$ than in a BCS superconductor. The behavior of the specific heat below $T_c$ is also rather unconventional, as specific heat approaches normal state form at $T \to 0$.

       We then solved the gap equation along the real axis, using $\Delta (\omega_m)$ as input. We obtained $\Delta (\omega)$ and used it to compute the DOS $N(\omega)$ and the spectral function $A(\omega)$.
      In a conventional BCS-type superconductor $A(\omega)$ and $N(\omega)$ are peaked at the gap value $\Delta (T)$, and the peak  position shifts to a smaller $\omega$ as temperature increases towards $T_c$ (the gap ``closes in").
      We found that at $N > N_{cr}$, the behavior is very different -- the position of the maximum in $N(\omega)$ increases linearly with $T$ and remains finite at $T_c$. The DOS remains finite at all frequencies, including $\omega =0$. At small $T$, $N(\omega)$ at small $\omega$ is reduced compared to $N(\omega) = N_0$ in the normal state, it displays a pseudogap behavior. As $T$ increases towards $T_c$, pseudogap just ``fills in".

       The  form of the spectral function $A(\omega)$  depends on the strength of the thermal contribution. In our model, thermal contribution diverged at a QCP.  In this limit, $A(\omega)$ has the same frequency dependence as the DOS $N(\omega)$.  Away from a QCP, when the bosonic  susceptibility $\chi (\Omega)$ is not singular at $\Omega =0$, thermal contribution does not diverge.  In the limit when the thermal contribution  is weak, $A(\omega)$ at when $N > N_{cr}$ has a single peak at $\omega =0$. At $N < N_{cr}$, it has  two sharp peaks
        at $\omega \approx \pm \Delta (0)$, when $T < T_{cross} (N)$, and a single peak at $\omega =0$ when $T > T_{cross} (N)$.

The issue we didn't discuss in this work is whether gap fluctuations (transverse and longitudinal) destroy long-range superconducting order in some $T$ range below $T_c (N)$.  Eliashberg theory, which we used,  neglects gap fluctuations.  It is very likely  that in some range below Eliashberg $T_c (N)$ long-range superconducting order gets destroyed, and the actual $T_{c,act} < T_c$.  We note in this regard that in our theory, the transformation from ``gap closing" to ``gap filling" in the DOS and the spectral function at $T \sim T_{cross} (N)$  is due to the fact that at $T> T_{cross} (N)$,  the feedback from the pairing on the fermionic self-energy is weak.  This last result does not actually rely on the existence of long-range superconducting order. If fluctuations destroy superconducting phase coherence, the feedback will be further reduced, but   we emphasize that the feedback on fermions as small above $T_{cross} (N)$ already within  the Eliashberg theory.  The same argument holds for the transformation  from two peaks at a finite frequency in $A(\omega)$ to a single peak at $\omega =0$.

The transformation from ``gap closing" behavior at small $T$  to ``gap filling" behavior at $T \sim T_c$ has been observed in high-$T_c$  cuprates, in the DOS~\cite{DOS}
 and  ARPES measurements of the spectral function in the antinodal region~\cite{DOS,dessau,kaminski,*Kaminski2,kanigel,*kanigel2,*kanigel3,norman_review,shen,*shen2,shen3,shen4,*kordyuk2,hoffman,Peng2013}.
   Symmetrized data of MDC ARPES measurements a  along particular direction of ${\bf k}$ in the near-nodal region showed the transformation from two peaks at a finite frequency to a single peak at $\omega =0$ (this is termed as the appearance of the Fermi arc).  These results are consistent with our microscopic analysis for the DOS and also for the spectral function,  if we assume that the thermal contribution is  stronger in the antinodal region than in the near-nodal region.  The strength of thermal contribution scales with the static bosonic susceptibility $\chi' (0)$.  Static $\chi'(0)$ is larger for antinodal fermions in, e.g., spin-fluctuation models~\cite{acs,scal,Scalapino_92},  where the interaction is peaked at momentum at or near $(\pi,\pi)$.

    A final remark.  In our analysis we didn't include the dependence of the pairing vertex and the gap on the angle along the Fermi surface, e.g., $\cos{2\theta}$ dependence for the $d-$wave gap in the cuprate superconductors.
    The $d-$wave form of the pairing gap  does not affect our results for the DOS and $A(\omega)$ in the antinodal region, as there the gap can be approximated by the constant. We note in this regard that our results for the development of the dip with increasing $T$ in the DOS and in the spectral function at large $P$,  (e.g., representative results in Fig.\ref{fig:DOSsummary} and right panel in Fig.\ref{fig:Asummary})  are quire consistent with DOS and ARPES data in the cuprates.  We associate our result for smaller $P$ (i.e., smaller thermal contribution) with the system behavior
     closer to the nodes. This association is valid if the pairing interaction in the cuprates is the strongest at momentum transfers connecting antinodal points and weaker at momentum transfer along the diagonals, like in spin-fluctuation scenario.   Our results then show (left panel in Fig.\ref{fig:Asummary}) that in the near-nodal regime, the two peaks, originally  separated by $2\Delta$ at a particular $k$-point on the Fermi surface   transform into a single peak at $\omega =0$ as  $T$ increases.  This effect is well known as the development of the Fermi arc.

\begin{figure}
  \begin{center}
    \includegraphics[width=12cm]{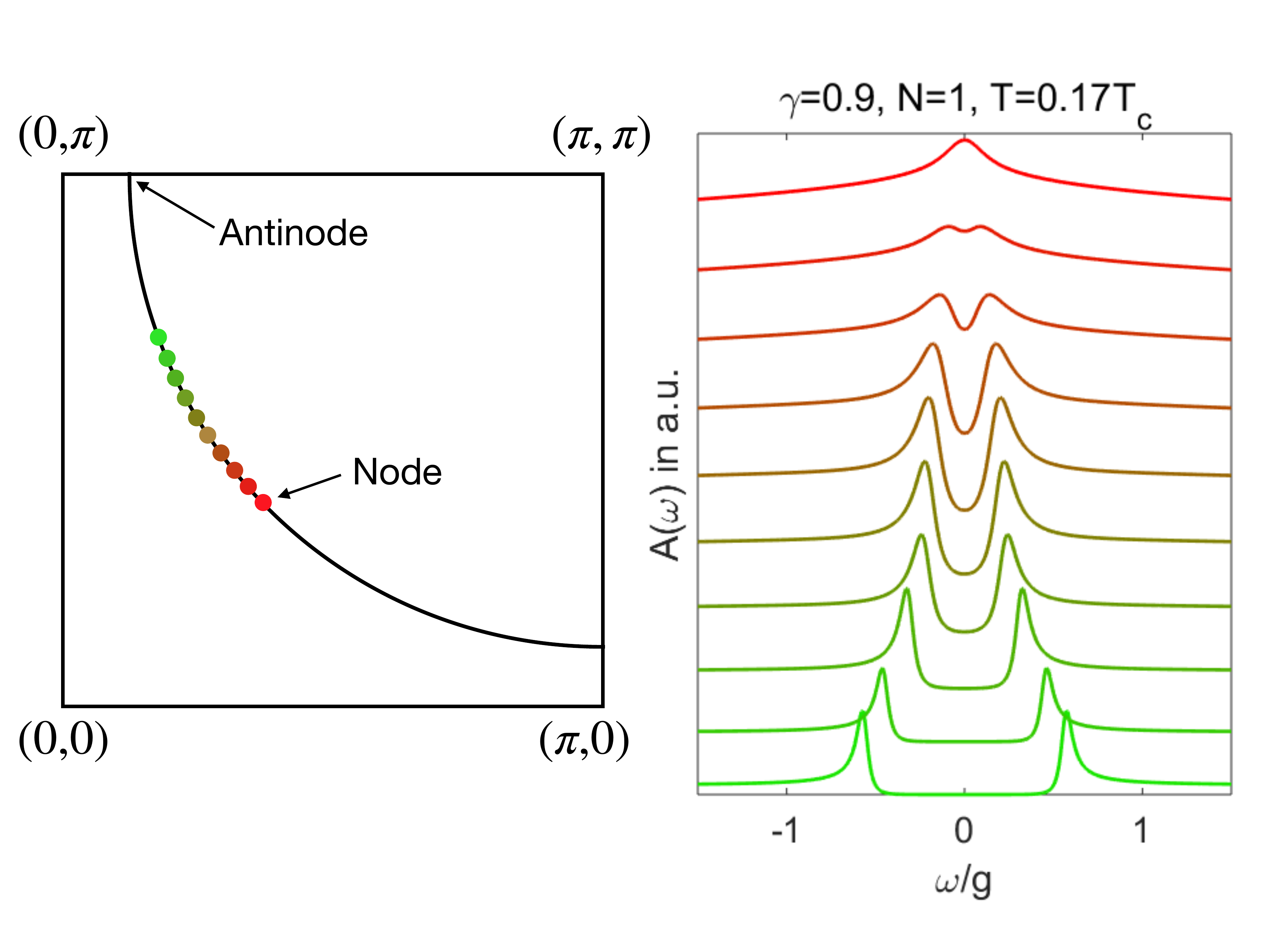}
    \caption{The spectral function $A(\omega)$ along the Fermi surface at $T < T_{cross}$. In the nodal region (Red) $A(\omega)$ has two closely located peaks, which merge at the node. In the antinodal region (Green), the two peaks of $A(\omega)$ are well separated.  We set
   $\gamma=0.9, N=1,$ and $ T=0.17T_c$. 
     }\label{fig:cuprates}
  \end{center}
\end{figure}

      The modeling of the angular dependence of the gap $\Delta (\theta)$ is needed for the analysis how the spectral function evolves as a function of $\theta$ at a given $T$. To obtain this dependence in our data, we added $\cos {2\theta}$ factor to $\Phi^* (\omega)$ and solved the Eliashberg equations at a given $P$, $T$, and $\gamma$.  At high $T > T_{cross}$, the evolution is similar to the one in Fig.\ref{fig:largeNspectral}.  Namely, near the node $A(\omega)$ has a single maximum at $\omega =0$, while in the antinodal region $A(\omega)$ has a dip at $\omega =0$ and a shallow maximum, whose frequency scales with $T$. At $T < T_{cross}$,  $A(\omega)$ has two
        weakly separated peaks in the nodal region and  strongly separated peaks in the antinodal region (Fig.\ref{fig:cuprates})  This behavior and the one in Fig. \ref{fig:smallNspectral} reproduce ARPES data in Refs. \cite{dessau,kanigel,kaminski,*Kaminski2,shen,*shen2,shen3,shen4,*kordyuk2,hoffman,Peng2013}.

  \acknowledgements
  We thank   D. Dessau, A. Millis, N. Prokofiev, S. Raghu, G. Torroba, and  A. Yazdani for useful discussions.   This work by Y. Wu and AVC was supported by the NSF DMR-1523036.

\section*{Appendix: Analytic continuation from Matsubara axis to real frequency axis}
\label{sec:appendix}
In this Appendix we show
the derivation of Eq.~\eqref{eq:anacon} for the pairing vertex $\Phi^* (\omega)$ and the self-energy $\Sigma^* (\omega)$ along real frequency axis.
 We follow Ref. \cite{Marsiglio_88} and  use spectral decomposition approach.  To avoid misunderstanding, here we explicitly keep the factors $i$ for
  Matsubara frequencies, i.e. define the interaction as $\chi(i\Omega)=\frac{g^\gamma}{|\Omega|^\gamma}$. For a general complex number $z$ the retarder
  $\chi (z)$ is
    we have
  \beq
  \chi(z)= \left(\frac{- g^2}{z^2} \right)^{\gamma/2}
  \label{dec_1}
  \eeq
  Along real frequency axis ($z = \omega + i \delta$) we have
  \bea
  &&\Ree \chi (\omega) = \frac{g^\gamma}{|\omega|^{\gamma}}\cos \frac{\pi\gamma}{2} \nonumber \\
    && \Imm  \chi (\omega) = \frac{g^\gamma \sgn(\omega)}{|\omega|^{\gamma}}\sin \frac{\pi\gamma}{2}.
\label{dec_2}
\eea

By Cauchi theorem, the susceptibility at arbitrary $z$ can be expressed via $\Imm \chi (x)$ as
\beq
\chi (z) = (2/\pi)  \int_0^\infty dx  \frac{\Imm\chi (x) x}{(x^2 - z^2)}
\label{dec_3}
\eeq
Along real frequency axis this reduces to KK relation $\Ree \chi (\omega) = (2/\pi) {\cal P} \int_0^\infty dx \Imm \chi (x) x/(x^2 - \omega^2)$.
In the calculations of $\Phi (\omega_m)$ and ${\tilde \Sigma} (\omega_m)$  we used the susceptibility $\chi (i\omega_m - i \omega_{m'})$, weighted with
${\Phi (\omega_{m'})}/{\sqrt{{\tilde \Sigma}^2 (\omega_{m'}) +\Phi^2 (\omega_{m'})}}$ and
   ${{\tilde \Sigma}(\omega_m)}/{\sqrt{{\tilde \Sigma}^2 (\omega_{m'})  +\Phi^2 (\omega_{m'})}}$, respectively, and summed up over $\omega_{m'}$ (Eq.
\ref{eq:gapeq}).  These expressions cannot be converted to real frequency $\omega$ by just replacing $z =i\omega_m$ by $z=\omega$ in $\chi (z- i \omega_{m'}) = (-g^2/(z-i \omega_{m'})^2)^{\gamma/2}$ because this $\chi(z)$ has branch cuts  in a complex plane of $z$ along $z = i\omega_{m'} +z_0$, where $z_0$ is a real variable (Ref. \cite{Marsiglio_88}).  Because of this complication,  we have to implement the full spectral decomposition procedure.
Namely, we depart from Eliashberg equations along Matsubara axis and use spectral representation to express $G(i\omega_m, \bm{k})$ and $\chi (i \omega_m - i \omega_{m'})$ via $\Imm G (x,\bm{k})$ and $\Imm \chi (x)$ along real axis as
\begin{equation}\label{eq:spectral}
  \begin{aligned}
     G(i\omega_m,\bm{k})&=\int_{-\infty}^{\infty} \frac{dx}{\pi}  \frac{{\Imm G^R (x,\bm{k})}}{x  - i\omega_m}\\
    \chi(i\omega_m-i\omega_{m'})&=\int_{-\infty}^{\infty} \frac{dy}{\pi}  \frac{\Imm \chi (y)}{y - i(\omega_m-\omega_{m'})}
  \end{aligned}
\end{equation}
where "R" stands for "retarded".
We then explicitly sum over $\omega_{m'}$ and integrate over ${\bf k}$ and obtain the expressions for ${\tilde \Sigma} (i\omega_m)$ and  $\Phi (i\omega_m)$, in which the dependence on $\omega_m$ is only via $1/(i\omega_m -x-y)$.  This form can be straightforwardly continued analytically to real frequency by just replacing  $i\omega_m$  by $\omega + i0^+$.

For compactness, we do the calculations  in Nambu formalism, in which one operates with the matrix Green's function
${\hat G} (i\omega_m, \bm{ k})$ and treats both $\Sigma(i\omega_m)$ and $\Phi(i\omega_m)$)
 as elements of matrix self-energy ${\hat \Sigma} (i\omega_m)$.
  The Eliashberg equation in Nambu formalism is
\begin{equation}\label{eq:self-energy}
 \hat \Sigma(i\omega_n)=- T\sum_{m} \int \frac{d^2 k}{(2\pi)^2} {\hat\tau_3}\hat G(i\omega_m,\bm{k}){\hat\tau_3}\chi(i\omega_n-i\omega_m),
\end{equation}
where ${\hat\tau_3}$ is a
Pauli matrix. $\hat\Sigma =  \Sigma{\hat\tau_0} -  \Phi  \hat\tau_1$,
 and the matrix $\hat G (i\omega_m,\bm{k}) = -(i\omega_m-\hat \Sigma(i\omega_m))^{-1}$. The diagonal and off-diagonal elements of $\hat G (i\omega_m,\bm{k})$ are  conventional  normal and anomalous Green's functions.

 Substituting the spectral representation (\ref{eq:spectral}) into (\ref{eq:self-energy})
 and performing the  summation over $\omega_m$, we obtain
\begin{equation}
  \begin{aligned}
    &\hat \Sigma(i\omega_n)= - T\sum_{m} \int \frac{d^2 k}{(2\pi)^2} \int \frac{dx dy}{\pi^2} {\hat\tau_3}\frac{\Imm {\hat G}^R (x,\bm{k})}{i\omega_m-x}{\hat\tau_3} \frac{\Imm \chi (y)}{i\omega_n-i\omega_m-y}\\
    &= - \int \frac{d^2 k}{(2\pi)^2} \int \frac{dx dy}{\pi^2} {\hat\tau_3} \Imm {\hat G}^R (x,\bm{k}) {\hat\tau_3} \Imm \chi (y)\frac{1}{i\omega_n-x-y}\left(n_F(x)+n_B(-y)\right)\\
  \end{aligned}
\end{equation}
Replacing $i\omega_n$ with $\omega+i0^+$ we obtain the self-energy at real frequencies
\begin{equation}
  \hat \Sigma(\omega)= - \int \frac{d^2 k}{(2\pi)^2} \int \frac{dx dy}{\pi^2}
 {\hat\tau_3}{\Imm {\hat G}^R (x,\bm{k})}{\hat\tau_3} \Imm \chi (y) \frac{1}{\omega-x-y+i0^+}\left(n_F(x)+n_B(-y)\right)
\end{equation}
We next express $\Imm {\hat G}^R (x, \bm{k})/(\omega-x-y+i0^+)$ via the full ${\hat G}^R (x, \bm{k})$ as
\begin{equation}
  \begin{aligned}
    \frac{2 \Imm {\hat G}^R (x,\bm{k})}{\omega-x-y+i0^+}&=\Imm\left[{\hat G}^R(x,\bm{k})\left(\frac{1}{\omega-x-y+i0^+}+\frac{1}{\omega-x-y-i0^+}\right)\right]\\
    &-i\Ree\left[{\hat G}^R(x,\bm{k}) \left(\frac{1}{\omega-x-y+i0^+}-\frac{1}{\omega-x-y-i0^+}\right)\right].
  \end{aligned}
\end{equation}
and integrate over $x$ by closing the integration contour over the upper half-plane of complex $x$
 Because ${\hat G}^R(x,\bm{k})$ is analytic in the upper half plane, the poles are
  at  $x = \omega-y+i0^+$ and $x = i(2n+1)\pi T$ [these are the poles coming from $n_F(x)$]. Using the residue theorem, we find
    \begin{align}
    &\hat \Sigma(\omega)=-\frac{1}{2} \int \frac{d^2 k}{(2\pi)^2} \int \frac{dy}{\pi} \Imm \chi (y) \nonumber\\
    &\times\left\{\Imm \left[\int \frac{dx}{\pi} \left(n_F(x)+n_B(-y)\right){\hat\tau_3}{\hat G}^R(x,\bm{k}){\hat\tau_3}\left(\frac{1}{\omega-x-y+i0^+}+\frac{1}{\omega-x-y-i0^+}\right)\right]\right. \nonumber\\
    &\left. -i\Ree
    \left[\int \frac{dx}{\pi} \left(n_F(x)+n_B(-y)\right){\hat\tau_3}{\hat G}^R(x,\bm{k}){\hat\tau_3}\left(\frac{1}{\omega-x-y+i0^+}-\frac{1}{\omega-x-y-i0^+}\right)\right]\right\} \nonumber\\
    =&\int \frac{d^2 k}{(2\pi)^2} \int \frac{dy}{\pi} \Imm \chi (y) T\sum_{\omega_n>0}\Imm\left[i{\hat\tau_3}{\hat G}^R(i\omega_n,\bm{k}){\hat\tau_3}\left(\frac{1}{\omega-i\omega_n-y}+\frac{1}{\omega-i\omega_n-y}\right)\right] \nonumber\\
    &+ \int \frac{d^2 k}{(2\pi)^2} \int \frac{dy}{\pi} \Imm \chi (y) \left(n_F(\omega-y)+n_B(-y)\right) \times\left[\Imm (i\hat{\tau}_3G^R(\omega-y,\bm{k})\hat{\tau}_3)-i\Ree(i\hat{\tau}_3G^R(\omega-y,\bm{k})\hat{\tau}_3)\right]\nonumber\\
    =& 2 T \sum_{\omega_n>0}  \int \frac{d^2 k}{(2\pi)^2} \int \frac{dy}{\pi}  \Imm \chi (y)  \Imm \frac{i{\hat\tau_3}{\hat G}^R(i\omega_n,\bm{k}){\hat\tau_3}}{\omega-i\omega_n-y} \nonumber \\
    &+ \int \frac{d^2 k}{(2\pi)^2} \int \frac{dy}{\pi}  \Imm \chi (y)  {\hat\tau_3}{\hat G}^R(\omega-y,\bm{k}){\hat\tau_3}\left(n_F(y-\omega)+n_B(y)\right),
  \end{align}
Using now $ (1/\pi) \int dy  \Imm \chi (y) /(\omega - i \omega_m -y) = -\chi (\omega-i\omega_m)$, we finally obtain
\begin{equation}\label{eq:imag}
  \begin{aligned}
    \hat \Sigma(\omega)&=-2T \sum_{\omega_n>0}  \int \frac{d^2 k}{(2\pi)^2} \Imm [i{\hat\tau_3}{\hat G}^R(i\omega_n,\bm{k}){\hat\tau_3}\chi(\omega-i\omega_n)]\\
    &- \int \frac{d^2 k}{(2\pi)^2} \int \frac{dy}{\pi} \Imm[\chi(y)] {\hat\tau_3}{\hat G}^R(\omega-y,\bm{k}){\hat\tau_3}\left(n_F(y-\omega)+n_B(y)\right).
  \end{aligned}
\end{equation}

Let's now spit this matrix equation  into the equations for the pairing vertex $\Phi (\omega)$ and conventional (non-anomalous) self-energy $\Sigma (\omega)$. Expressing $\hat \Sigma(\omega)$ as
\beq
\hat \Sigma(\omega)=\Sigma(\omega)\hat\tau_0-\Phi(\omega)\hat\tau_1
\eeq
and substituting into the Dyson equation, we obtain
\begin{equation}
  {\hat\tau_3}G(\omega,\bm{k}){\hat\tau_3}=\frac{1}{\xi_{\bm{k}}^2+\Phi(\omega)^2-(\omega+\Sigma(\omega) + i0^+)^2}\left(-(\omega+\Sigma(\omega))\hat\tau_0-
  \xi_{\bm{k}}{\hat\tau_3}+\Phi(\omega)\hat\tau_1\right)
\end{equation}
where $\xi_{\bm{k}}$ is the fermionic dispersion. Expressing next $\int d^2 k/(2\pi)^2 = N_0 \int d \xi_{\bm{k}}$, where $N_0$ is the DOS
in the normal state, and integrating over $\xi_{\bm{k}}$, we obtain
\begin{equation}
  \int \frac{d^2 k}{(2\pi)^2}
  {\hat\tau_3}G(\omega,\bm{k}){\hat\tau_3}=N_0\int_{-\infty}^{\infty}d\xi_{\bm{k}}{\hat\tau_3}G(\omega,\bm{k}){\hat\tau_3}=i\pi N_0\frac{-{\tilde \Sigma}(\omega))\hat\tau_0+\Phi(\omega)\hat\tau_1}{\sqrt{{\tilde \Sigma}^2(\omega) - \Phi(\omega)^2}}
\end{equation}
where ${\tilde \Sigma} (\omega) = \omega + \Sigma (\omega)$.
Absorbing the density of states $N_0$ into $\chi$ and splitting $\hat \Sigma(\omega)$ into normal and anomalous components, we obtain
\begin{equation} \label{eq:realsolution}
  \begin{aligned}
  {\tilde \Sigma} (\omega)&= \omega + i\pi T \sum_{\omega_m>0}\frac{{\tilde \Sigma}(\omega_m)}
     {\sqrt{\Phi^2 (\omega_m)+({\tilde \Sigma}^2 (\omega_m))}} \left( \chi(\omega_m +i\omega) - \chi(\omega_m -i\omega)\right)\\
    & + i \int dy \left[S_\Sigma (\omega-y) \Imm \chi (y) \left(n_F(y-\omega)+n_B(y)\right) \right]\\
    \Phi (\omega )&=\pi T \sum_{\omega_m >0}\frac{\Phi(\omega_m)}{\sqrt{\Phi^2(\omega_m) + {\tilde \Sigma}^2(\omega_m))}}
    \left(\chi(\omega_m +i\omega) + \chi(\omega_m -i\omega)\right)\\
    &+ i\int dy \left[S_\Phi (\omega -y)  \Imm \chi(y)  \left(n_F(y-\omega)+n_B(y)\right)\right] \\
  \end{aligned}
\end{equation}
where
\bea
S_\Phi (\omega) &=&  \frac{\Phi(\omega)}{\sqrt{{\tilde \Sigma}^2(\omega) - \Phi^2(\omega)}} = \frac{\Phi(\omega)}{{\tilde \Sigma}(\omega)} \frac{1}{\sqrt{1- \left(\frac{\Phi(\omega)}{{\tilde \Sigma}(\omega)}\right)^2}} \nonumber \\
S_\Sigma (\omega) &=&  \frac{{\tilde \Sigma}(\omega)}{\sqrt{{\tilde \Sigma}^2(\omega) - \Phi^2(\omega)}} =\frac{1}{\sqrt{1- \left(\frac{\Phi(\omega)}{{\tilde \Sigma}(\omega)}\right)^2}}
\label{aaa_1}
\eea
At a finite $T$ and small $y$, $n_B (y) \approx T/y$. At a QCP $\Imm \chi (y) n_B (y)$ then scales as  $ T/|y|^{1+\gamma}$, and integrals over $dy$ in (\ref{eq:realsolution}) diverge. The divergence, however, cancels out in the ratio $\Phi (\omega)/{\tilde \Sigma} (\omega) = \Delta(\omega)/\omega$ and  can be formally eliminated by introducing new $\Phi^* (\omega)$ and $\Sigma^* (\omega)$ in which the divergent pieces are subtracted:
\begin{equation}\label{eq:realsolution_1}
  \begin{aligned}
   &{\tilde \Sigma}^* (\omega)= \omega + i\pi T \sum_{\omega_m >0}\frac{{\tilde \Sigma}(\omega_m)}
     {\sqrt{(\Phi(\omega_m))^2+({\tilde \Sigma}(\omega_m))^2}}  \left(\chi(\omega_m +i\omega) - \chi(\omega_m -i\omega)\right)\\
    & + i \int dy \Imm \chi(y) \left[S_\Sigma (\omega-y)  \left(n_F(y-\omega)+n_B(y)\right) -S_\Sigma (\omega)  \frac{T}{y} \right]\\
    &\Phi^*(\omega)= \pi T \sum_{\omega_m >0}\frac{\Phi(\omega_m)}{\sqrt{(\Phi(\omega_m))^2 + ({\tilde \Sigma}(\omega_m))^2}}
    \left(\chi(\omega_m +i\omega) + \chi(\omega_m -i\omega)\right) \\
    &+ i\int dy  \Imm  \chi (y)  \left[S_\Phi (\omega -y)  \left(n_F(y-\omega)+n_B(y)\right)-S_\Phi (\omega)  \frac{T}{y} \right] \\
    \end{aligned}
\end{equation}
Comparing (\ref{eq:realsolution}) and (\ref{eq:realsolution_1}) we see that
\beq
\tilde{\Sigma}^*(\omega)=\tilde{\Sigma}(\omega)(1-Q(\omega)),~~\Phi^*(\omega)=\Phi(\omega)(1-Q(\omega)),
\label{dec_4}
\eeq
 where
\begin{equation}
  Q(\omega)=\frac{iP}{\sqrt{{\tilde\Sigma}^2(\omega)-\Phi^2(\omega)}}, \quad P=\int_{-\infty}^{\infty}d y \Imm \chi (y) \frac{T}{y}=\pi T\chi(0)
\end{equation}
The ratio $\Phi^* (\omega)/{\tilde \Sigma}^* (\omega)$ is the same as $\Phi (\omega)/{\tilde \Sigma} (\omega)$, i.e., the gap function $\Delta (\omega)$ can be equally expressed via non-singular  $\Phi^* (\omega)$ and ${\tilde \Sigma}^* (\omega)$. Furthermore, a little experimentation shows that
$S_\Phi (\omega)$ and $S_\Sigma (\omega)$, given by (\ref{aaa_1}),  can be equally expressed via $\Phi^* (\omega)$ and ${\tilde \Sigma}^* (\omega)$, as
 \bea
S_\Phi (\omega) &=&  \frac{\Phi^*(\omega)}{\sqrt{({\tilde \Sigma}^*(\omega))^2 - (\Phi^*(\omega))^2}} = \frac{\Phi^*(\omega)}{{\tilde \Sigma}^*(\omega)} \frac{1}{\sqrt{1- \left(\frac{\Phi^*(\omega)}{{\tilde \Sigma}^*(\omega)}\right)^2}} = \frac{\Delta(\omega)}{\sqrt{\omega^2 - (\Delta(\omega))^2}}
 \nonumber \\
S_\Sigma (\omega) &=&  \frac{{\tilde \Sigma}^*(\omega)}{\sqrt{({\tilde \Sigma}^*(\omega))^2 - (\Phi^*(\omega))^2}} =\frac{1}{\sqrt{1- \left(\frac{\Phi^*(\omega)}{{\tilde \Sigma}^*(\omega)}\right)^2}} =
 \frac{\omega}{\sqrt{\omega^2 -(\Delta(\omega))^2}}
\label{aaa_2}
\eea
By the same reason, $\Phi (\omega)$ and ${\tilde \Sigma} (\omega)$ can be expressed via  $\Phi^* (\omega)$ and ${\tilde \Sigma}^* (\omega)$ in a manner similar to Eq. (\ref{dec_4}):
  \beq
 \Phi (\omega) = \Phi^* (\omega) \left(1 + Q^* (\omega)\right), {\tilde \Sigma} (\omega) = {\tilde \Sigma}^* (\omega) \left(1 + Q^* (\omega)\right),
 \label{new_new_sm}
 \eeq
where
\beq
Q^* (\omega) = \frac{i\pi \chi (0)}{\sqrt{({\tilde \Sigma}^*)^2(\omega) - (\Phi^*)^2(\omega)}}
\label{exx__1_sm}
\eeq

Equations (\ref{eq:realsolution_1}) are free from divergencies and can be readily extended to $N \neq 1$, as we did in the main text.

Eqs. (\ref{eq:realsolution_1}) have been solved numerically by iterations. For  practical purposes, we found that in some cases the convergence is faster if we do calculations in two steps: first evaluate intermediate $\Phi^{**}$ and ${\tilde \Sigma}^{**}$, related to $\Phi$ and ${\tilde \Sigma}$ as in (\ref{dec_4}), but with  $P= \int_{-\delta}^{\delta}d y \Imm \chi (y)  \frac{T}{y}$, where $\delta$ is some finite number,  and then compute $\Phi^*$ and $\Sigma^*$ by adding the rest of the integral in $P$.  The best convergence is achieved by adjusting the value of $\delta$.

\bibliography{1stMref}

\begin{thebibliography}{119}%
\makeatletter
\providecommand \@ifxundefined [1]{%
 \@ifx{#1\undefined}
}%
\providecommand \@ifnum [1]{%
 \ifnum #1\expandafter \@firstoftwo
 \else \expandafter \@secondoftwo
 \fi
}%
\providecommand \@ifx [1]{%
 \ifx #1\expandafter \@firstoftwo
 \else \expandafter \@secondoftwo
 \fi
}%
\providecommand \natexlab [1]{#1}%
\providecommand \enquote  [1]{``#1''}%
\providecommand \bibnamefont  [1]{#1}%
\providecommand \bibfnamefont [1]{#1}%
\providecommand \citenamefont [1]{#1}%
\providecommand \href@noop [0]{\@secondoftwo}%
\providecommand \href [0]{\begingroup \@sanitize@url \@href}%
\providecommand \@href[1]{\@@startlink{#1}\@@href}%
\providecommand \@@href[1]{\endgroup#1\@@endlink}%
\providecommand \@sanitize@url [0]{\catcode `\\12\catcode `\$12\catcode
  `\&12\catcode `\#12\catcode `\^12\catcode `\_12\catcode `\%12\relax}%
\providecommand \@@startlink[1]{}%
\providecommand \@@endlink[0]{}%
\providecommand \url  [0]{\begingroup\@sanitize@url \@url }%
\providecommand \@url [1]{\endgroup\@href {#1}{\urlprefix }}%
\providecommand \urlprefix  [0]{URL }%
\providecommand \Eprint [0]{\href }%
\providecommand \doibase [0]{http://dx.doi.org/}%
\providecommand \selectlanguage [0]{\@gobble}%
\providecommand \bibinfo  [0]{\@secondoftwo}%
\providecommand \bibfield  [0]{\@secondoftwo}%
\providecommand \translation [1]{[#1]}%
\providecommand \BibitemOpen [0]{}%
\providecommand \bibitemStop [0]{}%
\providecommand \bibitemNoStop [0]{.\EOS\space}%
\providecommand \EOS [0]{\spacefactor3000\relax}%
\providecommand \BibitemShut  [1]{\csname bibitem#1\endcsname}%
\let\auto@bib@innerbib\@empty
\bibitem [{\citenamefont {Combescot}(1995)}]{combescot}%
  \BibitemOpen
  \bibfield  {author} {\bibinfo {author} {\bibfnamefont {R.}~\bibnamefont
  {Combescot}},\ }\href {\doibase 10.1103/PhysRevB.51.11625} {\bibfield
  {journal} {\bibinfo  {journal} {Phys. Rev. B}\ }\textbf {\bibinfo {volume}
  {51}},\ \bibinfo {pages} {11625} (\bibinfo {year} {1995})}\BibitemShut
  {NoStop}%
\bibitem [{Ber()}]{Bergmann}%
  \BibitemOpen
  \href {https://doi.org/10.1007/BF02351862} {\ }\bibinfo {note} {Bergmann, G.
  and Rainer, D. Z. Physik (1973) 263: 59.}\BibitemShut {Stop}%
\bibitem [{\citenamefont {Allen}\ and\ \citenamefont
  {Rainer}(1991)}]{Bergmann2}%
  \BibitemOpen
  \bibfield  {author} {\bibinfo {author} {\bibfnamefont {P.~B.}\ \bibnamefont
  {Allen}}\ and\ \bibinfo {author} {\bibfnamefont {D.}~\bibnamefont {Rainer}},\
  }\href {https://doi.org/10.1038/349396a0} {\bibfield  {journal} {\bibinfo
  {journal} {Nature}\ }\textbf {\bibinfo {volume} {349}},\ \bibinfo {pages}
  {396 EP } (\bibinfo {year} {1991})}\BibitemShut {NoStop}%
\bibitem [{\citenamefont {Allen}\ and\ \citenamefont {Dynes}(1975)}]{ad}%
  \BibitemOpen
  \bibfield  {author} {\bibinfo {author} {\bibfnamefont {P.~B.}\ \bibnamefont
  {Allen}}\ and\ \bibinfo {author} {\bibfnamefont {R.~C.}\ \bibnamefont
  {Dynes}},\ }\href {\doibase 10.1103/PhysRevB.12.905} {\bibfield  {journal}
  {\bibinfo  {journal} {Phys. Rev. B}\ }\textbf {\bibinfo {volume} {12}},\
  \bibinfo {pages} {905} (\bibinfo {year} {1975})}\BibitemShut {NoStop}%
\bibitem [{\citenamefont {Marsiglio}\ \emph {et~al.}(1988)\citenamefont
  {Marsiglio}, \citenamefont {Schossmann},\ and\ \citenamefont
  {Carbotte}}]{Marsiglio_88}%
  \BibitemOpen
  \bibfield  {author} {\bibinfo {author} {\bibfnamefont {F.}~\bibnamefont
  {Marsiglio}}, \bibinfo {author} {\bibfnamefont {M.}~\bibnamefont
  {Schossmann}}, \ and\ \bibinfo {author} {\bibfnamefont {J.~P.}\ \bibnamefont
  {Carbotte}},\ }\href {\doibase 10.1103/PhysRevB.37.4965} {\bibfield
  {journal} {\bibinfo  {journal} {Phys. Rev. B}\ }\textbf {\bibinfo {volume}
  {37}},\ \bibinfo {pages} {4965} (\bibinfo {year} {1988})}\BibitemShut
  {NoStop}%
\bibitem [{\citenamefont {Marsiglio}\ and\ \citenamefont
  {Carbotte}(1991)}]{Marsiglio_91}%
  \BibitemOpen
  \bibfield  {author} {\bibinfo {author} {\bibfnamefont {F.}~\bibnamefont
  {Marsiglio}}\ and\ \bibinfo {author} {\bibfnamefont {J.~P.}\ \bibnamefont
  {Carbotte}},\ }\href {\doibase 10.1103/PhysRevB.43.5355} {\bibfield
  {journal} {\bibinfo  {journal} {Phys. Rev. B}\ }\textbf {\bibinfo {volume}
  {43}},\ \bibinfo {pages} {5355} (\bibinfo {year} {1991})},\ \bibinfo {note}
  {for more recent results see F. Marsiglio and J.P. Carbotte,
  ``Electron-Phonon Superconductivity'', in ``The Physics of Conventional and
  Unconventional Superconductors'', Bennemann and Ketterson eds.,
  Springer-Verlag, (2006) and references therein.}\BibitemShut {Stop}%
\bibitem [{\citenamefont {Karakozov}\ \emph {et~al.}(1991)\citenamefont
  {Karakozov}, \citenamefont {Maksimov},\ and\ \citenamefont
  {Mikhailovsky}}]{Karakozov_91}%
  \BibitemOpen
  \bibfield  {author} {\bibinfo {author} {\bibfnamefont {A.}~\bibnamefont
  {Karakozov}}, \bibinfo {author} {\bibfnamefont {E.}~\bibnamefont {Maksimov}},
  \ and\ \bibinfo {author} {\bibfnamefont {A.}~\bibnamefont {Mikhailovsky}},\
  }\href {\doibase https://doi.org/10.1016/0038-1098(91)90556-B} {\bibfield
  {journal} {\bibinfo  {journal} {Solid State Communications}\ }\textbf
  {\bibinfo {volume} {79}},\ \bibinfo {pages} {329 } (\bibinfo {year}
  {1991})}\BibitemShut {NoStop}%
\bibitem [{\citenamefont {Bonesteel}\ \emph {et~al.}(1996)\citenamefont
  {Bonesteel}, \citenamefont {McDonald},\ and\ \citenamefont {Nayak}}]{nick_b}%
  \BibitemOpen
  \bibfield  {author} {\bibinfo {author} {\bibfnamefont {N.~E.}\ \bibnamefont
  {Bonesteel}}, \bibinfo {author} {\bibfnamefont {I.~A.}\ \bibnamefont
  {McDonald}}, \ and\ \bibinfo {author} {\bibfnamefont {C.}~\bibnamefont
  {Nayak}},\ }\href {\doibase 10.1103/PhysRevLett.77.3009} {\bibfield
  {journal} {\bibinfo  {journal} {Phys. Rev. Lett.}\ }\textbf {\bibinfo
  {volume} {77}},\ \bibinfo {pages} {3009} (\bibinfo {year}
  {1996})}\BibitemShut {NoStop}%
\bibitem [{\citenamefont {Abanov}\ \emph
  {et~al.}(2001{\natexlab{a}})\citenamefont {Abanov}, \citenamefont
  {Chubukov},\ and\ \citenamefont {Finkel'stein}}]{acf}%
  \BibitemOpen
  \bibfield  {author} {\bibinfo {author} {\bibfnamefont {A.}~\bibnamefont
  {Abanov}}, \bibinfo {author} {\bibfnamefont {A.~V.}\ \bibnamefont
  {Chubukov}}, \ and\ \bibinfo {author} {\bibfnamefont {A.~M.}\ \bibnamefont
  {Finkel'stein}},\ }\href {http://stacks.iop.org/0295-5075/54/i=4/a=488}
  {\bibfield  {journal} {\bibinfo  {journal} {EPL (Europhysics Letters)}\
  }\textbf {\bibinfo {volume} {54}},\ \bibinfo {pages} {488} (\bibinfo {year}
  {2001}{\natexlab{a}})}\BibitemShut {NoStop}%
\bibitem [{\citenamefont {Abanov}\ \emph {et~al.}(2003)\citenamefont {Abanov},
  \citenamefont {Chubukov},\ and\ \citenamefont {Schmalian}}]{acs}%
  \BibitemOpen
  \bibfield  {author} {\bibinfo {author} {\bibfnamefont {A.}~\bibnamefont
  {Abanov}}, \bibinfo {author} {\bibfnamefont {A.~V.}\ \bibnamefont
  {Chubukov}}, \ and\ \bibinfo {author} {\bibfnamefont {J.}~\bibnamefont
  {Schmalian}},\ }\href {\doibase 10.1080/0001873021000057123} {\bibfield
  {journal} {\bibinfo  {journal} {Advances in Physics}\ }\textbf {\bibinfo
  {volume} {52}},\ \bibinfo {pages} {119} (\bibinfo {year} {2003})}\BibitemShut
  {NoStop}%
\bibitem [{\citenamefont {Abanov}\ and\ \citenamefont
  {Chubukov}(1999{\natexlab{a}})}]{acs2}%
  \BibitemOpen
  \bibfield  {author} {\bibinfo {author} {\bibfnamefont {A.}~\bibnamefont
  {Abanov}}\ and\ \bibinfo {author} {\bibfnamefont {A.~V.}\ \bibnamefont
  {Chubukov}},\ }\href {\doibase 10.1103/PhysRevLett.83.1652} {\bibfield
  {journal} {\bibinfo  {journal} {Phys. Rev. Lett.}\ }\textbf {\bibinfo
  {volume} {83}},\ \bibinfo {pages} {1652} (\bibinfo {year}
  {1999}{\natexlab{a}})}\BibitemShut {NoStop}%
\bibitem [{\citenamefont {Abanov}\ \emph
  {et~al.}(2001{\natexlab{b}})\citenamefont {Abanov}, \citenamefont
  {Chubukov},\ and\ \citenamefont {J.}}]{finger_2001}%
  \BibitemOpen
  \bibfield  {author} {\bibinfo {author} {\bibfnamefont {A.}~\bibnamefont
  {Abanov}}, \bibinfo {author} {\bibfnamefont {A.~V.}\ \bibnamefont
  {Chubukov}}, \ and\ \bibinfo {author} {\bibfnamefont {S.}~\bibnamefont
  {J.}},\ }\href@noop {} {\bibfield  {journal} {\bibinfo  {journal} {Journal of
  Electron spectroscopy and related phenomena}\ }\textbf {\bibinfo {volume}
  {117}},\ \bibinfo {pages} {129} (\bibinfo {year}
  {2001}{\natexlab{b}})}\BibitemShut {NoStop}%
\bibitem [{\citenamefont {Son}(1999)}]{son}%
  \BibitemOpen
  \bibfield  {author} {\bibinfo {author} {\bibfnamefont {D.~T.}\ \bibnamefont
  {Son}},\ }\href {\doibase 10.1103/PhysRevD.59.094019} {\bibfield  {journal}
  {\bibinfo  {journal} {Phys. Rev. D}\ }\textbf {\bibinfo {volume} {59}},\
  \bibinfo {pages} {094019} (\bibinfo {year} {1999})}\BibitemShut {NoStop}%
\bibitem [{\citenamefont {Chubukov}\ and\ \citenamefont
  {Schmalian}(2005)}]{son2}%
  \BibitemOpen
  \bibfield  {author} {\bibinfo {author} {\bibfnamefont {A.~V.}\ \bibnamefont
  {Chubukov}}\ and\ \bibinfo {author} {\bibfnamefont {J.}~\bibnamefont
  {Schmalian}},\ }\href {\doibase 10.1103/PhysRevB.72.174520} {\bibfield
  {journal} {\bibinfo  {journal} {Phys. Rev. B}\ }\textbf {\bibinfo {volume}
  {72}},\ \bibinfo {pages} {174520} (\bibinfo {year} {2005})}\BibitemShut
  {NoStop}%
\bibitem [{\citenamefont {Lee}(2009)}]{sslee}%
  \BibitemOpen
  \bibfield  {author} {\bibinfo {author} {\bibfnamefont {S.-S.}\ \bibnamefont
  {Lee}},\ }\href {\doibase 10.1103/PhysRevB.80.165102} {\bibfield  {journal}
  {\bibinfo  {journal} {Phys. Rev. B}\ }\textbf {\bibinfo {volume} {80}},\
  \bibinfo {pages} {165102} (\bibinfo {year} {2009})}\BibitemShut {NoStop}%
\bibitem [{\citenamefont {Dalidovich}\ and\ \citenamefont
  {Lee}(2013)}]{sslee2}%
  \BibitemOpen
  \bibfield  {author} {\bibinfo {author} {\bibfnamefont {D.}~\bibnamefont
  {Dalidovich}}\ and\ \bibinfo {author} {\bibfnamefont {S.-S.}\ \bibnamefont
  {Lee}},\ }\href {\doibase 10.1103/PhysRevB.88.245106} {\bibfield  {journal}
  {\bibinfo  {journal} {Phys. Rev. B}\ }\textbf {\bibinfo {volume} {88}},\
  \bibinfo {pages} {245106} (\bibinfo {year} {2013})}\BibitemShut {NoStop}%
\bibitem [{\citenamefont {Sachdev}\ \emph {et~al.}(2009)\citenamefont
  {Sachdev}, \citenamefont {Metlitski}, \citenamefont {Qi},\ and\ \citenamefont
  {Xu}}]{subir}%
  \BibitemOpen
  \bibfield  {author} {\bibinfo {author} {\bibfnamefont {S.}~\bibnamefont
  {Sachdev}}, \bibinfo {author} {\bibfnamefont {M.~A.}\ \bibnamefont
  {Metlitski}}, \bibinfo {author} {\bibfnamefont {Y.}~\bibnamefont {Qi}}, \
  and\ \bibinfo {author} {\bibfnamefont {C.}~\bibnamefont {Xu}},\ }\href
  {\doibase 10.1103/PhysRevB.80.155129} {\bibfield  {journal} {\bibinfo
  {journal} {Phys. Rev. B}\ }\textbf {\bibinfo {volume} {80}},\ \bibinfo
  {pages} {155129} (\bibinfo {year} {2009})}\BibitemShut {NoStop}%
\bibitem [{\citenamefont {Moon}\ and\ \citenamefont {Sachdev}(2009)}]{subir2}%
  \BibitemOpen
  \bibfield  {author} {\bibinfo {author} {\bibfnamefont {E.~G.}\ \bibnamefont
  {Moon}}\ and\ \bibinfo {author} {\bibfnamefont {S.}~\bibnamefont {Sachdev}},\
  }\href {\doibase 10.1103/PhysRevB.80.035117} {\bibfield  {journal} {\bibinfo
  {journal} {Phys. Rev. B}\ }\textbf {\bibinfo {volume} {80}},\ \bibinfo
  {pages} {035117} (\bibinfo {year} {2009})}\BibitemShut {NoStop}%
\bibitem [{\citenamefont {Moon}\ and\ \citenamefont {Chubukov}(2010)}]{moon_2}%
  \BibitemOpen
  \bibfield  {author} {\bibinfo {author} {\bibfnamefont {E.-G.}\ \bibnamefont
  {Moon}}\ and\ \bibinfo {author} {\bibfnamefont {A.}~\bibnamefont
  {Chubukov}},\ }\href {\doibase 10.1007/s10909-010-0199-y} {\bibfield
  {journal} {\bibinfo  {journal} {Journal of Low Temperature Physics}\ }\textbf
  {\bibinfo {volume} {161}},\ \bibinfo {pages} {263} (\bibinfo {year}
  {2010})}\BibitemShut {NoStop}%
\bibitem [{\citenamefont {Metlitski}\ and\ \citenamefont
  {Sachdev}(2010{\natexlab{a}})}]{max}%
  \BibitemOpen
  \bibfield  {author} {\bibinfo {author} {\bibfnamefont {M.~A.}\ \bibnamefont
  {Metlitski}}\ and\ \bibinfo {author} {\bibfnamefont {S.}~\bibnamefont
  {Sachdev}},\ }\href {\doibase 10.1103/PhysRevB.82.075127} {\bibfield
  {journal} {\bibinfo  {journal} {Phys. Rev. B}\ }\textbf {\bibinfo {volume}
  {82}},\ \bibinfo {pages} {075127} (\bibinfo {year}
  {2010}{\natexlab{a}})}\BibitemShut {NoStop}%
\bibitem [{\citenamefont {Metlitski}\ and\ \citenamefont
  {Sachdev}(2010{\natexlab{b}})}]{max2}%
  \BibitemOpen
  \bibfield  {author} {\bibinfo {author} {\bibfnamefont {M.~A.}\ \bibnamefont
  {Metlitski}}\ and\ \bibinfo {author} {\bibfnamefont {S.}~\bibnamefont
  {Sachdev}},\ }\href {\doibase 10.1103/PhysRevB.82.075128} {\bibfield
  {journal} {\bibinfo  {journal} {Phys. Rev. B}\ }\textbf {\bibinfo {volume}
  {82}},\ \bibinfo {pages} {075128} (\bibinfo {year}
  {2010}{\natexlab{b}})}\BibitemShut {NoStop}%
\bibitem [{\citenamefont {Mross}\ \emph {et~al.}(2010)\citenamefont {Mross},
  \citenamefont {McGreevy}, \citenamefont {Liu},\ and\ \citenamefont
  {Senthil}}]{senthil}%
  \BibitemOpen
  \bibfield  {author} {\bibinfo {author} {\bibfnamefont {D.~F.}\ \bibnamefont
  {Mross}}, \bibinfo {author} {\bibfnamefont {J.}~\bibnamefont {McGreevy}},
  \bibinfo {author} {\bibfnamefont {H.}~\bibnamefont {Liu}}, \ and\ \bibinfo
  {author} {\bibfnamefont {T.}~\bibnamefont {Senthil}},\ }\href {\doibase
  10.1103/PhysRevB.82.045121} {\bibfield  {journal} {\bibinfo  {journal} {Phys.
  Rev. B}\ }\textbf {\bibinfo {volume} {82}},\ \bibinfo {pages} {045121}
  (\bibinfo {year} {2010})}\BibitemShut {NoStop}%
\bibitem [{\citenamefont {Mahajan}\ \emph {et~al.}(2013)\citenamefont
  {Mahajan}, \citenamefont {Ramirez}, \citenamefont {Kachru},\ and\
  \citenamefont {Raghu}}]{raghu}%
  \BibitemOpen
  \bibfield  {author} {\bibinfo {author} {\bibfnamefont {R.}~\bibnamefont
  {Mahajan}}, \bibinfo {author} {\bibfnamefont {D.~M.}\ \bibnamefont
  {Ramirez}}, \bibinfo {author} {\bibfnamefont {S.}~\bibnamefont {Kachru}}, \
  and\ \bibinfo {author} {\bibfnamefont {S.}~\bibnamefont {Raghu}},\ }\href
  {\doibase 10.1103/PhysRevB.88.115116} {\bibfield  {journal} {\bibinfo
  {journal} {Phys. Rev. B}\ }\textbf {\bibinfo {volume} {88}},\ \bibinfo
  {pages} {115116} (\bibinfo {year} {2013})}\BibitemShut {NoStop}%
\bibitem [{\citenamefont {Fitzpatrick}\ \emph {et~al.}(2013)\citenamefont
  {Fitzpatrick}, \citenamefont {Kachru}, \citenamefont {Kaplan},\ and\
  \citenamefont {Raghu}}]{raghu2}%
  \BibitemOpen
  \bibfield  {author} {\bibinfo {author} {\bibfnamefont {A.~L.}\ \bibnamefont
  {Fitzpatrick}}, \bibinfo {author} {\bibfnamefont {S.}~\bibnamefont {Kachru}},
  \bibinfo {author} {\bibfnamefont {J.}~\bibnamefont {Kaplan}}, \ and\ \bibinfo
  {author} {\bibfnamefont {S.}~\bibnamefont {Raghu}},\ }\href {\doibase
  10.1103/PhysRevB.88.125116} {\bibfield  {journal} {\bibinfo  {journal} {Phys.
  Rev. B}\ }\textbf {\bibinfo {volume} {88}},\ \bibinfo {pages} {125116}
  (\bibinfo {year} {2013})}\BibitemShut {NoStop}%
\bibitem [{\citenamefont {Fitzpatrick}\ \emph {et~al.}(2014)\citenamefont
  {Fitzpatrick}, \citenamefont {Kachru}, \citenamefont {Kaplan},\ and\
  \citenamefont {Raghu}}]{raghu3}%
  \BibitemOpen
  \bibfield  {author} {\bibinfo {author} {\bibfnamefont {A.~L.}\ \bibnamefont
  {Fitzpatrick}}, \bibinfo {author} {\bibfnamefont {S.}~\bibnamefont {Kachru}},
  \bibinfo {author} {\bibfnamefont {J.}~\bibnamefont {Kaplan}}, \ and\ \bibinfo
  {author} {\bibfnamefont {S.}~\bibnamefont {Raghu}},\ }\href {\doibase
  10.1103/PhysRevB.89.165114} {\bibfield  {journal} {\bibinfo  {journal} {Phys.
  Rev. B}\ }\textbf {\bibinfo {volume} {89}},\ \bibinfo {pages} {165114}
  (\bibinfo {year} {2014})}\BibitemShut {NoStop}%
\bibitem [{\citenamefont {Torroba}\ and\ \citenamefont {Wang}(2014)}]{raghu4}%
  \BibitemOpen
  \bibfield  {author} {\bibinfo {author} {\bibfnamefont {G.}~\bibnamefont
  {Torroba}}\ and\ \bibinfo {author} {\bibfnamefont {H.}~\bibnamefont {Wang}},\
  }\href {\doibase 10.1103/PhysRevB.90.165144} {\bibfield  {journal} {\bibinfo
  {journal} {Phys. Rev. B}\ }\textbf {\bibinfo {volume} {90}},\ \bibinfo
  {pages} {165144} (\bibinfo {year} {2014})}\BibitemShut {NoStop}%
\bibitem [{\citenamefont {Fitzpatrick}\ \emph {et~al.}(2015)\citenamefont
  {Fitzpatrick}, \citenamefont {Torroba},\ and\ \citenamefont {Wang}}]{raghu5}%
  \BibitemOpen
  \bibfield  {author} {\bibinfo {author} {\bibfnamefont {A.~L.}\ \bibnamefont
  {Fitzpatrick}}, \bibinfo {author} {\bibfnamefont {G.}~\bibnamefont
  {Torroba}}, \ and\ \bibinfo {author} {\bibfnamefont {H.}~\bibnamefont
  {Wang}},\ }\href {\doibase 10.1103/PhysRevB.91.195135} {\bibfield  {journal}
  {\bibinfo  {journal} {Phys. Rev. B}\ }\textbf {\bibinfo {volume} {91}},\
  \bibinfo {pages} {195135} (\bibinfo {year} {2015})},\ \bibinfo {note} {and
  references therein.}\BibitemShut {Stop}%
\bibitem [{\citenamefont {Monthoux}\ \emph {et~al.}(2007)\citenamefont
  {Monthoux}, \citenamefont {Pines},\ and\ \citenamefont {Lonzarich}}]{scal}%
  \BibitemOpen
  \bibfield  {author} {\bibinfo {author} {\bibfnamefont {P.}~\bibnamefont
  {Monthoux}}, \bibinfo {author} {\bibfnamefont {D.}~\bibnamefont {Pines}}, \
  and\ \bibinfo {author} {\bibfnamefont {G.~G.}\ \bibnamefont {Lonzarich}},\
  }\href {\doibase 10.1038/nature06480} {\bibfield  {journal} {\bibinfo
  {journal} {Nature}\ }\textbf {\bibinfo {volume} {450}},\ \bibinfo {pages}
  {1177} (\bibinfo {year} {2007})}\BibitemShut {NoStop}%
\bibitem [{\citenamefont {Scalapino}(2012{\natexlab{a}})}]{scal2}%
  \BibitemOpen
  \bibfield  {author} {\bibinfo {author} {\bibfnamefont {D.~J.}\ \bibnamefont
  {Scalapino}},\ }\href {\doibase 10.1103/RevModPhys.84.1383} {\bibfield
  {journal} {\bibinfo  {journal} {Rev. Mod. Phys.}\ }\textbf {\bibinfo {volume}
  {84}},\ \bibinfo {pages} {1383} (\bibinfo {year}
  {2012}{\natexlab{a}})}\BibitemShut {NoStop}%
\bibitem [{boo()}]{book1}%
  \BibitemOpen
  \href@noop {} {\ }\bibinfo {note} {M.R. Norman in ``Novel Superconductors",
  Bennemann and Ketterson eds., Oxford University Press (2014), and references
  therein.}\BibitemShut {Stop}%
\bibitem [{rev()}]{review4}%
  \BibitemOpen
  \href@noop {} {\ }\bibinfo {note} {S. Maiti and A. V. Chubukov in ``Novel
  Superconductors", Bennemann and Ketterson eds., Oxford University Press
  (2014), and references therein.}\BibitemShut {Stop}%
\bibitem [{\citenamefont {Fratino}\ \emph {et~al.}(2016)\citenamefont
  {Fratino}, \citenamefont {Sémon}, \citenamefont {Sordi},\ and\ \citenamefont
  {Tremblay}}]{review3}%
  \BibitemOpen
  \bibfield  {author} {\bibinfo {author} {\bibfnamefont {L.}~\bibnamefont
  {Fratino}}, \bibinfo {author} {\bibfnamefont {P.}~\bibnamefont {Sémon}},
  \bibinfo {author} {\bibfnamefont {G.}~\bibnamefont {Sordi}}, \ and\ \bibinfo
  {author} {\bibfnamefont {A.-M.~S.}\ \bibnamefont {Tremblay}},\ }\href
  {\doibase doi.org/10.1038/srep22715} {\bibfield  {journal} {\bibinfo
  {journal} {Scientific Reports}\ }\textbf {\bibinfo {volume} {6}},\ \bibinfo
  {pages} {22715} (\bibinfo {year} {2016})}\BibitemShut {NoStop}%
\bibitem [{\citenamefont {Metlitski}\ \emph {et~al.}(2015)\citenamefont
  {Metlitski}, \citenamefont {Mross}, \citenamefont {Sachdev},\ and\
  \citenamefont {Senthil}}]{max_last}%
  \BibitemOpen
  \bibfield  {author} {\bibinfo {author} {\bibfnamefont {M.~A.}\ \bibnamefont
  {Metlitski}}, \bibinfo {author} {\bibfnamefont {D.~F.}\ \bibnamefont
  {Mross}}, \bibinfo {author} {\bibfnamefont {S.}~\bibnamefont {Sachdev}}, \
  and\ \bibinfo {author} {\bibfnamefont {T.}~\bibnamefont {Senthil}},\ }\href
  {\doibase 10.1103/PhysRevB.91.115111} {\bibfield  {journal} {\bibinfo
  {journal} {Phys. Rev. B}\ }\textbf {\bibinfo {volume} {91}},\ \bibinfo
  {pages} {115111} (\bibinfo {year} {2015})}\BibitemShut {NoStop}%
\bibitem [{\citenamefont {Efetov}\ \emph {et~al.}(2013)\citenamefont {Efetov},
  \citenamefont {Meier},\ and\ \citenamefont {Pepin}}]{efetov}%
  \BibitemOpen
  \bibfield  {author} {\bibinfo {author} {\bibfnamefont {K.~B.}\ \bibnamefont
  {Efetov}}, \bibinfo {author} {\bibfnamefont {H.}~\bibnamefont {Meier}}, \
  and\ \bibinfo {author} {\bibfnamefont {C.}~\bibnamefont {Pepin}},\ }\href
  {\doibase 10.1038/nphys2641} {\bibfield  {journal} {\bibinfo  {journal}
  {Nature Physics}\ }\textbf {\bibinfo {volume} {9}},\ \bibinfo {pages} {442}
  (\bibinfo {year} {2013})}\BibitemShut {NoStop}%
\bibitem [{\citenamefont {Wang}\ and\ \citenamefont
  {Chubukov}(2013{\natexlab{a}})}]{wang}%
  \BibitemOpen
  \bibfield  {author} {\bibinfo {author} {\bibfnamefont {Y.}~\bibnamefont
  {Wang}}\ and\ \bibinfo {author} {\bibfnamefont {A.~V.}\ \bibnamefont
  {Chubukov}},\ }\href {\doibase 10.1103/PhysRevLett.110.127001} {\bibfield
  {journal} {\bibinfo  {journal} {Phys. Rev. Lett.}\ }\textbf {\bibinfo
  {volume} {110}},\ \bibinfo {pages} {127001} (\bibinfo {year}
  {2013}{\natexlab{a}})}\BibitemShut {NoStop}%
\bibitem [{\citenamefont {Chubukov}\ and\ \citenamefont
  {W\"olfle}(2014)}]{wang2}%
  \BibitemOpen
  \bibfield  {author} {\bibinfo {author} {\bibfnamefont {A.~V.}\ \bibnamefont
  {Chubukov}}\ and\ \bibinfo {author} {\bibfnamefont {P.}~\bibnamefont
  {W\"olfle}},\ }\href {\doibase 10.1103/PhysRevB.89.045108} {\bibfield
  {journal} {\bibinfo  {journal} {Phys. Rev. B}\ }\textbf {\bibinfo {volume}
  {89}},\ \bibinfo {pages} {045108} (\bibinfo {year} {2014})}\BibitemShut
  {NoStop}%
\bibitem [{\citenamefont {Raghu}\ \emph {et~al.}(2015)\citenamefont {Raghu},
  \citenamefont {Torroba},\ and\ \citenamefont {Wang}}]{raghu_15}%
  \BibitemOpen
  \bibfield  {author} {\bibinfo {author} {\bibfnamefont {S.}~\bibnamefont
  {Raghu}}, \bibinfo {author} {\bibfnamefont {G.}~\bibnamefont {Torroba}}, \
  and\ \bibinfo {author} {\bibfnamefont {H.}~\bibnamefont {Wang}},\ }\href
  {\doibase 10.1103/PhysRevB.92.205104} {\bibfield  {journal} {\bibinfo
  {journal} {Phys. Rev. B}\ }\textbf {\bibinfo {volume} {92}},\ \bibinfo
  {pages} {205104} (\bibinfo {year} {2015})}\BibitemShut {NoStop}%
\bibitem [{\citenamefont {Wang}\ \emph {et~al.}(2016)\citenamefont {Wang},
  \citenamefont {Abanov}, \citenamefont {Altshuler}, \citenamefont
  {Yuzbashyan},\ and\ \citenamefont {Chubukov}}]{Wang2016}%
  \BibitemOpen
  \bibfield  {author} {\bibinfo {author} {\bibfnamefont {Y.}~\bibnamefont
  {Wang}}, \bibinfo {author} {\bibfnamefont {A.}~\bibnamefont {Abanov}},
  \bibinfo {author} {\bibfnamefont {B.~L.}\ \bibnamefont {Altshuler}}, \bibinfo
  {author} {\bibfnamefont {E.~A.}\ \bibnamefont {Yuzbashyan}}, \ and\ \bibinfo
  {author} {\bibfnamefont {A.~V.}\ \bibnamefont {Chubukov}},\ }\href {\doibase
  10.1103/PhysRevLett.117.157001} {\bibfield  {journal} {\bibinfo  {journal}
  {Phys. Rev. Lett.}\ }\textbf {\bibinfo {volume} {117}},\ \bibinfo {pages}
  {157001} (\bibinfo {year} {2016})}\BibitemShut {NoStop}%
\bibitem [{\citenamefont {Lederer}\ \emph {et~al.}(2015)\citenamefont
  {Lederer}, \citenamefont {Schattner}, \citenamefont {Berg},\ and\
  \citenamefont {Kivelson}}]{steve_sam}%
  \BibitemOpen
  \bibfield  {author} {\bibinfo {author} {\bibfnamefont {S.}~\bibnamefont
  {Lederer}}, \bibinfo {author} {\bibfnamefont {Y.}~\bibnamefont {Schattner}},
  \bibinfo {author} {\bibfnamefont {E.}~\bibnamefont {Berg}}, \ and\ \bibinfo
  {author} {\bibfnamefont {S.~A.}\ \bibnamefont {Kivelson}},\ }\href {\doibase
  10.1103/PhysRevLett.114.097001} {\bibfield  {journal} {\bibinfo  {journal}
  {Phys. Rev. Lett.}\ }\textbf {\bibinfo {volume} {114}},\ \bibinfo {pages}
  {097001} (\bibinfo {year} {2015})}\BibitemShut {NoStop}%
\bibitem [{\citenamefont {Tsvelik}(2017)}]{tsvelik}%
  \BibitemOpen
  \bibfield  {author} {\bibinfo {author} {\bibfnamefont {A.~M.}\ \bibnamefont
  {Tsvelik}},\ }\href {\doibase 10.1103/PhysRevB.95.201112} {\bibfield
  {journal} {\bibinfo  {journal} {Phys. Rev. B}\ }\textbf {\bibinfo {volume}
  {95}},\ \bibinfo {pages} {201112} (\bibinfo {year} {2017})}\BibitemShut
  {NoStop}%
\bibitem [{\citenamefont {Vojta}\ and\ \citenamefont {Sachdev}(1999)}]{vojta}%
  \BibitemOpen
  \bibfield  {author} {\bibinfo {author} {\bibfnamefont {M.}~\bibnamefont
  {Vojta}}\ and\ \bibinfo {author} {\bibfnamefont {S.}~\bibnamefont
  {Sachdev}},\ }\href {\doibase 10.1103/PhysRevLett.83.3916} {\bibfield
  {journal} {\bibinfo  {journal} {Phys. Rev. Lett.}\ }\textbf {\bibinfo
  {volume} {83}},\ \bibinfo {pages} {3916} (\bibinfo {year}
  {1999})}\BibitemShut {NoStop}%
\bibitem [{\citenamefont {Fradkin}\ \emph {et~al.}(2010)\citenamefont
  {Fradkin}, \citenamefont {Kivelson}, \citenamefont {Lawler}, \citenamefont
  {Eisenstein},\ and\ \citenamefont {Mackenzie}}]{mack}%
  \BibitemOpen
  \bibfield  {author} {\bibinfo {author} {\bibfnamefont {E.}~\bibnamefont
  {Fradkin}}, \bibinfo {author} {\bibfnamefont {S.~A.}\ \bibnamefont
  {Kivelson}}, \bibinfo {author} {\bibfnamefont {M.~J.}\ \bibnamefont
  {Lawler}}, \bibinfo {author} {\bibfnamefont {J.~P.}\ \bibnamefont
  {Eisenstein}}, \ and\ \bibinfo {author} {\bibfnamefont {A.~P.}\ \bibnamefont
  {Mackenzie}},\ }\href {\doibase 10.1146/annurev-conmatphys-070909-103925}
  {\bibfield  {journal} {\bibinfo  {journal} {Annual Review of Condensed Matter
  Physics}\ }\textbf {\bibinfo {volume} {1}},\ \bibinfo {pages} {153} (\bibinfo
  {year} {2010})}\BibitemShut {NoStop}%
\bibitem [{\citenamefont {Bok}\ \emph {et~al.}(2016)\citenamefont {Bok},
  \citenamefont {Bae}, \citenamefont {Choi}, \citenamefont {Varma},
  \citenamefont {Zhang}, \citenamefont {He}, \citenamefont {Zhang},
  \citenamefont {Yu},\ and\ \citenamefont {Zhou}}]{varma}%
  \BibitemOpen
  \bibfield  {author} {\bibinfo {author} {\bibfnamefont {J.~M.}\ \bibnamefont
  {Bok}}, \bibinfo {author} {\bibfnamefont {J.~J.}\ \bibnamefont {Bae}},
  \bibinfo {author} {\bibfnamefont {H.-Y.}\ \bibnamefont {Choi}}, \bibinfo
  {author} {\bibfnamefont {C.~M.}\ \bibnamefont {Varma}}, \bibinfo {author}
  {\bibfnamefont {W.}~\bibnamefont {Zhang}}, \bibinfo {author} {\bibfnamefont
  {J.}~\bibnamefont {He}}, \bibinfo {author} {\bibfnamefont {Y.}~\bibnamefont
  {Zhang}}, \bibinfo {author} {\bibfnamefont {L.}~\bibnamefont {Yu}}, \ and\
  \bibinfo {author} {\bibfnamefont {X.~J.}\ \bibnamefont {Zhou}},\ }\href
  {\doibase 10.1126/sciadv.1501329} {\bibfield  {journal} {\bibinfo  {journal}
  {Science Advances}\ }\textbf {\bibinfo {volume} {2}} (\bibinfo {year}
  {2016}),\ 10.1126/sciadv.1501329}\BibitemShut {NoStop}%
\bibitem [{\citenamefont {Shibauchi}\ \emph {et~al.}(2014)\citenamefont
  {Shibauchi}, \citenamefont {Carrington},\ and\ \citenamefont
  {Matsuda}}]{matsuda}%
  \BibitemOpen
  \bibfield  {author} {\bibinfo {author} {\bibfnamefont {T.}~\bibnamefont
  {Shibauchi}}, \bibinfo {author} {\bibfnamefont {A.}~\bibnamefont
  {Carrington}}, \ and\ \bibinfo {author} {\bibfnamefont {Y.}~\bibnamefont
  {Matsuda}},\ }\href@noop {} {\bibfield  {journal} {\bibinfo  {journal}
  {Annual Review of Condensed Matter Physics}\ }\textbf {\bibinfo {volume}
  {5}},\ \bibinfo {pages} {113} (\bibinfo {year} {2014})}\BibitemShut {NoStop}%
\bibitem [{\citenamefont {Vilardi}\ \emph {et~al.}(2018)\citenamefont
  {Vilardi}, \citenamefont {Taranto},\ and\ \citenamefont {Metzner}}]{metzner}%
  \BibitemOpen
  \bibfield  {author} {\bibinfo {author} {\bibfnamefont {D.}~\bibnamefont
  {Vilardi}}, \bibinfo {author} {\bibfnamefont {C.}~\bibnamefont {Taranto}}, \
  and\ \bibinfo {author} {\bibfnamefont {W.}~\bibnamefont {Metzner}},\
  }\href@noop {} {\bibfield  {journal} {\bibinfo  {journal} {arXiv:1810.02290
  and references therein}\ } (\bibinfo {year} {2018})}\BibitemShut {NoStop}%
\bibitem [{\citenamefont {Gerlach}\ \emph {et~al.}(2017)\citenamefont
  {Gerlach}, \citenamefont {Schattner}, \citenamefont {Berg},\ and\
  \citenamefont {Trebst}}]{berg}%
  \BibitemOpen
  \bibfield  {author} {\bibinfo {author} {\bibfnamefont {M.~H.}\ \bibnamefont
  {Gerlach}}, \bibinfo {author} {\bibfnamefont {Y.}~\bibnamefont {Schattner}},
  \bibinfo {author} {\bibfnamefont {E.}~\bibnamefont {Berg}}, \ and\ \bibinfo
  {author} {\bibfnamefont {S.}~\bibnamefont {Trebst}},\ }\href {\doibase
  10.1103/PhysRevB.95.035124} {\bibfield  {journal} {\bibinfo  {journal} {Phys.
  Rev. B}\ }\textbf {\bibinfo {volume} {95}},\ \bibinfo {pages} {035124}
  (\bibinfo {year} {2017})}\BibitemShut {NoStop}%
\bibitem [{\citenamefont {Schattner}\ \emph {et~al.}(2016)\citenamefont
  {Schattner}, \citenamefont {Lederer}, \citenamefont {Kivelson},\ and\
  \citenamefont {Berg}}]{berg_2}%
  \BibitemOpen
  \bibfield  {author} {\bibinfo {author} {\bibfnamefont {Y.}~\bibnamefont
  {Schattner}}, \bibinfo {author} {\bibfnamefont {S.}~\bibnamefont {Lederer}},
  \bibinfo {author} {\bibfnamefont {S.~A.}\ \bibnamefont {Kivelson}}, \ and\
  \bibinfo {author} {\bibfnamefont {E.}~\bibnamefont {Berg}},\ }\href {\doibase
  10.1103/PhysRevX.6.031028} {\bibfield  {journal} {\bibinfo  {journal} {Phys.
  Rev. X}\ }\textbf {\bibinfo {volume} {6}},\ \bibinfo {pages} {031028}
  (\bibinfo {year} {2016})}\BibitemShut {NoStop}%
\bibitem [{\citenamefont {Wang}\ \emph {et~al.}(2017)\citenamefont {Wang},
  \citenamefont {Schattner}, \citenamefont {Berg},\ and\ \citenamefont
  {Fernandes}}]{berg_3}%
  \BibitemOpen
  \bibfield  {author} {\bibinfo {author} {\bibfnamefont {X.}~\bibnamefont
  {Wang}}, \bibinfo {author} {\bibfnamefont {Y.}~\bibnamefont {Schattner}},
  \bibinfo {author} {\bibfnamefont {E.}~\bibnamefont {Berg}}, \ and\ \bibinfo
  {author} {\bibfnamefont {R.~M.}\ \bibnamefont {Fernandes}},\ }\href {\doibase
  10.1103/PhysRevB.95.174520} {\bibfield  {journal} {\bibinfo  {journal} {Phys.
  Rev. B}\ }\textbf {\bibinfo {volume} {95}},\ \bibinfo {pages} {174520}
  (\bibinfo {year} {2017})}\BibitemShut {NoStop}%
\bibitem [{\citenamefont {Haule}\ and\ \citenamefont
  {Kotliar}(2007)}]{kotliar}%
  \BibitemOpen
  \bibfield  {author} {\bibinfo {author} {\bibfnamefont {K.}~\bibnamefont
  {Haule}}\ and\ \bibinfo {author} {\bibfnamefont {G.}~\bibnamefont
  {Kotliar}},\ }\href {\doibase 10.1103/PhysRevB.76.104509} {\bibfield
  {journal} {\bibinfo  {journal} {Phys. Rev. B}\ }\textbf {\bibinfo {volume}
  {76}},\ \bibinfo {pages} {104509} (\bibinfo {year} {2007})}\BibitemShut
  {NoStop}%
\bibitem [{\citenamefont {Xu}\ \emph {et~al.}(2017)\citenamefont {Xu},
  \citenamefont {Kotliar},\ and\ \citenamefont {Tsvelik}}]{kotliar2}%
  \BibitemOpen
  \bibfield  {author} {\bibinfo {author} {\bibfnamefont {W.}~\bibnamefont
  {Xu}}, \bibinfo {author} {\bibfnamefont {G.}~\bibnamefont {Kotliar}}, \ and\
  \bibinfo {author} {\bibfnamefont {A.~M.}\ \bibnamefont {Tsvelik}},\ }\href
  {\doibase 10.1103/PhysRevB.95.121113} {\bibfield  {journal} {\bibinfo
  {journal} {Phys. Rev. B}\ }\textbf {\bibinfo {volume} {95}},\ \bibinfo
  {pages} {121113} (\bibinfo {year} {2017})}\BibitemShut {NoStop}%
\bibitem [{\citenamefont {Sordi}\ \emph {et~al.}(2012)\citenamefont {Sordi},
  \citenamefont {Simon}, \citenamefont {Haule},\ and\ \citenamefont
  {Tremblay}}]{tremblay_2}%
  \BibitemOpen
  \bibfield  {author} {\bibinfo {author} {\bibfnamefont {G.}~\bibnamefont
  {Sordi}}, \bibinfo {author} {\bibfnamefont {P.}~\bibnamefont {Simon}},
  \bibinfo {author} {\bibfnamefont {K.}~\bibnamefont {Haule}}, \ and\ \bibinfo
  {author} {\bibfnamefont {A.-M.~S.}\ \bibnamefont {Tremblay}},\ }\href
  {\doibase 10.1103/PhysRevLett.108.216401} {\bibfield  {journal} {\bibinfo
  {journal} {Phys. Rev. Lett.}\ }\textbf {\bibinfo {volume} {108}},\ \bibinfo
  {pages} {216401} (\bibinfo {year} {2012})}\BibitemShut {NoStop}%
\bibitem [{\citenamefont {Georges}\ \emph {et~al.}(2013)\citenamefont
  {Georges}, \citenamefont {Medici},\ and\ \citenamefont {Mravlje}}]{georges}%
  \BibitemOpen
  \bibfield  {author} {\bibinfo {author} {\bibfnamefont {A.}~\bibnamefont
  {Georges}}, \bibinfo {author} {\bibfnamefont {L.~d.}\ \bibnamefont {Medici}},
  \ and\ \bibinfo {author} {\bibfnamefont {J.}~\bibnamefont {Mravlje}},\
  }\href@noop {} {\bibfield  {journal} {\bibinfo  {journal} {Annu. Rev.
  Condens. Matter Phys.}\ }\textbf {\bibinfo {volume} {4}},\ \bibinfo {pages}
  {137} (\bibinfo {year} {2013})}\BibitemShut {NoStop}%
\bibitem [{\citenamefont {Wu}\ \emph {et~al.}(2017)\citenamefont {Wu},
  \citenamefont {Ferrero}, \citenamefont {Georges},\ and\ \citenamefont
  {Kozik}}]{georges2}%
  \BibitemOpen
  \bibfield  {author} {\bibinfo {author} {\bibfnamefont {W.}~\bibnamefont
  {Wu}}, \bibinfo {author} {\bibfnamefont {M.}~\bibnamefont {Ferrero}},
  \bibinfo {author} {\bibfnamefont {A.}~\bibnamefont {Georges}}, \ and\
  \bibinfo {author} {\bibfnamefont {E.}~\bibnamefont {Kozik}},\ }\href
  {\doibase 10.1103/PhysRevB.96.041105} {\bibfield  {journal} {\bibinfo
  {journal} {Phys. Rev. B}\ }\textbf {\bibinfo {volume} {96}},\ \bibinfo
  {pages} {041105} (\bibinfo {year} {2017})}\BibitemShut {NoStop}%
\bibitem [{\citenamefont {Khveshchenko}\ and\ \citenamefont
  {Shively}(2006)}]{khvesh}%
  \BibitemOpen
  \bibfield  {author} {\bibinfo {author} {\bibfnamefont {D.~V.}\ \bibnamefont
  {Khveshchenko}}\ and\ \bibinfo {author} {\bibfnamefont {W.~F.}\ \bibnamefont
  {Shively}},\ }\href {\doibase 10.1103/PhysRevB.73.115104} {\bibfield
  {journal} {\bibinfo  {journal} {Phys. Rev. B}\ }\textbf {\bibinfo {volume}
  {73}},\ \bibinfo {pages} {115104} (\bibinfo {year} {2006})}\BibitemShut
  {NoStop}%
\bibitem [{\citenamefont {Lee}\ \emph {et~al.}(2018)\citenamefont {Lee},
  \citenamefont {Chubukov}, \citenamefont {Miao},\ and\ \citenamefont
  {Kotliar}}]{Kotliar2018}%
  \BibitemOpen
  \bibfield  {author} {\bibinfo {author} {\bibfnamefont {T.-H.}\ \bibnamefont
  {Lee}}, \bibinfo {author} {\bibfnamefont {A.}~\bibnamefont {Chubukov}},
  \bibinfo {author} {\bibfnamefont {H.}~\bibnamefont {Miao}}, \ and\ \bibinfo
  {author} {\bibfnamefont {G.}~\bibnamefont {Kotliar}},\ }\href
  {https://arxiv.org/abs/1805.10280} {\bibfield  {journal} {\bibinfo  {journal}
  {arXiv}\ }\textbf {\bibinfo {volume} {1805}},\ \bibinfo {pages} {10280}
  (\bibinfo {year} {2018})}\BibitemShut {NoStop}%
\bibitem [{we_()}]{we_last_D}%
  \BibitemOpen
  \href@noop {} {\ }\bibinfo {note} {Y-M Wu, Ar. Abanov, and A. V. Chubukov,
  arXiv:1811.02087}\BibitemShut {NoStop}%
\bibitem [{\citenamefont {Rech}\ \emph {et~al.}(2006)\citenamefont {Rech},
  \citenamefont {P\'epin},\ and\ \citenamefont {Chubukov}}]{pepin}%
  \BibitemOpen
  \bibfield  {author} {\bibinfo {author} {\bibfnamefont {J.}~\bibnamefont
  {Rech}}, \bibinfo {author} {\bibfnamefont {C.}~\bibnamefont {P\'epin}}, \
  and\ \bibinfo {author} {\bibfnamefont {A.~V.}\ \bibnamefont {Chubukov}},\
  }\href {\doibase 10.1103/PhysRevB.74.195126} {\bibfield  {journal} {\bibinfo
  {journal} {Phys. Rev. B}\ }\textbf {\bibinfo {volume} {74}},\ \bibinfo
  {pages} {195126} (\bibinfo {year} {2006})}\BibitemShut {NoStop}%
\bibitem [{q=0()}]{q=0}%
  \BibitemOpen
  \href@noop {} {\ }\bibinfo {note} {There is a large body of literature on QCP
  with $q=0$. For recent works, see}\BibitemShut {NoStop}%
\bibitem [{\citenamefont {Sur}\ and\ \citenamefont {Lee}(2015)}]{q=01}%
  \BibitemOpen
  \bibfield  {author} {\bibinfo {author} {\bibfnamefont {S.}~\bibnamefont
  {Sur}}\ and\ \bibinfo {author} {\bibfnamefont {S.-S.}\ \bibnamefont {Lee}},\
  }\href {\doibase 10.1103/PhysRevB.91.125136} {\bibfield  {journal} {\bibinfo
  {journal} {Phys. Rev. B}\ }\textbf {\bibinfo {volume} {91}},\ \bibinfo
  {pages} {125136} (\bibinfo {year} {2015})}\BibitemShut {NoStop}%
\bibitem [{\citenamefont {Punk}(2015)}]{q=02}%
  \BibitemOpen
  \bibfield  {author} {\bibinfo {author} {\bibfnamefont {M.}~\bibnamefont
  {Punk}},\ }\href {\doibase 10.1103/PhysRevB.91.115131} {\bibfield  {journal}
  {\bibinfo  {journal} {Phys. Rev. B}\ }\textbf {\bibinfo {volume} {91}},\
  \bibinfo {pages} {115131} (\bibinfo {year} {2015})},\ \bibinfo {note} {and
  references therein.}\BibitemShut {Stop}%
\bibitem [{\citenamefont {Abanov}\ \emph {et~al.}(2008)\citenamefont {Abanov},
  \citenamefont {Chubukov},\ and\ \citenamefont {Norman}}]{acn}%
  \BibitemOpen
  \bibfield  {author} {\bibinfo {author} {\bibfnamefont {A.}~\bibnamefont
  {Abanov}}, \bibinfo {author} {\bibfnamefont {A.~V.}\ \bibnamefont
  {Chubukov}}, \ and\ \bibinfo {author} {\bibfnamefont {M.~R.}\ \bibnamefont
  {Norman}},\ }\href {\doibase 10.1103/PhysRevB.78.220507} {\bibfield
  {journal} {\bibinfo  {journal} {Phys. Rev. B}\ }\textbf {\bibinfo {volume}
  {78}},\ \bibinfo {pages} {220507} (\bibinfo {year} {2008})}\BibitemShut
  {NoStop}%
\bibitem [{\citenamefont {Altshuler}\ \emph {et~al.}(1995)\citenamefont
  {Altshuler}, \citenamefont {Ioffe},\ and\ \citenamefont {Millis}}]{2kf}%
  \BibitemOpen
  \bibfield  {author} {\bibinfo {author} {\bibfnamefont {B.~L.}\ \bibnamefont
  {Altshuler}}, \bibinfo {author} {\bibfnamefont {L.~B.}\ \bibnamefont
  {Ioffe}}, \ and\ \bibinfo {author} {\bibfnamefont {A.~J.}\ \bibnamefont
  {Millis}},\ }\href {\doibase 10.1103/PhysRevB.52.5563} {\bibfield  {journal}
  {\bibinfo  {journal} {Phys. Rev. B}\ }\textbf {\bibinfo {volume} {52}},\
  \bibinfo {pages} {5563} (\bibinfo {year} {1995})}\BibitemShut {NoStop}%
\bibitem [{\citenamefont {Bergeron}\ \emph {et~al.}(2012)\citenamefont
  {Bergeron}, \citenamefont {Chowdhury}, \citenamefont {Punk}, \citenamefont
  {Sachdev},\ and\ \citenamefont {Tremblay}}]{2kf2}%
  \BibitemOpen
  \bibfield  {author} {\bibinfo {author} {\bibfnamefont {D.}~\bibnamefont
  {Bergeron}}, \bibinfo {author} {\bibfnamefont {D.}~\bibnamefont {Chowdhury}},
  \bibinfo {author} {\bibfnamefont {M.}~\bibnamefont {Punk}}, \bibinfo {author}
  {\bibfnamefont {S.}~\bibnamefont {Sachdev}}, \ and\ \bibinfo {author}
  {\bibfnamefont {A.-M.~S.}\ \bibnamefont {Tremblay}},\ }\href {\doibase
  10.1103/PhysRevB.86.155123} {\bibfield  {journal} {\bibinfo  {journal} {Phys.
  Rev. B}\ }\textbf {\bibinfo {volume} {86}},\ \bibinfo {pages} {155123}
  (\bibinfo {year} {2012})}\BibitemShut {NoStop}%
\bibitem [{\citenamefont {Wang}\ and\ \citenamefont
  {Chubukov}(2013{\natexlab{b}})}]{2kf3}%
  \BibitemOpen
  \bibfield  {author} {\bibinfo {author} {\bibfnamefont {Y.}~\bibnamefont
  {Wang}}\ and\ \bibinfo {author} {\bibfnamefont {A.}~\bibnamefont
  {Chubukov}},\ }\href {\doibase 10.1103/PhysRevB.88.024516} {\bibfield
  {journal} {\bibinfo  {journal} {Phys. Rev. B}\ }\textbf {\bibinfo {volume}
  {88}},\ \bibinfo {pages} {024516} (\bibinfo {year}
  {2013}{\natexlab{b}})}\BibitemShut {NoStop}%
\bibitem [{\citenamefont {Scalapino}(2012{\natexlab{b}})}]{review}%
  \BibitemOpen
  \bibfield  {author} {\bibinfo {author} {\bibfnamefont {D.~J.}\ \bibnamefont
  {Scalapino}},\ }\href {\doibase 10.1103/RevModPhys.84.1383} {\bibfield
  {journal} {\bibinfo  {journal} {Rev. Mod. Phys.}\ }\textbf {\bibinfo {volume}
  {84}},\ \bibinfo {pages} {1383} (\bibinfo {year}
  {2012}{\natexlab{b}})}\BibitemShut {NoStop}%
\bibitem [{\citenamefont {Lederer}\ \emph {et~al.}(2017)\citenamefont
  {Lederer}, \citenamefont {Schattner}, \citenamefont {Berg},\ and\
  \citenamefont {Kivelson}}]{review2}%
  \BibitemOpen
  \bibfield  {author} {\bibinfo {author} {\bibfnamefont {S.}~\bibnamefont
  {Lederer}}, \bibinfo {author} {\bibfnamefont {Y.}~\bibnamefont {Schattner}},
  \bibinfo {author} {\bibfnamefont {E.}~\bibnamefont {Berg}}, \ and\ \bibinfo
  {author} {\bibfnamefont {S.~A.}\ \bibnamefont {Kivelson}},\ }\href {\doibase
  10.1073/pnas.1620651114} {\bibfield  {journal} {\bibinfo  {journal}
  {Proceedings of the National Academy of Sciences}\ }\textbf {\bibinfo
  {volume} {114}},\ \bibinfo {pages} {4905} (\bibinfo {year}
  {2017})}\BibitemShut {NoStop}%
\bibitem [{\citenamefont {Anderson}(1959)}]{anderson}%
  \BibitemOpen
  \bibfield  {author} {\bibinfo {author} {\bibfnamefont {P.}~\bibnamefont
  {Anderson}},\ }\href {\doibase https://doi.org/10.1016/0022-3697(59)90036-8}
  {\bibfield  {journal} {\bibinfo  {journal} {Journal of Physics and Chemistry
  of Solids}\ }\textbf {\bibinfo {volume} {11}},\ \bibinfo {pages} {26 }
  (\bibinfo {year} {1959})}\BibitemShut {NoStop}%
\bibitem [{\citenamefont {Abrikosov}\ and\ \citenamefont
  {Gor'kov}(1959)}]{abr-gor}%
  \BibitemOpen
  \bibfield  {author} {\bibinfo {author} {\bibfnamefont {A.~A.}\ \bibnamefont
  {Abrikosov}}\ and\ \bibinfo {author} {\bibfnamefont {L.~P.}\ \bibnamefont
  {Gor'kov}},\ }\href {http://www.jetp.ac.ru/cgi-bin/e/index/e/9/1/p220?a=list}
  {\bibfield  {journal} {\bibinfo  {journal} {JETP}\ }\textbf {\bibinfo
  {volume} {9}},\ \bibinfo {pages} {220} (\bibinfo {year} {1959})}\BibitemShut
  {NoStop}%
\bibitem [{\citenamefont {Fischer}\ \emph {et~al.}(2007)\citenamefont
  {Fischer}, \citenamefont {Kugler}, \citenamefont {Maggio-Aprile},
  \citenamefont {Berthod},\ and\ \citenamefont {Renner}}]{DOS}%
  \BibitemOpen
  \bibfield  {author} {\bibinfo {author} {\bibfnamefont {O.}~\bibnamefont
  {Fischer}}, \bibinfo {author} {\bibfnamefont {M.}~\bibnamefont {Kugler}},
  \bibinfo {author} {\bibfnamefont {I.}~\bibnamefont {Maggio-Aprile}}, \bibinfo
  {author} {\bibfnamefont {C.}~\bibnamefont {Berthod}}, \ and\ \bibinfo
  {author} {\bibfnamefont {C.}~\bibnamefont {Renner}},\ }\href {\doibase
  10.1103/RevModPhys.79.353} {\bibfield  {journal} {\bibinfo  {journal} {Rev.
  Mod. Phys.}\ }\textbf {\bibinfo {volume} {79}},\ \bibinfo {pages} {353}
  (\bibinfo {year} {2007})}\BibitemShut {NoStop}%
\bibitem [{\citenamefont {Reber}\ \emph {et~al.}(2012)\citenamefont {Reber},
  \citenamefont {Plumb}, \citenamefont {Sun}, \citenamefont {Cao},
  \citenamefont {Wang}, \citenamefont {McElroy}, \citenamefont {Iwasawa},
  \citenamefont {Arita}, \citenamefont {Wen}, \citenamefont {Xu}, \citenamefont
  {Gu}, \citenamefont {Yoshida}, \citenamefont {Eisaki}, \citenamefont
  {Aiura},\ and\ \citenamefont {Dessau}}]{dessau}%
  \BibitemOpen
  \bibfield  {author} {\bibinfo {author} {\bibfnamefont {T.~J.}\ \bibnamefont
  {Reber}}, \bibinfo {author} {\bibfnamefont {N.~C.}\ \bibnamefont {Plumb}},
  \bibinfo {author} {\bibfnamefont {Z.}~\bibnamefont {Sun}}, \bibinfo {author}
  {\bibfnamefont {Y.}~\bibnamefont {Cao}}, \bibinfo {author} {\bibfnamefont
  {Q.}~\bibnamefont {Wang}}, \bibinfo {author} {\bibfnamefont {K.}~\bibnamefont
  {McElroy}}, \bibinfo {author} {\bibfnamefont {H.}~\bibnamefont {Iwasawa}},
  \bibinfo {author} {\bibfnamefont {M.}~\bibnamefont {Arita}}, \bibinfo
  {author} {\bibfnamefont {J.~S.}\ \bibnamefont {Wen}}, \bibinfo {author}
  {\bibfnamefont {Z.~J.}\ \bibnamefont {Xu}}, \bibinfo {author} {\bibfnamefont
  {G.}~\bibnamefont {Gu}}, \bibinfo {author} {\bibfnamefont {Y.}~\bibnamefont
  {Yoshida}}, \bibinfo {author} {\bibfnamefont {H.}~\bibnamefont {Eisaki}},
  \bibinfo {author} {\bibfnamefont {Y.}~\bibnamefont {Aiura}}, \ and\ \bibinfo
  {author} {\bibfnamefont {D.~S.}\ \bibnamefont {Dessau}},\ }\href
  {https://doi.org/10.1038/nphys2352} {\bibfield  {journal} {\bibinfo
  {journal} {Nature Physics}\ }\textbf {\bibinfo {volume} {8}},\ \bibinfo
  {pages} {606 EP } (\bibinfo {year} {2012})}\BibitemShut {NoStop}%
\bibitem [{\citenamefont {Kondo}\ \emph {et~al.}(2013)\citenamefont {Kondo},
  \citenamefont {Palczewski}, \citenamefont {Hamaya}, \citenamefont {Takeuchi},
  \citenamefont {Wen}, \citenamefont {Xu}, \citenamefont {Gu},\ and\
  \citenamefont {Kaminski}}]{kaminski}%
  \BibitemOpen
  \bibfield  {author} {\bibinfo {author} {\bibfnamefont {T.}~\bibnamefont
  {Kondo}}, \bibinfo {author} {\bibfnamefont {A.~D.}\ \bibnamefont
  {Palczewski}}, \bibinfo {author} {\bibfnamefont {Y.}~\bibnamefont {Hamaya}},
  \bibinfo {author} {\bibfnamefont {T.}~\bibnamefont {Takeuchi}}, \bibinfo
  {author} {\bibfnamefont {J.~S.}\ \bibnamefont {Wen}}, \bibinfo {author}
  {\bibfnamefont {Z.~J.}\ \bibnamefont {Xu}}, \bibinfo {author} {\bibfnamefont
  {G.}~\bibnamefont {Gu}}, \ and\ \bibinfo {author} {\bibfnamefont
  {A.}~\bibnamefont {Kaminski}},\ }\href {\doibase
  10.1103/PhysRevLett.111.157003} {\bibfield  {journal} {\bibinfo  {journal}
  {Phys. Rev. Lett.}\ }\textbf {\bibinfo {volume} {111}},\ \bibinfo {pages}
  {157003} (\bibinfo {year} {2013})}\BibitemShut {NoStop}%
\bibitem [{\citenamefont {Kondo}\ \emph {et~al.}(2008)\citenamefont {Kondo},
  \citenamefont {Santander-Syro}, \citenamefont {Copie}, \citenamefont {Liu},
  \citenamefont {Tillman}, \citenamefont {Mun}, \citenamefont {Schmalian},
  \citenamefont {Bud'ko}, \citenamefont {Tanatar}, \citenamefont {Canfield},\
  and\ \citenamefont {Kaminski}}]{Kaminski2}%
  \BibitemOpen
  \bibfield  {author} {\bibinfo {author} {\bibfnamefont {T.}~\bibnamefont
  {Kondo}}, \bibinfo {author} {\bibfnamefont {A.~F.}\ \bibnamefont
  {Santander-Syro}}, \bibinfo {author} {\bibfnamefont {O.}~\bibnamefont
  {Copie}}, \bibinfo {author} {\bibfnamefont {C.}~\bibnamefont {Liu}}, \bibinfo
  {author} {\bibfnamefont {M.~E.}\ \bibnamefont {Tillman}}, \bibinfo {author}
  {\bibfnamefont {E.~D.}\ \bibnamefont {Mun}}, \bibinfo {author} {\bibfnamefont
  {J.}~\bibnamefont {Schmalian}}, \bibinfo {author} {\bibfnamefont {S.~L.}\
  \bibnamefont {Bud'ko}}, \bibinfo {author} {\bibfnamefont {M.~A.}\
  \bibnamefont {Tanatar}}, \bibinfo {author} {\bibfnamefont {P.~C.}\
  \bibnamefont {Canfield}}, \ and\ \bibinfo {author} {\bibfnamefont
  {A.}~\bibnamefont {Kaminski}},\ }\href {\doibase
  10.1103/PhysRevLett.101.147003} {\bibfield  {journal} {\bibinfo  {journal}
  {Phys. Rev. Lett.}\ }\textbf {\bibinfo {volume} {101}},\ \bibinfo {pages}
  {147003} (\bibinfo {year} {2008})}\BibitemShut {NoStop}%
\bibitem [{\citenamefont {Kanigel}\ \emph {et~al.}(2006)\citenamefont
  {Kanigel}, \citenamefont {Norman}, \citenamefont {Randeria}, \citenamefont
  {Chatterjee}, \citenamefont {Souma}, \citenamefont {Kaminski}, \citenamefont
  {Fretwell}, \citenamefont {Rosenkranz}, \citenamefont {Shi}, \citenamefont
  {Sato}, \citenamefont {Takahashi}, \citenamefont {Li}, \citenamefont {Raffy},
  \citenamefont {Kadowaki}, \citenamefont {Hinks}, \citenamefont {Ozyuzer},\
  and\ \citenamefont {Campuzano}}]{kanigel}%
  \BibitemOpen
  \bibfield  {author} {\bibinfo {author} {\bibfnamefont {A.}~\bibnamefont
  {Kanigel}}, \bibinfo {author} {\bibfnamefont {M.~R.}\ \bibnamefont {Norman}},
  \bibinfo {author} {\bibfnamefont {M.}~\bibnamefont {Randeria}}, \bibinfo
  {author} {\bibfnamefont {U.}~\bibnamefont {Chatterjee}}, \bibinfo {author}
  {\bibfnamefont {S.}~\bibnamefont {Souma}}, \bibinfo {author} {\bibfnamefont
  {A.}~\bibnamefont {Kaminski}}, \bibinfo {author} {\bibfnamefont {H.~M.}\
  \bibnamefont {Fretwell}}, \bibinfo {author} {\bibfnamefont {S.}~\bibnamefont
  {Rosenkranz}}, \bibinfo {author} {\bibfnamefont {M.}~\bibnamefont {Shi}},
  \bibinfo {author} {\bibfnamefont {T.}~\bibnamefont {Sato}}, \bibinfo {author}
  {\bibfnamefont {T.}~\bibnamefont {Takahashi}}, \bibinfo {author}
  {\bibfnamefont {Z.~Z.}\ \bibnamefont {Li}}, \bibinfo {author} {\bibfnamefont
  {H.}~\bibnamefont {Raffy}}, \bibinfo {author} {\bibfnamefont
  {K.}~\bibnamefont {Kadowaki}}, \bibinfo {author} {\bibfnamefont
  {D.}~\bibnamefont {Hinks}}, \bibinfo {author} {\bibfnamefont
  {L.}~\bibnamefont {Ozyuzer}}, \ and\ \bibinfo {author} {\bibfnamefont
  {J.~C.}\ \bibnamefont {Campuzano}},\ }\href
  {https://doi.org/10.1038/nphys334} {\bibfield  {journal} {\bibinfo  {journal}
  {Nature Physics}\ }\textbf {\bibinfo {volume} {2}},\ \bibinfo {pages} {447 EP
  } (\bibinfo {year} {2006})}\BibitemShut {NoStop}%
\bibitem [{\citenamefont {Kanigel}\ \emph {et~al.}(2007)\citenamefont
  {Kanigel}, \citenamefont {Chatterjee}, \citenamefont {Randeria},
  \citenamefont {Norman}, \citenamefont {Souma}, \citenamefont {Shi},
  \citenamefont {Li}, \citenamefont {Raffy},\ and\ \citenamefont
  {Campuzano}}]{kanigel2}%
  \BibitemOpen
  \bibfield  {author} {\bibinfo {author} {\bibfnamefont {A.}~\bibnamefont
  {Kanigel}}, \bibinfo {author} {\bibfnamefont {U.}~\bibnamefont {Chatterjee}},
  \bibinfo {author} {\bibfnamefont {M.}~\bibnamefont {Randeria}}, \bibinfo
  {author} {\bibfnamefont {M.~R.}\ \bibnamefont {Norman}}, \bibinfo {author}
  {\bibfnamefont {S.}~\bibnamefont {Souma}}, \bibinfo {author} {\bibfnamefont
  {M.}~\bibnamefont {Shi}}, \bibinfo {author} {\bibfnamefont {Z.~Z.}\
  \bibnamefont {Li}}, \bibinfo {author} {\bibfnamefont {H.}~\bibnamefont
  {Raffy}}, \ and\ \bibinfo {author} {\bibfnamefont {J.~C.}\ \bibnamefont
  {Campuzano}},\ }\href {\doibase 10.1103/PhysRevLett.99.157001} {\bibfield
  {journal} {\bibinfo  {journal} {Phys. Rev. Lett.}\ }\textbf {\bibinfo
  {volume} {99}},\ \bibinfo {pages} {157001} (\bibinfo {year}
  {2007})}\BibitemShut {NoStop}%
\bibitem [{\citenamefont {Ding}\ \emph {et~al.}(1995)\citenamefont {Ding},
  \citenamefont {Campuzano}, \citenamefont {Bellman}, \citenamefont {Yokoya},
  \citenamefont {Norman}, \citenamefont {Randeria}, \citenamefont {Takahashi},
  \citenamefont {Katayama-Yoshida}, \citenamefont {Mochiku}, \citenamefont
  {Kadowaki},\ and\ \citenamefont {Jennings}}]{kanigel3}%
  \BibitemOpen
  \bibfield  {author} {\bibinfo {author} {\bibfnamefont {H.}~\bibnamefont
  {Ding}}, \bibinfo {author} {\bibfnamefont {J.~C.}\ \bibnamefont {Campuzano}},
  \bibinfo {author} {\bibfnamefont {A.~F.}\ \bibnamefont {Bellman}}, \bibinfo
  {author} {\bibfnamefont {T.}~\bibnamefont {Yokoya}}, \bibinfo {author}
  {\bibfnamefont {M.~R.}\ \bibnamefont {Norman}}, \bibinfo {author}
  {\bibfnamefont {M.}~\bibnamefont {Randeria}}, \bibinfo {author}
  {\bibfnamefont {T.}~\bibnamefont {Takahashi}}, \bibinfo {author}
  {\bibfnamefont {H.}~\bibnamefont {Katayama-Yoshida}}, \bibinfo {author}
  {\bibfnamefont {T.}~\bibnamefont {Mochiku}}, \bibinfo {author} {\bibfnamefont
  {K.}~\bibnamefont {Kadowaki}}, \ and\ \bibinfo {author} {\bibfnamefont
  {G.}~\bibnamefont {Jennings}},\ }\href {\doibase 10.1103/PhysRevLett.74.2784}
  {\bibfield  {journal} {\bibinfo  {journal} {Phys. Rev. Lett.}\ }\textbf
  {\bibinfo {volume} {74}},\ \bibinfo {pages} {2784} (\bibinfo {year}
  {1995})}\BibitemShut {NoStop}%
\bibitem [{nor()}]{norman_review}%
  \BibitemOpen
  \href@noop {} {\ }\bibinfo {note} {J.C. Campuzano, M.R. Norman, and M.
  Randeria "Photoemission in the high-$T_c$ supercondutors", in "Novel
  Superconductors", v.2, K.H. Bennemann and J.B. Ketterson eds, Springer
  2008.}\BibitemShut {Stop}%
\bibitem [{\citenamefont {Damascelli}\ \emph {et~al.}(2003)\citenamefont
  {Damascelli}, \citenamefont {Hussain},\ and\ \citenamefont {Shen}}]{shen}%
  \BibitemOpen
  \bibfield  {author} {\bibinfo {author} {\bibfnamefont {A.}~\bibnamefont
  {Damascelli}}, \bibinfo {author} {\bibfnamefont {Z.}~\bibnamefont {Hussain}},
  \ and\ \bibinfo {author} {\bibfnamefont {Z.-X.}\ \bibnamefont {Shen}},\
  }\href {\doibase 10.1103/RevModPhys.75.473} {\bibfield  {journal} {\bibinfo
  {journal} {Rev. Mod. Phys.}\ }\textbf {\bibinfo {volume} {75}},\ \bibinfo
  {pages} {473} (\bibinfo {year} {2003})}\BibitemShut {NoStop}%
\bibitem [{\citenamefont {Hashimoto}\ \emph {et~al.}(2014)\citenamefont
  {Hashimoto}, \citenamefont {Vishik}, \citenamefont {He}, \citenamefont
  {Devereaux},\ and\ \citenamefont {Shen}}]{shen2}%
  \BibitemOpen
  \bibfield  {author} {\bibinfo {author} {\bibfnamefont {M.}~\bibnamefont
  {Hashimoto}}, \bibinfo {author} {\bibfnamefont {I.~M.}\ \bibnamefont
  {Vishik}}, \bibinfo {author} {\bibfnamefont {R.-H.}\ \bibnamefont {He}},
  \bibinfo {author} {\bibfnamefont {T.~P.}\ \bibnamefont {Devereaux}}, \ and\
  \bibinfo {author} {\bibfnamefont {Z.-X.}\ \bibnamefont {Shen}},\ }\href
  {https://doi.org/10.1038/nphys3009} {\bibfield  {journal} {\bibinfo
  {journal} {Nature Physics}\ }\textbf {\bibinfo {volume} {10}},\ \bibinfo
  {pages} {483 EP } (\bibinfo {year} {2014})},\ \bibinfo {note} {review
  Article}\BibitemShut {NoStop}%
\bibitem [{\citenamefont {Johnson}\ \emph {et~al.}(2001)\citenamefont
  {Johnson}, \citenamefont {Valla}, \citenamefont {Fedorov}, \citenamefont
  {Yusof}, \citenamefont {Wells}, \citenamefont {Li}, \citenamefont
  {Moodenbaugh}, \citenamefont {Gu}, \citenamefont {Koshizuka}, \citenamefont
  {Kendziora}, \citenamefont {Jian},\ and\ \citenamefont {Hinks}}]{shen3}%
  \BibitemOpen
  \bibfield  {author} {\bibinfo {author} {\bibfnamefont {P.~D.}\ \bibnamefont
  {Johnson}}, \bibinfo {author} {\bibfnamefont {T.}~\bibnamefont {Valla}},
  \bibinfo {author} {\bibfnamefont {A.~V.}\ \bibnamefont {Fedorov}}, \bibinfo
  {author} {\bibfnamefont {Z.}~\bibnamefont {Yusof}}, \bibinfo {author}
  {\bibfnamefont {B.~O.}\ \bibnamefont {Wells}}, \bibinfo {author}
  {\bibfnamefont {Q.}~\bibnamefont {Li}}, \bibinfo {author} {\bibfnamefont
  {A.~R.}\ \bibnamefont {Moodenbaugh}}, \bibinfo {author} {\bibfnamefont
  {G.~D.}\ \bibnamefont {Gu}}, \bibinfo {author} {\bibfnamefont
  {N.}~\bibnamefont {Koshizuka}}, \bibinfo {author} {\bibfnamefont
  {C.}~\bibnamefont {Kendziora}}, \bibinfo {author} {\bibfnamefont
  {S.}~\bibnamefont {Jian}}, \ and\ \bibinfo {author} {\bibfnamefont {D.~G.}\
  \bibnamefont {Hinks}},\ }\href {\doibase 10.1103/PhysRevLett.87.177007}
  {\bibfield  {journal} {\bibinfo  {journal} {Phys. Rev. Lett.}\ }\textbf
  {\bibinfo {volume} {87}},\ \bibinfo {pages} {177007} (\bibinfo {year}
  {2001})}\BibitemShut {NoStop}%
\bibitem [{\citenamefont {Kordyuk}\ and\ \citenamefont
  {Borisenko}(2006)}]{shen4}%
  \BibitemOpen
  \bibfield  {author} {\bibinfo {author} {\bibfnamefont {A.~A.}\ \bibnamefont
  {Kordyuk}}\ and\ \bibinfo {author} {\bibfnamefont {S.~V.}\ \bibnamefont
  {Borisenko}},\ }\href {\doibase 10.1063/1.2199429} {\bibfield  {journal}
  {\bibinfo  {journal} {Low Temperature Physics}\ }\textbf {\bibinfo {volume}
  {32}},\ \bibinfo {pages} {298} (\bibinfo {year} {2006})}\BibitemShut
  {NoStop}%
\bibitem [{\citenamefont {Kordyuk}(2015)}]{kordyuk2}%
  \BibitemOpen
  \bibfield  {author} {\bibinfo {author} {\bibfnamefont {A.~A.}\ \bibnamefont
  {Kordyuk}},\ }\href {\doibase 10.1063/1.4919371} {\bibfield  {journal}
  {\bibinfo  {journal} {Low Temperature Physics}\ }\textbf {\bibinfo {volume}
  {41}},\ \bibinfo {pages} {319} (\bibinfo {year} {2015})}\BibitemShut
  {NoStop}%
\bibitem [{\citenamefont {He}\ \emph {et~al.}(2014)\citenamefont {He},
  \citenamefont {Yin}, \citenamefont {Zech}, \citenamefont {Soumyanarayanan},
  \citenamefont {Yee}, \citenamefont {Williams}, \citenamefont {Boyer},
  \citenamefont {Chatterjee}, \citenamefont {Wise}, \citenamefont {Zeljkovic},
  \citenamefont {Kondo}, \citenamefont {Takeuchi}, \citenamefont {Ikuta},
  \citenamefont {Mistark}, \citenamefont {Markiewicz}, \citenamefont {Bansil},
  \citenamefont {Sachdev}, \citenamefont {Hudson},\ and\ \citenamefont
  {Hoffman}}]{hoffman}%
  \BibitemOpen
  \bibfield  {author} {\bibinfo {author} {\bibfnamefont {Y.}~\bibnamefont
  {He}}, \bibinfo {author} {\bibfnamefont {Y.}~\bibnamefont {Yin}}, \bibinfo
  {author} {\bibfnamefont {M.}~\bibnamefont {Zech}}, \bibinfo {author}
  {\bibfnamefont {A.}~\bibnamefont {Soumyanarayanan}}, \bibinfo {author}
  {\bibfnamefont {M.~M.}\ \bibnamefont {Yee}}, \bibinfo {author} {\bibfnamefont
  {T.}~\bibnamefont {Williams}}, \bibinfo {author} {\bibfnamefont {M.~C.}\
  \bibnamefont {Boyer}}, \bibinfo {author} {\bibfnamefont {K.}~\bibnamefont
  {Chatterjee}}, \bibinfo {author} {\bibfnamefont {W.~D.}\ \bibnamefont
  {Wise}}, \bibinfo {author} {\bibfnamefont {I.}~\bibnamefont {Zeljkovic}},
  \bibinfo {author} {\bibfnamefont {T.}~\bibnamefont {Kondo}}, \bibinfo
  {author} {\bibfnamefont {T.}~\bibnamefont {Takeuchi}}, \bibinfo {author}
  {\bibfnamefont {H.}~\bibnamefont {Ikuta}}, \bibinfo {author} {\bibfnamefont
  {P.}~\bibnamefont {Mistark}}, \bibinfo {author} {\bibfnamefont {R.~S.}\
  \bibnamefont {Markiewicz}}, \bibinfo {author} {\bibfnamefont
  {A.}~\bibnamefont {Bansil}}, \bibinfo {author} {\bibfnamefont
  {S.}~\bibnamefont {Sachdev}}, \bibinfo {author} {\bibfnamefont {E.~W.}\
  \bibnamefont {Hudson}}, \ and\ \bibinfo {author} {\bibfnamefont {J.~E.}\
  \bibnamefont {Hoffman}},\ }\href {\doibase 10.1126/science.1248221}
  {\bibfield  {journal} {\bibinfo  {journal} {Science}\ }\textbf {\bibinfo
  {volume} {344}},\ \bibinfo {pages} {608} (\bibinfo {year}
  {2014})}\BibitemShut {NoStop}%
\bibitem [{\citenamefont {Peng}\ \emph {et~al.}(2013)\citenamefont {Peng},
  \citenamefont {Meng}, \citenamefont {Mou}, \citenamefont {He}, \citenamefont
  {Zhao}, \citenamefont {Wu}, \citenamefont {Liu}, \citenamefont {Dong},
  \citenamefont {He}, \citenamefont {Zhang}, \citenamefont {Wang},
  \citenamefont {Peng}, \citenamefont {Wang}, \citenamefont {Zhang},
  \citenamefont {Yang}, \citenamefont {Chen}, \citenamefont {Xu}, \citenamefont
  {Lee},\ and\ \citenamefont {Zhou}}]{Peng2013}%
  \BibitemOpen
  \bibfield  {author} {\bibinfo {author} {\bibfnamefont {Y.}~\bibnamefont
  {Peng}}, \bibinfo {author} {\bibfnamefont {J.}~\bibnamefont {Meng}}, \bibinfo
  {author} {\bibfnamefont {D.}~\bibnamefont {Mou}}, \bibinfo {author}
  {\bibfnamefont {J.}~\bibnamefont {He}}, \bibinfo {author} {\bibfnamefont
  {L.}~\bibnamefont {Zhao}}, \bibinfo {author} {\bibfnamefont {Y.}~\bibnamefont
  {Wu}}, \bibinfo {author} {\bibfnamefont {G.}~\bibnamefont {Liu}}, \bibinfo
  {author} {\bibfnamefont {X.}~\bibnamefont {Dong}}, \bibinfo {author}
  {\bibfnamefont {S.}~\bibnamefont {He}}, \bibinfo {author} {\bibfnamefont
  {J.}~\bibnamefont {Zhang}}, \bibinfo {author} {\bibfnamefont
  {X.}~\bibnamefont {Wang}}, \bibinfo {author} {\bibfnamefont {Q.}~\bibnamefont
  {Peng}}, \bibinfo {author} {\bibfnamefont {Z.}~\bibnamefont {Wang}}, \bibinfo
  {author} {\bibfnamefont {S.}~\bibnamefont {Zhang}}, \bibinfo {author}
  {\bibfnamefont {F.}~\bibnamefont {Yang}}, \bibinfo {author} {\bibfnamefont
  {C.}~\bibnamefont {Chen}}, \bibinfo {author} {\bibfnamefont {Z.}~\bibnamefont
  {Xu}}, \bibinfo {author} {\bibfnamefont {T.~K.}\ \bibnamefont {Lee}}, \ and\
  \bibinfo {author} {\bibfnamefont {X.~J.}\ \bibnamefont {Zhou}},\ }\href
  {https://doi.org/10.1038/ncomms3459} {\bibfield  {journal} {\bibinfo
  {journal} {Nature Communications}\ }\textbf {\bibinfo {volume} {4}},\
  \bibinfo {pages} {2459 EP } (\bibinfo {year} {2013})},\ \bibinfo {note}
  {article}\BibitemShut {NoStop}%
\bibitem [{\citenamefont {Balatsky}\ \emph {et~al.}(2006)\citenamefont
  {Balatsky}, \citenamefont {Vekhter},\ and\ \citenamefont
  {Zhu}}]{imp_general}%
  \BibitemOpen
  \bibfield  {author} {\bibinfo {author} {\bibfnamefont {A.~V.}\ \bibnamefont
  {Balatsky}}, \bibinfo {author} {\bibfnamefont {I.}~\bibnamefont {Vekhter}}, \
  and\ \bibinfo {author} {\bibfnamefont {J.-X.}\ \bibnamefont {Zhu}},\ }\href
  {\doibase 10.1103/RevModPhys.78.373} {\bibfield  {journal} {\bibinfo
  {journal} {Rev. Mod. Phys.}\ }\textbf {\bibinfo {volume} {78}},\ \bibinfo
  {pages} {373} (\bibinfo {year} {2006})}\BibitemShut {NoStop}%
\bibitem [{\citenamefont {Kuli\ifmmode~\acute{c}\else \'{c}\fi{}}\ and\
  \citenamefont {Dolgov}(1999)}]{imp_general2}%
  \BibitemOpen
  \bibfield  {author} {\bibinfo {author} {\bibfnamefont {M.~L.}\ \bibnamefont
  {Kuli\ifmmode~\acute{c}\else \'{c}\fi{}}}\ and\ \bibinfo {author}
  {\bibfnamefont {O.~V.}\ \bibnamefont {Dolgov}},\ }\href {\doibase
  10.1103/PhysRevB.60.13062} {\bibfield  {journal} {\bibinfo  {journal} {Phys.
  Rev. B}\ }\textbf {\bibinfo {volume} {60}},\ \bibinfo {pages} {13062}
  (\bibinfo {year} {1999})}\BibitemShut {NoStop}%
\bibitem [{\citenamefont {Kemper}\ \emph {et~al.}(2009)\citenamefont {Kemper},
  \citenamefont {Doluweera}, \citenamefont {Maier}, \citenamefont {Jarrell},
  \citenamefont {Hirschfeld},\ and\ \citenamefont {Cheng}}]{imp_general3}%
  \BibitemOpen
  \bibfield  {author} {\bibinfo {author} {\bibfnamefont {A.~F.}\ \bibnamefont
  {Kemper}}, \bibinfo {author} {\bibfnamefont {D.~G. S.~P.}\ \bibnamefont
  {Doluweera}}, \bibinfo {author} {\bibfnamefont {T.~A.}\ \bibnamefont
  {Maier}}, \bibinfo {author} {\bibfnamefont {M.}~\bibnamefont {Jarrell}},
  \bibinfo {author} {\bibfnamefont {P.~J.}\ \bibnamefont {Hirschfeld}}, \ and\
  \bibinfo {author} {\bibfnamefont {H.-P.}\ \bibnamefont {Cheng}},\ }\href
  {\doibase 10.1103/PhysRevB.79.104502} {\bibfield  {journal} {\bibinfo
  {journal} {Phys. Rev. B}\ }\textbf {\bibinfo {volume} {79}},\ \bibinfo
  {pages} {104502} (\bibinfo {year} {2009})}\BibitemShut {NoStop}%
\bibitem [{\citenamefont {Norman}\ \emph
  {et~al.}(1998{\natexlab{a}})\citenamefont {Norman}, \citenamefont {Ding},
  \citenamefont {Randeria}, \citenamefont {Campuzano}, \citenamefont {Yokoya},
  \citenamefont {Takeuchi}, \citenamefont {Takahashi}, \citenamefont {Mochiku},
  \citenamefont {Kadowaki}, \citenamefont {Guptasarma},\ and\ \citenamefont
  {Hinks}}]{imp_cuprates}%
  \BibitemOpen
  \bibfield  {author} {\bibinfo {author} {\bibfnamefont {M.~R.}\ \bibnamefont
  {Norman}}, \bibinfo {author} {\bibfnamefont {H.}~\bibnamefont {Ding}},
  \bibinfo {author} {\bibfnamefont {M.}~\bibnamefont {Randeria}}, \bibinfo
  {author} {\bibfnamefont {J.~C.}\ \bibnamefont {Campuzano}}, \bibinfo {author}
  {\bibfnamefont {T.}~\bibnamefont {Yokoya}}, \bibinfo {author} {\bibfnamefont
  {T.}~\bibnamefont {Takeuchi}}, \bibinfo {author} {\bibfnamefont
  {T.}~\bibnamefont {Takahashi}}, \bibinfo {author} {\bibfnamefont
  {T.}~\bibnamefont {Mochiku}}, \bibinfo {author} {\bibfnamefont
  {K.}~\bibnamefont {Kadowaki}}, \bibinfo {author} {\bibfnamefont
  {P.}~\bibnamefont {Guptasarma}}, \ and\ \bibinfo {author} {\bibfnamefont
  {D.~G.}\ \bibnamefont {Hinks}},\ }\href {https://doi.org/10.1038/32366}
  {\bibfield  {journal} {\bibinfo  {journal} {Nature}\ }\textbf {\bibinfo
  {volume} {392}},\ \bibinfo {pages} {157 EP } (\bibinfo {year}
  {1998}{\natexlab{a}})}\BibitemShut {NoStop}%
\bibitem [{\citenamefont {Norman}\ \emph
  {et~al.}(1998{\natexlab{b}})\citenamefont {Norman}, \citenamefont {Randeria},
  \citenamefont {Ding},\ and\ \citenamefont {Campuzano}}]{imp_cuprates2}%
  \BibitemOpen
  \bibfield  {author} {\bibinfo {author} {\bibfnamefont {M.~R.}\ \bibnamefont
  {Norman}}, \bibinfo {author} {\bibfnamefont {M.}~\bibnamefont {Randeria}},
  \bibinfo {author} {\bibfnamefont {H.}~\bibnamefont {Ding}}, \ and\ \bibinfo
  {author} {\bibfnamefont {J.~C.}\ \bibnamefont {Campuzano}},\ }\href {\doibase
  10.1103/PhysRevB.57.R11093} {\bibfield  {journal} {\bibinfo  {journal} {Phys.
  Rev. B}\ }\textbf {\bibinfo {volume} {57}},\ \bibinfo {pages} {R11093}
  (\bibinfo {year} {1998}{\natexlab{b}})}\BibitemShut {NoStop}%
\bibitem [{\citenamefont {Franz}\ and\ \citenamefont
  {Millis}(1998)}]{imp_cuprates3}%
  \BibitemOpen
  \bibfield  {author} {\bibinfo {author} {\bibfnamefont {M.}~\bibnamefont
  {Franz}}\ and\ \bibinfo {author} {\bibfnamefont {A.~J.}\ \bibnamefont
  {Millis}},\ }\href {\doibase 10.1103/PhysRevB.58.14572} {\bibfield  {journal}
  {\bibinfo  {journal} {Phys. Rev. B}\ }\textbf {\bibinfo {volume} {58}},\
  \bibinfo {pages} {14572} (\bibinfo {year} {1998})}\BibitemShut {NoStop}%
\bibitem [{\citenamefont {Paramekanti}\ and\ \citenamefont
  {Zhao}(2007)}]{imp_cuprates4}%
  \BibitemOpen
  \bibfield  {author} {\bibinfo {author} {\bibfnamefont {A.}~\bibnamefont
  {Paramekanti}}\ and\ \bibinfo {author} {\bibfnamefont {E.}~\bibnamefont
  {Zhao}},\ }\href {\doibase 10.1103/PhysRevB.75.140507} {\bibfield  {journal}
  {\bibinfo  {journal} {Phys. Rev. B}\ }\textbf {\bibinfo {volume} {75}},\
  \bibinfo {pages} {140507} (\bibinfo {year} {2007})}\BibitemShut {NoStop}%
\bibitem [{\citenamefont {Berg}\ and\ \citenamefont
  {Altman}(2007)}]{imp_cuprates5}%
  \BibitemOpen
  \bibfield  {author} {\bibinfo {author} {\bibfnamefont {E.}~\bibnamefont
  {Berg}}\ and\ \bibinfo {author} {\bibfnamefont {E.}~\bibnamefont {Altman}},\
  }\href {\doibase 10.1103/PhysRevLett.99.247001} {\bibfield  {journal}
  {\bibinfo  {journal} {Phys. Rev. Lett.}\ }\textbf {\bibinfo {volume} {99}},\
  \bibinfo {pages} {247001} (\bibinfo {year} {2007})},\ \bibinfo {note} {and
  references therein.}\BibitemShut {Stop}%
\bibitem [{\citenamefont {Chubukov}\ \emph {et~al.}(2007)\citenamefont
  {Chubukov}, \citenamefont {Norman}, \citenamefont {Millis},\ and\
  \citenamefont {Abrahams}}]{elihu}%
  \BibitemOpen
  \bibfield  {author} {\bibinfo {author} {\bibfnamefont {A.~V.}\ \bibnamefont
  {Chubukov}}, \bibinfo {author} {\bibfnamefont {M.~R.}\ \bibnamefont
  {Norman}}, \bibinfo {author} {\bibfnamefont {A.~J.}\ \bibnamefont {Millis}},
  \ and\ \bibinfo {author} {\bibfnamefont {E.}~\bibnamefont {Abrahams}},\
  }\href {\doibase 10.1103/PhysRevB.76.180501} {\bibfield  {journal} {\bibinfo
  {journal} {Phys. Rev. B}\ }\textbf {\bibinfo {volume} {76}},\ \bibinfo
  {pages} {180501} (\bibinfo {year} {2007})}\BibitemShut {NoStop}%
\bibitem [{\citenamefont {Abrikosov}\ and\ \citenamefont
  {Gor'kov}(1961)}]{abr_gor_1}%
  \BibitemOpen
  \bibfield  {author} {\bibinfo {author} {\bibfnamefont {A.~A.}\ \bibnamefont
  {Abrikosov}}\ and\ \bibinfo {author} {\bibfnamefont {L.~P.}\ \bibnamefont
  {Gor'kov}},\ }\href
  {http://www.jetp.ac.ru/cgi-bin/e/index/e/12/2/p337?a=list} {\bibfield
  {journal} {\bibinfo  {journal} {JETP}\ }\textbf {\bibinfo {volume} {12}},\
  \bibinfo {pages} {227} (\bibinfo {year} {1961})}\BibitemShut {NoStop}%
\bibitem [{\citenamefont {Maki}(1964)}]{maki}%
  \BibitemOpen
  \bibfield  {author} {\bibinfo {author} {\bibfnamefont {K.}~\bibnamefont
  {Maki}},\ }\href {\doibase 10.1103/PhysicsPhysiqueFizika.1.21} {\bibfield
  {journal} {\bibinfo  {journal} {Physics Physique Fizika}\ }\textbf {\bibinfo
  {volume} {1}},\ \bibinfo {pages} {21} (\bibinfo {year} {1964})}\BibitemShut
  {NoStop}%
\bibitem [{\citenamefont {Fulde}\ and\ \citenamefont {Maki}(1965)}]{maki2}%
  \BibitemOpen
  \bibfield  {author} {\bibinfo {author} {\bibfnamefont {P.}~\bibnamefont
  {Fulde}}\ and\ \bibinfo {author} {\bibfnamefont {K.}~\bibnamefont {Maki}},\
  }\href {\doibase 10.1103/PhysRevLett.15.675} {\bibfield  {journal} {\bibinfo
  {journal} {Phys. Rev. Lett.}\ }\textbf {\bibinfo {volume} {15}},\ \bibinfo
  {pages} {675} (\bibinfo {year} {1965})}\BibitemShut {NoStop}%
\bibitem [{\citenamefont {WADA}(1964)}]{maki3}%
  \BibitemOpen
  \bibfield  {author} {\bibinfo {author} {\bibfnamefont {Y.}~\bibnamefont
  {WADA}},\ }\href {\doibase 10.1103/RevModPhys.36.253} {\bibfield  {journal}
  {\bibinfo  {journal} {Rev. Mod. Phys.}\ }\textbf {\bibinfo {volume} {36}},\
  \bibinfo {pages} {253} (\bibinfo {year} {1964})}\BibitemShut {NoStop}%
\bibitem [{\citenamefont {Scalapino}\ \emph {et~al.}(1965)\citenamefont
  {Scalapino}, \citenamefont {Wada},\ and\ \citenamefont {Swihart}}]{maki4}%
  \BibitemOpen
  \bibfield  {author} {\bibinfo {author} {\bibfnamefont {D.~J.}\ \bibnamefont
  {Scalapino}}, \bibinfo {author} {\bibfnamefont {Y.}~\bibnamefont {Wada}}, \
  and\ \bibinfo {author} {\bibfnamefont {J.~C.}\ \bibnamefont {Swihart}},\
  }\href {\doibase 10.1103/PhysRevLett.14.102} {\bibfield  {journal} {\bibinfo
  {journal} {Phys. Rev. Lett.}\ }\textbf {\bibinfo {volume} {14}},\ \bibinfo
  {pages} {102} (\bibinfo {year} {1965})}\BibitemShut {NoStop}%
\bibitem [{\citenamefont {Fink}\ \emph {et~al.}(2006)\citenamefont {Fink},
  \citenamefont {Koitzsch}, \citenamefont {Geck}, \citenamefont {Zabolotnyy},
  \citenamefont {Knupfer}, \citenamefont {B\"uchner}, \citenamefont
  {Chubukov},\ and\ \citenamefont {Berger}}]{fink}%
  \BibitemOpen
  \bibfield  {author} {\bibinfo {author} {\bibfnamefont {J.}~\bibnamefont
  {Fink}}, \bibinfo {author} {\bibfnamefont {A.}~\bibnamefont {Koitzsch}},
  \bibinfo {author} {\bibfnamefont {J.}~\bibnamefont {Geck}}, \bibinfo {author}
  {\bibfnamefont {V.}~\bibnamefont {Zabolotnyy}}, \bibinfo {author}
  {\bibfnamefont {M.}~\bibnamefont {Knupfer}}, \bibinfo {author} {\bibfnamefont
  {B.}~\bibnamefont {B\"uchner}}, \bibinfo {author} {\bibfnamefont
  {A.}~\bibnamefont {Chubukov}}, \ and\ \bibinfo {author} {\bibfnamefont
  {H.}~\bibnamefont {Berger}},\ }\href {\doibase 10.1103/PhysRevB.74.165102}
  {\bibfield  {journal} {\bibinfo  {journal} {Phys. Rev. B}\ }\textbf {\bibinfo
  {volume} {74}},\ \bibinfo {pages} {165102} (\bibinfo {year}
  {2006})}\BibitemShut {NoStop}%
\bibitem [{\citenamefont {Abanov}\ and\ \citenamefont
  {Chubukov}(1999{\natexlab{b}})}]{abanov_ch}%
  \BibitemOpen
  \bibfield  {author} {\bibinfo {author} {\bibfnamefont {A.}~\bibnamefont
  {Abanov}}\ and\ \bibinfo {author} {\bibfnamefont {A.~V.}\ \bibnamefont
  {Chubukov}},\ }\href {\doibase 10.1103/PhysRevLett.83.1652} {\bibfield
  {journal} {\bibinfo  {journal} {Phys. Rev. Lett.}\ }\textbf {\bibinfo
  {volume} {83}},\ \bibinfo {pages} {1652} (\bibinfo {year}
  {1999}{\natexlab{b}})}\BibitemShut {NoStop}%
\bibitem [{\citenamefont {Eschrig}(2006)}]{eschrig}%
  \BibitemOpen
  \bibfield  {author} {\bibinfo {author} {\bibfnamefont {M.}~\bibnamefont
  {Eschrig}},\ }\href {\doibase 10.1080/00018730600645636} {\bibfield
  {journal} {\bibinfo  {journal} {Advances in Physics}\ }\textbf {\bibinfo
  {volume} {55}},\ \bibinfo {pages} {47} (\bibinfo {year} {2006})}\BibitemShut
  {NoStop}%
\bibitem [{\citenamefont {Luttinger}\ and\ \citenamefont {Ward}(1960)}]{lw}%
  \BibitemOpen
  \bibfield  {author} {\bibinfo {author} {\bibfnamefont {J.~M.}\ \bibnamefont
  {Luttinger}}\ and\ \bibinfo {author} {\bibfnamefont {J.~C.}\ \bibnamefont
  {Ward}},\ }\href {\doibase 10.1103/PhysRev.118.1417} {\bibfield  {journal}
  {\bibinfo  {journal} {Phys. Rev.}\ }\textbf {\bibinfo {volume} {118}},\
  \bibinfo {pages} {1417} (\bibinfo {year} {1960})}\BibitemShut {NoStop}%
\bibitem [{\citenamefont {Haslinger}\ and\ \citenamefont
  {Chubukov}(2003)}]{haslinger}%
  \BibitemOpen
  \bibfield  {author} {\bibinfo {author} {\bibfnamefont {R.}~\bibnamefont
  {Haslinger}}\ and\ \bibinfo {author} {\bibfnamefont {A.~V.}\ \bibnamefont
  {Chubukov}},\ }\href {\doibase 10.1103/PhysRevB.68.214508} {\bibfield
  {journal} {\bibinfo  {journal} {Phys. Rev. B}\ }\textbf {\bibinfo {volume}
  {68}},\ \bibinfo {pages} {214508} (\bibinfo {year} {2003})}\BibitemShut
  {NoStop}%
\bibitem [{\citenamefont {Millis}\ \emph {et~al.}(1988)\citenamefont {Millis},
  \citenamefont {Sachdev},\ and\ \citenamefont {Varma}}]{msv}%
  \BibitemOpen
  \bibfield  {author} {\bibinfo {author} {\bibfnamefont {A.~J.}\ \bibnamefont
  {Millis}}, \bibinfo {author} {\bibfnamefont {S.}~\bibnamefont {Sachdev}}, \
  and\ \bibinfo {author} {\bibfnamefont {C.~M.}\ \bibnamefont {Varma}},\ }\href
  {\doibase 10.1103/PhysRevB.37.4975} {\bibfield  {journal} {\bibinfo
  {journal} {Phys. Rev. B}\ }\textbf {\bibinfo {volume} {37}},\ \bibinfo
  {pages} {4975} (\bibinfo {year} {1988})}\BibitemShut {NoStop}%
\bibitem [{agd()}]{agd}%
  \BibitemOpen
  \href@noop {} {\ }\bibinfo {note} {A.\ A.\ Abrikosov, L.\ P.\ Gorkov, and I.\
  E.\ Dzyaloshinski, \textit{Methods of Quantum Feld Theory in Statistical
  Physics}, 2nd edition, (Pergamon Oxford, 1965).}\BibitemShut {Stop}%
\bibitem [{\citenamefont {Wang}\ \emph {et~al.}(2001)\citenamefont {Wang},
  \citenamefont {Mao},\ and\ \citenamefont {Bedell}}]{triplet}%
  \BibitemOpen
  \bibfield  {author} {\bibinfo {author} {\bibfnamefont {Z.}~\bibnamefont
  {Wang}}, \bibinfo {author} {\bibfnamefont {W.}~\bibnamefont {Mao}}, \ and\
  \bibinfo {author} {\bibfnamefont {K.}~\bibnamefont {Bedell}},\ }\href
  {\doibase 10.1103/PhysRevLett.87.257001} {\bibfield  {journal} {\bibinfo
  {journal} {Phys. Rev. Lett.}\ }\textbf {\bibinfo {volume} {87}},\ \bibinfo
  {pages} {257001} (\bibinfo {year} {2001})}\BibitemShut {NoStop}%
\bibitem [{\citenamefont {Roussev}\ and\ \citenamefont
  {Millis}(2001)}]{triplet2}%
  \BibitemOpen
  \bibfield  {author} {\bibinfo {author} {\bibfnamefont {R.}~\bibnamefont
  {Roussev}}\ and\ \bibinfo {author} {\bibfnamefont {A.~J.}\ \bibnamefont
  {Millis}},\ }\href {\doibase 10.1103/PhysRevB.63.140504} {\bibfield
  {journal} {\bibinfo  {journal} {Phys. Rev. B}\ }\textbf {\bibinfo {volume}
  {63}},\ \bibinfo {pages} {140504} (\bibinfo {year} {2001})}\BibitemShut
  {NoStop}%
\bibitem [{\citenamefont {Chubukov}\ \emph {et~al.}(2003)\citenamefont
  {Chubukov}, \citenamefont {Finkel'stein}, \citenamefont {Haslinger},\ and\
  \citenamefont {Morr}}]{triplet3}%
  \BibitemOpen
  \bibfield  {author} {\bibinfo {author} {\bibfnamefont {A.~V.}\ \bibnamefont
  {Chubukov}}, \bibinfo {author} {\bibfnamefont {A.~M.}\ \bibnamefont
  {Finkel'stein}}, \bibinfo {author} {\bibfnamefont {R.}~\bibnamefont
  {Haslinger}}, \ and\ \bibinfo {author} {\bibfnamefont {D.~K.}\ \bibnamefont
  {Morr}},\ }\href {\doibase 10.1103/PhysRevLett.90.077002} {\bibfield
  {journal} {\bibinfo  {journal} {Phys. Rev. Lett.}\ }\textbf {\bibinfo
  {volume} {90}},\ \bibinfo {pages} {077002} (\bibinfo {year}
  {2003})}\BibitemShut {NoStop}%
\bibitem [{\citenamefont {Millis}(1992)}]{millis_05}%
  \BibitemOpen
  \bibfield  {author} {\bibinfo {author} {\bibfnamefont {A.~J.}\ \bibnamefont
  {Millis}},\ }\href {\doibase 10.1103/PhysRevB.45.13047} {\bibfield  {journal}
  {\bibinfo  {journal} {Phys. Rev. B}\ }\textbf {\bibinfo {volume} {45}},\
  \bibinfo {pages} {13047} (\bibinfo {year} {1992})}\BibitemShut {NoStop}%
\bibitem [{\citenamefont {Castellani}\ \emph {et~al.}(1995)\citenamefont
  {Castellani}, \citenamefont {Di~Castro},\ and\ \citenamefont
  {Grilli}}]{ital}%
  \BibitemOpen
  \bibfield  {author} {\bibinfo {author} {\bibfnamefont {C.}~\bibnamefont
  {Castellani}}, \bibinfo {author} {\bibfnamefont {C.}~\bibnamefont
  {Di~Castro}}, \ and\ \bibinfo {author} {\bibfnamefont {M.}~\bibnamefont
  {Grilli}},\ }\href {\doibase 10.1103/PhysRevLett.75.4650} {\bibfield
  {journal} {\bibinfo  {journal} {Phys. Rev. Lett.}\ }\textbf {\bibinfo
  {volume} {75}},\ \bibinfo {pages} {4650} (\bibinfo {year}
  {1995})}\BibitemShut {NoStop}%
\bibitem [{\citenamefont {Perali}\ \emph {et~al.}(1996)\citenamefont {Perali},
  \citenamefont {Castellani}, \citenamefont {Di~Castro},\ and\ \citenamefont
  {Grilli}}]{ital2}%
  \BibitemOpen
  \bibfield  {author} {\bibinfo {author} {\bibfnamefont {A.}~\bibnamefont
  {Perali}}, \bibinfo {author} {\bibfnamefont {C.}~\bibnamefont {Castellani}},
  \bibinfo {author} {\bibfnamefont {C.}~\bibnamefont {Di~Castro}}, \ and\
  \bibinfo {author} {\bibfnamefont {M.}~\bibnamefont {Grilli}},\ }\href
  {\doibase 10.1103/PhysRevB.54.16216} {\bibfield  {journal} {\bibinfo
  {journal} {Phys. Rev. B}\ }\textbf {\bibinfo {volume} {54}},\ \bibinfo
  {pages} {16216} (\bibinfo {year} {1996})}\BibitemShut {NoStop}%
\bibitem [{\citenamefont {Andergassen}\ \emph {et~al.}(2001)\citenamefont
  {Andergassen}, \citenamefont {Caprara}, \citenamefont {Di~Castro},\ and\
  \citenamefont {Grilli}}]{ital3}%
  \BibitemOpen
  \bibfield  {author} {\bibinfo {author} {\bibfnamefont {S.}~\bibnamefont
  {Andergassen}}, \bibinfo {author} {\bibfnamefont {S.}~\bibnamefont
  {Caprara}}, \bibinfo {author} {\bibfnamefont {C.}~\bibnamefont {Di~Castro}},
  \ and\ \bibinfo {author} {\bibfnamefont {M.}~\bibnamefont {Grilli}},\ }\href
  {\doibase 10.1103/PhysRevLett.87.056401} {\bibfield  {journal} {\bibinfo
  {journal} {Phys. Rev. Lett.}\ }\textbf {\bibinfo {volume} {87}},\ \bibinfo
  {pages} {056401} (\bibinfo {year} {2001})}\BibitemShut {NoStop}%
\bibitem [{\citenamefont {Chowdhury}\ and\ \citenamefont
  {Sachdev}(2014{\natexlab{a}})}]{wang_2}%
  \BibitemOpen
  \bibfield  {author} {\bibinfo {author} {\bibfnamefont {D.}~\bibnamefont
  {Chowdhury}}\ and\ \bibinfo {author} {\bibfnamefont {S.}~\bibnamefont
  {Sachdev}},\ }\href {\doibase 10.1103/PhysRevB.90.134516} {\bibfield
  {journal} {\bibinfo  {journal} {Phys. Rev. B}\ }\textbf {\bibinfo {volume}
  {90}},\ \bibinfo {pages} {134516} (\bibinfo {year}
  {2014}{\natexlab{a}})}\BibitemShut {NoStop}%
\bibitem [{\citenamefont {Chowdhury}\ and\ \citenamefont
  {Sachdev}(2014{\natexlab{b}})}]{wang_22}%
  \BibitemOpen
  \bibfield  {author} {\bibinfo {author} {\bibfnamefont {D.}~\bibnamefont
  {Chowdhury}}\ and\ \bibinfo {author} {\bibfnamefont {S.}~\bibnamefont
  {Sachdev}},\ }\href {\doibase 10.1103/PhysRevB.90.245136} {\bibfield
  {journal} {\bibinfo  {journal} {Phys. Rev. B}\ }\textbf {\bibinfo {volume}
  {90}},\ \bibinfo {pages} {245136} (\bibinfo {year}
  {2014}{\natexlab{b}})}\BibitemShut {NoStop}%
\bibitem [{\citenamefont {Wang}\ and\ \citenamefont {Chubukov}(2015)}]{wang23}%
  \BibitemOpen
  \bibfield  {author} {\bibinfo {author} {\bibfnamefont {Y.}~\bibnamefont
  {Wang}}\ and\ \bibinfo {author} {\bibfnamefont {A.~V.}\ \bibnamefont
  {Chubukov}},\ }\href {\doibase 10.1103/PhysRevB.92.125108} {\bibfield
  {journal} {\bibinfo  {journal} {Phys. Rev. B}\ }\textbf {\bibinfo {volume}
  {92}},\ \bibinfo {pages} {125108} (\bibinfo {year} {2015})}\BibitemShut
  {NoStop}%
\bibitem [{\citenamefont {Bardeen}\ and\ \citenamefont
  {Stephen}(1964)}]{Bardeen}%
  \BibitemOpen
  \bibfield  {author} {\bibinfo {author} {\bibfnamefont {J.}~\bibnamefont
  {Bardeen}}\ and\ \bibinfo {author} {\bibfnamefont {M.}~\bibnamefont
  {Stephen}},\ }\href {\doibase 10.1103/PhysRev.136.A1485} {\bibfield
  {journal} {\bibinfo  {journal} {Phys. Rev.}\ }\textbf {\bibinfo {volume}
  {136}},\ \bibinfo {pages} {A1485} (\bibinfo {year} {1964})}\BibitemShut
  {NoStop}%
\bibitem [{\citenamefont {Eliashberg}(1960)}]{eliashberg}%
  \BibitemOpen
  \bibfield  {author} {\bibinfo {author} {\bibfnamefont {G.~M.}\ \bibnamefont
  {Eliashberg}},\ }\href
  {http://www.jetp.ac.ru/cgi-bin/e/index/e/11/3/p696?a=list} {\bibfield
  {journal} {\bibinfo  {journal} {JETP}\ }\textbf {\bibinfo {volume} {11}},\
  \bibinfo {pages} {696} (\bibinfo {year} {1960})}\BibitemShut {NoStop}%
\bibitem [{foo()}]{footnote}%
  \BibitemOpen
  \href@noop {} {\ }\bibinfo {note} {Note that $\Delta^R(\omega) \propto
  i\omega$ should not be confused with odd-frequency pairing. When expressed in
  time-ordered operators at real frequencies, $\Delta(\omega) \propto
  i|\omega|$ which is even in frequency.}\BibitemShut {Stop}%
\bibitem [{\citenamefont {Chubukov}\ \emph {et~al.}(2016)\citenamefont
  {Chubukov}, \citenamefont {Eremin},\ and\ \citenamefont {Efremov}}]{cee}%
  \BibitemOpen
  \bibfield  {author} {\bibinfo {author} {\bibfnamefont {A.~V.}\ \bibnamefont
  {Chubukov}}, \bibinfo {author} {\bibfnamefont {I.}~\bibnamefont {Eremin}}, \
  and\ \bibinfo {author} {\bibfnamefont {D.~V.}\ \bibnamefont {Efremov}},\
  }\href {\doibase 10.1103/PhysRevB.93.174516} {\bibfield  {journal} {\bibinfo
  {journal} {Phys. Rev. B}\ }\textbf {\bibinfo {volume} {93}},\ \bibinfo
  {pages} {174516} (\bibinfo {year} {2016})}\BibitemShut {NoStop}%
\bibitem [{Sca()}]{Scalapino_92}%
  \BibitemOpen
  \href@noop {} {\ }\bibinfo {note} {D.J. Scalapino in
  Ref.\onlinecite{review}.}\BibitemShut {Stop}%
\end{thebibliography}%

\end{document}